\documentclass[10pt,aps,pra,amsmath,amssymb,notitlepage,superscriptaddress]{revtex4-2}
\usepackage{tabularx}
\usepackage{eucal}
\usepackage{mathrsfs}
\usepackage[T1]{fontenc}
\usepackage[utf8x]{inputenc}
\usepackage[english]{babel}

\usepackage[sfdefault,lf]{carlito}
\usepackage[parfill]{parskip}

\usepackage{fancyhdr}
\usepackage{blindtext}

\usepackage{amsmath}
\usepackage{graphicx}
\usepackage{booktabs}
\usepackage[a4paper,top=3cm,bottom=2cm,left=3cm,right=3cm,marginparwidth=1.75cm]{geometry}

\usepackage[colorinlistoftodos]{todonotes}
\usepackage{braket}
\usepackage{appendix}
\usepackage{amssymb}
\usepackage{bbm}
\usepackage{mathtools}

\DeclarePairedDelimiter\floor{\lfloor}{\rfloor}

\newcommand{\A}{\mathscr{A}}
\newcommand{\F}{\mathscr{F}}
\newcommand{\C}{\mathscr{C}}
\newcommand{\G}{\mathscr{G}}

\usepackage{chngcntr}

\usepackage[compact]{titlesec}         
\titlespacing{\section}{5pt}{5pt}{5pt} % this reduces space between (sub)sections to 0pt, for example
\AtBeginDocument{%% this will reduce spaces between parts (above and below) of texts within a (sub)section to 0pt, for example - like between an 'eqnarray' and text
\setlength\abovedisplayskip{4pt}
\setlength\belowdisplayskip{4pt}}

\usepackage{physics}
\usepackage{siunitx}
\usepackage[colorlinks=true, allcolors=blue]{hyperref}

\begin{document}

\author{C. Javaherian}
\affiliation{Centre for Quantum Software and Information, University of Technology Sydney, Australia} 

%\author{G. A. Paz-Silva}
%\affiliation{Centre for Quantum Dynamics, Griffith University, Brisbane, Australia}

\author{C. Ferrie}
\affiliation{Centre for Quantum Software and Information, University of Technology Sydney, Australia} 

\title{Quantum noise spectroscopy by qudit spectators}

\begin{abstract}
Quantum noise spectroscopy is a well-known method for detecting environmental noise spectrum and has various applications in quantum sensing, quantum network design, and quantum computing. In this work, a protocol for quantum noise spectroscopy by qudit spectators (arbitrary d-level quantum systems) is presented for noises with stochastic and stationary Gaussian distributions. The protocol uses Weyl basis decomposition, consisting of generalized Pauli matrices, as the theoretical tool, and the Alvarez-Suter noise spectroscopy method, by which the qudit is being exposed to a series of pulse sequences, and the noise polyspectra is extracted by multiple measurements of an observable of the qudit.
%in which the qudit is exposed to a series of spectroscopy pulse sequences, and an observable of the qudit is measured consequently, to reveal the noise polyspectra surrounding the qudit.  
The formalism is further developed and simulated for three theoretical frameworks:
The first two frameworks utilize a simplified Weyl basis and are capable of detecting Z-type dephasing noise. 
One is specifically modeled for a qutrit (d=3), and the second framework is designed for all qudits and has been tested through simulations of a ququad (d=4) and a quoct (d=8).
The third framework employs the full Weyl basis, enabling it to identify all types of XYZ-dephasing noise. The simulations for this framework have been conducted for a quoct (d=8) implemented by the Antimony atom.
The analytical protocol for noise spectroscopy that has been introduced transforms qudit systems into quantum spectators, and the presented simulations demonstrate the accuracy and reliability of this noise spectroscopy protocol. 
\end{abstract}

\maketitle

%\tableofcontents

\section{Introduction}
The surrounding of a quantum device has inevitable influence on its performance. Knowing the environmental effects allows them to be canceled or manipulated to optimize the system's efficiency. For example, in noise assisted transport, the quantum system’s environment is engineered to maximize the transport efficiency through the system \cite{Plenio_2008, PhysRevA.90.042313}. On the other hand, one could prevent or counteract the known environmental effects to sustain the efficiency of a designed system. This is a prerequisite for achieving noise optimized design in fault-tolerant quantum information processing (QIP) architectures \cite{PhysRevA.79.032318, PhysRevLett.110.010502} and quantum error correction \cite{Boixo2018, Johnson2017}. Acquiring knowledge about the influence of the environment on a quantum system could also be used for sensing purposes through measuring the effect of the surrounding material on the energy levels or, equivalently, interacting frequencies of a quantum spectator \cite{ PhysRevLett.92.117905}. Quantum sensing and metrology have a broad range of applications \cite{PhysRevLett.96.010401, PhysRevLett.104.133601, PhysRevA.87.022324, PhysRevA.92.010302, PhysRevApplied.5.014007}. \\
%\textit{Solutions and methods}- 

Various quantum noise spectroscopy (QNS) protocols have been introduced to extract information about a quantum system’s environment. They mostly consist of the following parts: preparing the quantum system in a known state, letting it evolve under the influence of the environment (bath) and a series of engineered pulse sequences, measuring some system measurable operators, and, according to the measurement results, extracting information about the bath \cite{PhysRevA.95.022121}.  
Due to the nature of measurements in quantum mechanics, where expectation values are achieved by repeatedly measuring a given observable, the aforementioned  information on the bath is at best statistical in nature. That is, the impact of the largely unknown and inaccessible bath is captured through suitably defined correlation functions of the operators describing the characteristics of the system influenced by the environment. In the simplifying case in which the noise is Gaussian, for example, one is formally interested in characterizing the first and second order cumulants of the noise process, as they fully describe the desired statistical properties.

In parametric and non-parametric spectral estimation, some or no prior information about the noise spectra are known. Examples of non-parametric QNS are multitaper and Slepian-based spectral estimation \cite{1456701, Fray2017, Fray2020} which use discrete prolate spheroidal sequences to envelope the applied fields and optimally reduce the spectral leakage \cite{6773659,  6773660, 6773467, 6773515, 6771595, 1083556}. Dynamical decoupling or DD-based QNS techniques like $N$-pulse CPMG spectroscopy are well known approaches for QIP applications. The frequency-comb technique is formed by repeating a fixed DD sequence such that its corresponding filter function approximates a frequency comb. In the original Alvarez-Suter QNS approach, the qubit decay rate for different frequency-comb sequences is measured. This results in a linear inversion problem, with the solution of the noise spectrum \cite {PhysRevLett.107.230501,PhysRevA.98.032315}.
%\textit{Quantum spectators}- 

The QNS protocols based on qubits have been investigated through several theoretical and experimental models and platforms: Experimentally, the one qubit QNS protocols reconstructing Gaussian (Z-type) dephasing noise spectra have been implemented for example in solid-state NMR \cite{PhysRevLett.107.230501}, superconducting and spin qubits \cite{PhysRevB.89.020503, PhysRevLett.110.146804} and nitrogen vacancy centers in diamond \cite{ PhysRevA.90.032319, PhysRevLett.114.017601}. Different theoretical QNS protocols have been also introduced for  Z-type dephasing nose, for  example the QNS protocols for the Gaussian/non-Gaussian stationary dephasing environment around a single qubit \cite{PhysRevLett.116.150503}, or a Gaussian dephasing spectra around two or multi-qubit systems \cite{PhysRevA.94.012109, PhysRevA.95.022121} that besides noise spectra could sense the bosonic bath temperature and the coupling of the qubits to the oscillatory bath \cite{ PhysRevA.95.022121}. \\

In the present work, our aim is to expand upon existing comb-based QNS protocols. Our method of achieving this is through the use of the Weyl basis decomposition. This mathematical approach allows for a more detailed characterization of environmental noises that affect quantum systems. One significant advantage of employing the Weyl decomposition in our mathematical framework is its flexibility. While previous methods primarily focused on identifying and analyzing (Z-type) dephasing noise, our approach, with the aid of Weyl decomposition, offers the capacity to detect and understand a broader range of noise types. Specifically, we can now address all categories of XYZ-dephasing noise including Z-type dephasing, encompassing a comprehensive range of disturbances that might affect quantum operations.

Another characteristic of our QNS protocol is that the physical system under consideration is a qu\emph{dit} --- a d-level quantum system that despite its presence in the quantum world, is not investigated thoroughly in the QNS literature. By employing qudits, we are able to identify high-order noise spectra, termed as polyspectra \cite{PhysRevLett.116.150503}. Their enhanced sensitivity to environmental noises is attributed to the d distinct energy levels of qudits. These levels can engage with a broader range of environmental frequencies than a solitary qubit or even a d-qubit system. In addition to amplifying the sensitivity through the use of qudit systems, we increase the compactness of the QNS system when contrasted with using a chain of various two-level qubit spectators. For instance, all the energy levels of an atomic qudit are concentrated in a single location, unlike their multi-qubit counterparts which are dispersed. This centralized nature can offer advantages on specific platforms and devices.

Our frequency-comb QNS protocol comes with certain assumptions and considerations. Firstly, within our QNS framework, we supposed that the noise spectra are stationary with zero-mean Gaussian distributions. This assumption aligns well with quantum systems that have weak interactions with non-sparse environments \cite{1083556}. Another aspect to specify is the desired resolution of the extracted noise spectra. In our method, enhancing the resolution can be achieved by decreasing the duration of the reference pulse sequence utilized during QNS. If there are concerns about identifying fine spectral functions or pinpointing spur-like components within the comb-frequency formalism \cite{PhysRevA.97.032101}, it is advisable to magnify the resolution in our protocol to adequately capture these structures. Moreover, it is essential to set the upper limit of detectable frequency (or the passband) sufficiently high in our protocol to prevent spectral leakage, an issue worth noting \cite{PhysRevX.5.021009}.\\

This paper is organized as follows. Section II provides a theoretical modeling of our QNS protocol for qudit systems using the Complete Weyl basis decomposition. This section also derives the switching functions associated with a reference pulse sequence and presents measurement results in the Weyl basis that relate to unidentified noise spectra, specifically the XYZ-dephasing noises which are stochastic and have a zero-mean Gaussian distribution. Detailed calculations related to this section are provided in the appendices. In Section III, we introduce a streamlined protocol using the Reduced Weyl basis tailored for qutrit systems, alongside simulations of retrieved Z-type dephasing noise spectra for a sample qutrit. Section IV brings another version of the protocol through Reduced Weyl decomposition for all qudits, emphasizing on Z-type dephasing noises, and is tested on both a ququad (d=4) and a quoct (d=8). Section V showcases the protocol in its most expansive form using the Complete Weyl basis, applied to the Antimony spin quoct system, followed by a simulation of environmental XYZ-dephasing noise polyspectra. Finally, Section VI provides a summary of the findings.\\

\section{Modeling: the qudit, noises, and pulse sequences}
Suppose a qudit characterized by $d$ distinguished energy levels or eigenstates $\{\ket{i},\:i \in S=\{0,...,d-1\}\}$ is influenced by specific environmental noises. 
We will consider a unitary evolution for the closed system of noise and qudit, thus excluding any assumptions of particle loss, such as dissipation or thermal noises. 
In our model of noise spectroscopy, we account for stochastic noises affecting the eigenenergies and coherences. Such decoherence or dephasing noise follows zero-mean Gaussian probability distribution and result from, for instance, elastic collisions between an ion qudit and environmental particles or quasiparticles. 
On the other hand, inelastic collisions, which may cause dissipative or thermal noises, lead to th the loss of particles or energy from the system.
The net effect is a qudit whose internal energies and coherences are fluctuating and time-dependent, as captured by the effective Hamiltonian in the computational basis:
\begin{equation}\label{Hcomputationalbasis}
H=\sum_{n,m\in S}\varepsilon_{nm}(t)\ket{n}\bra{m},  
\end{equation}
where the energies and coherences $\varepsilon_{nm}(t)=\varepsilon_{nm}^0+\delta\varepsilon_{nm}(t)$ include a static term $\varepsilon_{nm}^0$  %-- related to the structure of qudit's internal energies and the time-independent effects of the surrounding noises-- 
and an stochastic term $\delta\varepsilon_{nm}(t)$. % induced by the noise processes. 
In the classical noise model, while the evolution is unitary for every noise trajectory, decoherence originates when one estimates an expectation value by taking the average $\langle \cdot \rangle$ over many such trajectories. 
% This simplifies calculating the expectation values of the probe system's observables in the quantum noise spectroscopy problem.
It will be convenient to write this Hamiltonian in Weyl's basis. This basis, $\{Z^aX^b\}_{a,b \in S}$, is composed of 
$Z$ and $X$ operators that are defined by their action on the $d$-dimensional computational basis $\{\ket{i},\:i \in S\},$ namely:
\begin{align}\label{generalizedPauliMatrices}
\begin{split}
Z^{a}\ket{i}&=\xi^{ai}\ket{i},\\
X^a\ket{i} &=\ket{i\oplus a},    
\end{split}
\end{align}
where $ \xi = e^{2\pi I/d}$ is the $d^{\rm th}$ root of unity, $I = \sqrt{-1}$, and $i\oplus a=(i+a)$ \textit{modulo} $d$. The Weyl basis can be interpreted as the direct higher dimension generalization of the Pauli basis for qubits, as its generators satisfy the commutation relation $Z^a X^b=\xi ^{ab} X^b Z^a$. 
In this language, expanding the Hamiltonian of a qu\textbf{\textit{d}}it surrounded by only  (Z-type) dephasing noise leads to an expansion solely on $Z^a$ operators, which can be seen as a generalization of a typical qu\textbf{\textit{b}}it (Z-type) dephasing model, i.e. $H=\beta(t) \sigma_z$ for a spin qubit experiencing dephasing noise with $\sigma_z$ as Pauli $Z$ operator. For convenience in what follows, we will call the group $\{Z^a\}_{a\in S}$ the \textit{reduced} Weyl basis (RW) with Z-type dephasing noise, and the general Weyl basis (W) as the group $\{Z^aX^b\}_{a,b \in S}$ with the XYZ-dephasing noise.  Writing the Hamiltonian of Eq.\eqref{Hcomputationalbasis} in Reduced Weyl and full Weyl basis, one finds that 
\begin{align}\label{HWRW}
\begin{split}
H^{RW} &=\sum_{a \in S}\beta_{a}(t)Z^a,\\
H^{W}  &=\sum_{a,b \in S}\beta_{ab}(t)Z^a X^b
\end{split}
\end{align}
where $\beta_{a}(t)$ and $\beta_{ab}(t)$ are stochastic scalar functions representing the effect of noises in time domain. 
For consistency with future notation, we unify the above formulas as follows:
\begin{align}\label{HfullWeyldephasing}
\begin{split}
H^{\mathscr{A}^n_H}_{q,n}(t) =& \sum_{\mathscr{A}^n_{\sum}\in S} \beta_{\mathscr{A}^n_{\beta}}(t) Z^{\mathscr{A}^n_z} X^{\mathscr{A}^n_x},
\end{split}
\end{align}
where the calligraphic $A_{var}$ $(\mathscr{A}_{\rm var})$ is the index or power of the variable $var$, and the upper-index $n$ in $\mathscr{A}^n$ indicates that only \textit{\textbf{n}oise}, and not any pulse, is affecting the qudit. 
Table \ref{table1} defines the $\mathscr{A}_{\rm var}^n$ indices/powers with $var \in \{  H,X,Z,\beta,\sum   \}$, and the columns W (Weyl) and RW (Reduced Weyl) correspond to Eq.\eqref{HWRW} (lower part) with $(a,b) \rightarrow (i_1,i_0)$ and Eq.\eqref{HWRW} (upper part) with $(a) \rightarrow (i_0)$, respectively. \\
\begin{table}[ht!]
\centering
%\printglossary[title={Abbreviations},type=acronym,style=long]
%\setcounter{table}{0}
\caption{Indices/Powers $\mathscr{A}^n_i$}
\begin{tabularx}{0.5\textwidth} 
{
  | >{\centering\arraybackslash}X 
  | >{\centering\arraybackslash}X
  | >{\centering\arraybackslash}X
  | >{\centering\arraybackslash}X | }
\hline 
$\mathscr{A}^n_H$ & $W$ & $RW$ \\
\hline
$\mathscr{A}^n_X$ & $i_0$ &  $0$ \\
\hline
$\mathscr{A}^n_Z$     & $i_1$  & $i_0$  \\
\hline
$\mathscr{A}^n_{\beta}$  &  $i_0,i_1$   &  $i_0$ \\
\hline
$\mathscr{A}^n_{\sum}$   &  $i_0,i_1$    &  $i_0$ \\
\hline
\end{tabularx}
\label{table1}
\end{table}

Due to the Hermitian characteristic of the Hamiltonian, the $\beta_{\A_{\beta}}(t)$ dephasing noise function follows the relation (appendix \eqref{apdx.3.6}, Eq.\eqref{beta_condition-0}):
\begin{align}\label{betaHermitian}
\begin{split}
\beta_{\mathscr{A}^n_{\beta}}(t) &=
\xi^{-{\A}^n_{Z} {\A}^n_{X} } \beta^*_{-\mathscr{A}^n_{\beta}}(t)
\end{split}
\end{align}
Equating Eqs.\eqref{Hcomputationalbasis} and \eqref{HfullWeyldephasing}, we find that $\beta_{\A_{\beta}}(t)$ is related to the time dependent energies of the noisy qudit $\varepsilon_{nm}(t)$ as follows: (appendix \ref{apdx.3.1}, Eqs.\eqref{C1-betaEpsilonMain} and \eqref{C1-EpsilonBetaMain})
\begin{align}\label{betaEnery}
\begin{split}
\beta_{ab}(t) &= \frac{1}{d}\sum_{i,j \in S} \xi^{-a i} \delta_{b,i-j} \varepsilon_{ij}(t),\:\:\:\:\: \varepsilon_{nm}(t) = 
\sum_{a,b \in S} \xi^{an}  \delta_{n,m \oplus b} \beta_{ab}(t)
\end{split}
\end{align}

We continue the noise spectroscopy modeling by following the long tradition going back to the early days of NMR, we extract information about the qudit and its environment with sequences of control pulses. That is, we suppose that the qudit surrounded by dephasing noise possessing the Hamiltonian of Eq.\eqref{HfullWeyldephasing} is prepared in a known initial state. 
We will assume the ability to apply resonant pulses 
\begin{equation}\label{unitarypulse}
P_{(i,j)} =\ket{i}\bra{j} + \ket{j}\bra{i} + \mathbbm{1} - \ket{i}\bra{i} - \ket{j}\bra{j},  
\end{equation}
which implement a transition between states $\ket{i}$ and $\ket{j}$, at arbitrary time $t$. More specifically, we will consider
$2\gamma$ resonant pulses 
$\{P_{(i_k,j_k)}\}_{k=1,...,2\gamma}$ at time instances $\{ t_k\}_{k=1,...,2\gamma}$, where $t_{1}=0$ and $t_{2\gamma}=T$ is the total duration of the reference pulse sequence. The indices $i_k$ and $j_k$ mark the energy levels of the transition being addressed. Resonant and unitary pulses of Eq.\eqref{unitarypulse} have the following characteristics:
\begin{equation}\label{PulseCharacteristics}
P_{(i_k,j_k)}^{-1} = P_{(i_k,j_k)}^{*} = P_{(j_k,i_k)} = P_{(i_k,j_k)} 
\end{equation}
We assume a reference pulse sequence for our formalism that is \textit{symmetric} such that the sequence is made up of pairs of identical pulses as follows:
\begin{equation}\label{symmetricsequence0}
\{ ( P^{-1}_{(i_k,j_k)}  ,  P_{(i_k,j_k)}  )  \}_{k=1,...,\gamma}=\{ ( P_{(i_1,j_1)}^{-1},P_{(i_1,j_1)} ) ,\: (P_{(i_2,j_2)}^{-1},P_{(i_2,j_2)} ),\: \ldots,\: (P_{(i_{\gamma},j_{\gamma})}^{-1},P_{(i_{\gamma},j_{\gamma})} )\}  
\end{equation}
where the pair pulses are applied at time instances $\tilde{t}_{k}$ and $t_k$, respectively, separated with interval $\tau_k=t_k  - \tilde{t}_{k} $, and the second pulse of each pair is simultaneous with the first pulse of the next pair, i.e. $t_k = \tilde{t}_{k+1}$. \\
From now, we refer to the above reference pulse sequence by the following simplified format:
\begin{equation}\label{symmetricsequence}
\{  \tilde{P}_{(i_k,j_k)}   \}_{k=1,...,\gamma}=\{ ( \tilde{P}_{(i_1,j_1)} ,\: \tilde{P}_{(i_2,j_2)} ,\: \ldots,\: \tilde{P}_{(i_{\gamma},j_{\gamma})}\}  
\end{equation}
We also consider the condition $j_k=i_{k+1}$, which leads to the relationship,
\begin{equation}\label{ResonanPulseCondition}
P_{(i_k,j_k)} P^{-1}_{(j_k,j_{k+1})}
=P_{(i_k,j_{k+1})}    
\end{equation}
As one illustration, we will take into account the condition of assuming $i_k \rightarrow k: 0,...,d-1$ 
%and utilizing in our noise spectroscopy of d-level qudit, 
that transforms  Eq.\eqref{symmetricsequence} into a reference sequence of $d$ resonant pulse pairs as follows:
\begin{equation}\label{ReferencePulseSequence}
\{\tilde{P}_{(i,i+1)}\}_{i=0,...,d-1}=\{\tilde{P}_{0,1},\tilde{P}_{1,2},\tilde{P}_{2,3},...,\tilde{P}_{d-1,0}\},
\end{equation}
As the second illustration, we assume the following reference pulse sequence:
\begin{equation}\label{ReferencePulseSequence1}
\{\tilde{P}_{(i_r,j_r)}\}_{r=0,...,d-1}   
\end{equation}
where $(i_r,j_r)$ are randomly chosen from the set $S$, while maintaining the following condition:
\begin{equation}\label{Sequencecondition3}
P_{(i_0,j_0)} = P_{(j_{d-1},i_{d-1})}   
\end{equation}

Now we find the effective Hamiltonian of a qudit, noise and pulse sequence of the form of Eqs.\eqref{ReferencePulseSequence} or \eqref{ReferencePulseSequence1} for the general Weyl decomposed case (W) at interval $\tau_r$ as follows (appendix \ref{apdx.A.1}):
\begin{align}\label{1.0}
\begin{split}
H^{W}(t\in \tau_r) 
&=\sum_{a,b,a',b'\in S}\beta_{ab}(t) y_{a'b'}^{a b}(t) Z^{a'}X^{b'},
\end{split}
\end{align}
where $y_{a'b'}^{a b}(t)$, the switching function, is the Weyl coefficient corresponding the pulses and has constant values within each interval of the duration of reference sequence as below:
\begin{align}\label{1.0.switchingfunction}
\begin{split}
y_{a'b'}^{a b} &(t\in \tau_r) = y_{a' b'}^{i_r j_r a b} = y_{a' b'}^{i j a b},\\
y_{a',0}^{i j a b} =& \frac{-1}{d}(\xi^{-a' i}+\xi^{-a' j})(\delta_{j,i+ b}\xi^{a j}+\delta_{i,j+ b}\xi^{a i}),\\
y_{a',i-j}^{i j a b} =& \frac{1}{d}\xi^{-a' i}(\delta_{j,i +b} \xi^{a j}+\delta_{i,j +b}\xi^{a i}),\\
y_{a',j-i}^{i j a b} =& \frac{1}{d}\xi^{-a' j}(\delta_{j,i +b} \xi^{a j}+\delta_{i,j +b}\xi^{a i}),\\
y_{a',i-j+b}^{i j a b} =& \frac{1}{d}(\xi^{(a-a')(i+b)}+\xi^{-a'i+aj}),\\
y_{a',j-i +b}^{i j a b} =& \frac{1}{d}(\xi^{(a-a')(j+b)}+\xi^{-a'j+ai}),\\
y_{a',b}^{i j a b} =& \frac{1}{d}(-(\xi^{(a-a')i}+\xi^{(a-a')j})(\xi^{(a-a')b}+1)+\xi^{-a'i}d\:\delta_{a,a'}),\\
y_{a',b'\notin S'}^{i j a b} =& 0,\:S'=\{ 0,i-j, j-i, i-j+b, j-i+b, b \}\in S
\end{split}
\end{align}
The indices $i_r,j_r$ or $i,j$ in the first line correspond to the pulse $P_{i_r,j_r}$ applied in the beginning of the interval $\tau_r$. For the Reduced Weyl case (RW), the effective Hamiltonian of the qudit, noise, and pulses within the $r^{th}$ interval is found as following (appendix \ref{apdx.A.2}): 
\begin{align}\label{HRW}
\begin{split}
H^{RW}(t\in \tau_r) & = P_{i_r,j_r}^{-1} H P_{i_r,j_r} = \sum_{a,m\in S}\beta_{a}(t) y_{m,a}(t)Z^m,\\
y_{m,a}(t\in \tau_r) &= \delta_{m,a} +\frac{1}{d}(\xi^{-mj_r} - \xi^{-mi_r})(\xi^{a i_r} - \xi^{a j_r}).
\end{split}    
\end{align}
The scalar switching function $y_{m,a}(t)$ contains information about the characteristics of the reference pulse sequence and its constant value varies for each interval $\tau_r$.
It is shown that at least for a qutrit system ($d=3$) noted as (QT), the effective Hamiltonian and the switching function in the Reduced Weyl  case has the following format (appendix \ref{apdx.A.3}):
\begin{align}\label{HQtrit}
\begin{split}
H^{QT}(t\in \tau_r)=&\sum_{a \in S}\beta_{-a}(t)y_{-a}(t)Z^a,\\
y_{a} (t\in \tau_1) = 1,\:&y_{a}(t\in \tau_2)=\xi^{-a},\:y_{a} (t\in \tau_3) = \xi^{a}.
\end{split}    
\end{align}

We assume the following indexing rule to unify the different representations of the effective Hamiltonian of qudit, noise, pulses in Wyel, Reduced Weyl, and Qutrit forms as follows:
\begin{align}\label{Hgeneral}
\begin{split}
H^{\mathscr{A}_H}_{(q,n,p)}(t) =& \sum_{\mathscr{A}_{\sum}\in S} \beta_{\mathscr{A}_{\beta}}(t) y_{\mathscr{A}_y}(t) Z^{\mathscr{A}_z} X^{\mathscr{A}_x},
\end{split}
\end{align}
where $(q,n,p)$, the lower index of $H$, indicates that the qudit (q) is affected by noise (n) and pulses (p). 
Table \ref{table2} defines the $\mathscr{A}_i$ indices/powers with $i\in\{H,X,Z,y,\beta,\sum \}$, and  the columns W  (Weyl), RW  (Reduced Weyl), and RW (Qutrit) in the table  correspond to Eq.\eqref{1.0} with $(a,b,a',b') \rightarrow (i_3,i_2,i_1,i_0)$, Eq.\eqref{HRW} with $(a,m)\rightarrow (i_1,i_0)$, and Eq.\eqref{HQtrit} with $(a)\rightarrow (i_0)$, respectively.\\
\begin{table}[ht!]
\centering
\caption{Indices/Powers $\mathscr{A}_i$}
\begin{tabularx}{0.5\textwidth} 
{ 
  | >{\centering\arraybackslash}X 
  | >{\centering\arraybackslash}X
  | >{\centering\arraybackslash}X
  | >{\centering\arraybackslash}X | }
\hline 
$\mathscr{A}_H$ & $W$ & $RW$ & $Qutrit$ \\
\hline
$\mathscr{A}_X$ & $i_0$ &  $0$ & $0$ \\
\hline
$\mathscr{A}_Z$     & $i_1$  & $i_0$ & $i_0$ \\
\hline
$\mathscr{A}_y$  & $i_0,i_1,i_2,i_3$  & $i_0,i_1$ & $-i_0$ \\
\hline
$\mathscr{A}_{\beta}$  &  $i_2,i_3$   &  $i_1$ & $-i_0$\\
\hline
$\mathscr{A}_{\sum}$   &  $i_0,i_1,i_2,i_3$    &  $i_0,i_1$ & $i_0$\\
\hline
\end{tabularx}
\label{table2}
\end{table}

The Hermitian characteristic of the Hamiltonian of Eq.\eqref{Hgeneral} implies the following conditions on the switching functions (appendix \ref{apdx.3.6}, Eq.\eqref{yconjugate}):
\begin{align}\label{y-conditions}
\begin{split}
y_{\mathscr{A}_y}(t) &= \xi^{ {\A}_{Z}^n {\A}_{X}^n - {\A}_{Z} {\A}_{Z}} y^*_{-\mathscr{A}_y}(t).
\end{split}
\end{align}
Now we find the expectation value of an arbitrary qudit observable $\langle\hat{O}(t)\rangle$ in terms of noise and switching functions. Here is the general Weyl decomposition of the arbitrary measurable observable and the initial density matrix of the qudit:
\begin{equation}\label{Weyldecompositions}
\langle\hat{O}(t)\rangle=\sum_{m,n \in S} O_{mn}\langle Z^m X^n\rangle,\: \hat{\rho}(0)=\sum_{p,q\in S} V_{pq}Z^pX^q.
\end{equation}
Given the effective Hamiltonian for the unitary evolution of qudit, the expectation value of Weyl operators are found using Dyson series (appendix \ref{apdx.2.1}, Eq.\eqref{B.1.2}):
\begin{equation}\label{1.2}
\begin{split}
\langle\langle Z^m X^n\rangle\rangle= G_{mn} -\frac{1}{2}\int_0^t\int_0^t dt' dt^{\prime\prime}\sum_{\A_{\sum},\tilde{\A}_{\sum}\in S} J_{m n\A_J}
\langle\beta_{\A_{\beta}}(t')\beta_{\tilde{\A}_{\beta}}(t^{\prime\prime})\rangle y_{\A_y}(t')y_{\tilde{\A}_y}(t^{\prime\prime}),
\end{split}
\end{equation}
where, 
\begin{align}\label{GJ}
\begin{split}
G_{mn} &= d\xi^{mn}V_{id-m,id-n},\\
J_{m n \A_J} &= d \sum_{p,q\in S} 
(1-\xi^{-m\A_X+n\A_Z})
(1-\xi^{\tilde{\A}_Z n-\tilde{\A}_X m})
\xi^{-\tilde{\A}_Z \A_X -mq + 
(p+m)(q+n)}\times\\
& V_{pq}
\delta_{p,-m-\A_Z-\tilde{\A}_Z}
\delta_{p,-n-\A_X-\tilde{\A}_X},\\
& i\in \mathbbm{Z},\:\A_J=\A_Z,\A_X,\tilde{\A}_Z,\tilde{\A}_x\\ &\{id-m-\A_Z-\tilde{\A}_Z,id-m,id-n-\A_X-\tilde{\A}_X \}\in S
\end{split}
\end{align}

In above formula the extra average $\langle . \rangle$ is taken over many \textit{classical} realizations of the stochastic noise functions. 
We assume that stochastic function $\beta_{\A_{\beta}}(t)$ has zero-mean Gaussian probability distribution and is stationary. The stationary condition imposes the following relation to the correlation functions:
\begin{align}\label{stationary}
\langle\beta_{\A_{\beta}}(t')\beta_{\tilde{\A}_{\beta}}(t'')\rangle=\langle\beta_{\A_{\beta}}(0)\beta_{\tilde{\A}_{\beta}}(t''-t')\rangle,\:t''>t'.
\end{align}
Using quantum noise spectroscopy, it is possible to infer information about the above self ($\A_{\beta}=\tilde{\A}_{\beta}$) and cross ($\A_{\beta}\ne \tilde{\A}_{\beta}$) noise correlation functions, i.e. finding the variance that is the deviation of Gaussian probability function from the zero mean value. Here a frequency-comb approach is used that was first introduced by G. A. Alvarez and D. Suter for qubit systems \cite{PhysRevLett.107.230501}:
The system is prepared in a known initial state for $N$ rounds $(\rho_r(0);\:r=1,...,N)$. Then for each round, a different pulse series is applied to the qudit system, where each pulse series consist of $M$ consecutive repetitions of a reference pulse sequence of $\{ \tilde{P}_{(i_k,j_k)}\}_{k=1,...,\gamma}$ that is resonant and symmetric as Eqs.\eqref{unitarypulse} and \eqref{symmetricsequence} with the fixed duration $T/r,\:r=1,...,N$. So the total duration of $M$ repetitions of the $r^{th}$ pulse series is $MT/r$.
At the end of each round, the qudit's observable $\hat{O}$ is measured resulting in $N$ different outcomes.
Substituting Eqs.\eqref{1.2}, and \eqref{GJ} in Eq.\eqref{Weyldecompositions}, we find a relation among the measured qudit's observable, reference pulse characteristics, and the noise functions as follows (appendix \ref{apdx.2.1}, Eqs.\eqref{mainO},\eqref{mainOetalambda1}):
\begin{align}\label{qudit}
\begin{split}
\langle\langle \hat{O}(t=MT/r) \rangle\rangle^r
& = \eta -\frac{1}{2}\int_0^t\int_0^t dt'dt''\sum_{\A_{\sum},\tilde{\A}_{\sum} \in S} \lambda_{\A_{\lambda}}
y_{\A_y}(t')y_{\tilde{\A}_y}(t'') \langle\beta_{\A_{\beta}}(t')\beta_{\tilde{\A}_{\beta}}(t'')\rangle , \\
& = \eta -\frac{1}{4\pi}\int_{-\infty}^{\infty}d\omega \sum_{\A_{\sum},\tilde{\A}_{\sum}\in S} \lambda_{\A_{\lambda}}
F_{\A_F}(\omega,t) F_{-\tilde{\A}_F}^*(\omega,t) S_{\A_S}(\omega),\\
& \approx \eta -
\frac{M}{2T/r} \sum_{k=-\infty}^{\infty} \sum_{\A_{\sum},\tilde{\A}_{\sum}\in S} \lambda_{\A_{\lambda}} F_{\A_F}(rk\omega_0,T/r) F_{-\tilde{\A}_F}^*(rk\omega_0,T/r) S_{\A_S}(kr\omega_0),
\end{split}
\end{align}
where,
\begin{align}\label{quditetalambda}
\begin{split}
\eta
& = 
d \sum_{m,n\in S}
O_{mn}\xi^{mn}V_{id-m,id-n}, \\ 
\lambda_{\A_{\lambda}}
& = \sum_{m,n \in S} O_{mn} \tilde{J}_{mn\A_J}\\
& = d\sum_{m,n,p,q \in S} O_{mn} 
 V_{p,q}
 \delta_
 {-p,+m+\A_Z+\tilde{\A}_Z-id}\delta_{-q,+n+\A_X+\tilde{\A}_X-id} \\
& \times (1-\xi^{-m\A_X+n\A_Z})
(1-\xi^{\tilde{\A}_Z n-\tilde{\A}_X m})
\xi^{-\tilde{\A}_Z \A_X
-mq +(m+p)(n+q)} \xi^{\tilde{\A}^n_{Z}\tilde{\A}^n_{X}-\tilde{\A}_{Z}\tilde{\A}_{X}}.
\end{split}
\end{align} 
The index $\A_{\lambda}$ is a function of other indices $\{\A_z,\A_x,\tilde{\A}_z,\tilde{\A}_x,\tilde{\A}^n_{Z}\tilde{\A}^n_{X}\}$ introduced in the appendix, and $\omega_0=2\pi/T$.  
The following function is called filter function and is the Fourier transform of the switching function $y_{\A_y}(t)$: 
\begin{align}\label{Fdefinitions}
\begin{split}
F_{\A_F}(\omega,t) &= \int_{0}^{t}dt'y_{\A_y}(t')e^{I\omega t'},\\
\end{split}
\end{align}
and the below function is the noise power spectral densities or the noise polyspectra:
\begin{align}\label{Sdefinitions}
\begin{split}
S_{\A_S,\rightarrow}(\omega) &= 
\int_{-\infty}^{\infty}dt 
\langle 
\beta_{\A_{\beta}}(0)
\beta_{\tilde{\A_{\beta}}}(t)
\rangle 
e^{I\omega t},
\end{split}
\end{align}
that is the Fourier transform of the statistical average of the auto or cross correlations of the dephasing noise functions in Weyl basis. The arrow symbol indicates the order of variables, so that the inverse arrow corresponds to the Fourier transform of  $\langle \beta_{\tilde{\A_{\beta}}}(0)
\beta_{\A_{\beta}}(t) \rangle $.

The relation of these noise Polyspectra and that of in the computational basis, which is the Fourier transform of energy fluctuations, is further discussed in appendix \ref{energy-correlations}, and it is shown that the total noise power spectral density that is detected by the qudit spectator is related to the noise polyspectra in Weyl basis $S_{ab}^{a'b'}(\omega)$, as follows (appendix \ref{energy-correlations}, Eq.\eqref{NoiseSpectrum-DifferentRepresentations}):
\begin{align}\label{ActualNoiseSpectrum-Weyl}
\begin{split}
S'(\omega) = 
 \sum_{mnm'n'a a' \in S}
\xi^{a m  + a' m'}
S_{a,n-m}^{a',n'-m'}(\omega),
\end{split}
\end{align}
where we have applied $b=m-n$ and $b'=m'-n'$. The indices $\{m,n,m',n'\} \in S$ are the corresponding energy levels for auto or cross correlations of the qudit coherences that are discussed in appendix \ref{energy-correlations}. 
In the Reduced Weyl basis, the corresponding formula of Eq.\eqref{ActualNoiseSpectrum-Weyl} is achieved by considering $b=0$ and $b'=0$ or equivalently $m=n$ and $m'=n'$, as follows (appendix \ref{energy-correlations}, Eq.\eqref{NoiseSpectrum-DifferentRepresentations-reducedWeyl}):
\begin{align}\label{ActualNoiseSpectrum-ReducedWeyl}
\begin{split}
S'(\omega) = 
\sum_{mm'a a' \in S}
\xi^{a m  + a' m'}
S_{a,a'}(\omega),
\end{split}
\end{align}
The unified form of Eqs.\eqref{ActualNoiseSpectrum-Weyl} and \eqref{ActualNoiseSpectrum-ReducedWeyl} can be expressed as:
\begin{align}\label{ActualNoiseSpectrum-UnifiedForm}
\begin{split}
S'(\omega) = 
\sum_{\A_{\sum} \in S}
\xi^{\A_{\xi}} 
S_{{\A}_S}(\omega),
\end{split}
\end{align}
where $\mathscr{A}_{var}$ indices/powers are defined in Table \ref{table3} with $ var \in \{ S, \xi,\sum \}$. The columns W (Weyl) and RW (Reduced Weyl) of Table \ref{table3} correspond to 
Eq.\eqref{ActualNoiseSpectrum-Weyl} with $(a,m,n,a',m',n') \rightarrow (i_0,i_1,i_2,j_0,j_1,j_2)$  and 
Eq.\eqref{ActualNoiseSpectrum-ReducedWeyl} with $(a,m,a',m') \rightarrow (i_0,i_1,j_0,j_1)$, respectively.\\
\begin{table}[ht!]
\centering
\caption{Indices/Powers $\mathscr{A}_{var}$}
\begin{tabularx}{0.5\textwidth} 
{
  | >{\centering\arraybackslash}X 
  | >{\centering\arraybackslash}X
  | >{\centering\arraybackslash}X | }
\hline 
$\mathscr{A}_{var}$ & $W$ & $RW$ \\
\hline 
$\mathscr{A}_S$    &    $i_0, i_1-i_2$ , $j_0,j_1-j_2$    &     $i_0,j_0$ \\
\hline
$\mathscr{A}_{\xi}$     &    $i_0i_1+j_0j_1$      &      $i_0i_1 + j_0 j_1$ \\
\hline
$\mathscr{A}_{\sum}$   &     $i_0,i_1,i_2,j_0,j_1,j_2$    &  $i_0,i_1,j_0,j_1$ \\
\hline
\end{tabularx}
\label{table3}
\end{table}

In the following sections, we develop and simulate our qudit noise spectroscopy protocol for the Reduced Weyl forms simulated for a qutrit, ququad, and quoct, and the general Weyl form simulated for the Antimony quoct system.\\

In these parts, we consider certain true noise spectral densities in the Weyl basis and simulate our qudit noise spectroscopy protocol to recover the initially assumed true noise polyspectra in the same basis. Since we use true noise polyspectra rather than experimentally measured data, we do not derive or present the true total noise spectrum that generated them, and consequently, we do not use or further develop Eq.\eqref{ActualNoiseSpectrum-UnifiedForm} in our simulations.\\

\section{Simulation: qutrit noise spectator with Reduced Weyl Hamiltonian}\label{SecIII}

In this section we simulate our noise spectroscopy protocol for a qutrit ($d=3$) in Reduced Weyl basis with Z-type dephasing noise.
Suppose a reference pulse sequence of the form of Eq.\eqref{ReferencePulseSequence} as follows:
\begin{equation}\label{qutritpulse}
\{ \tilde{P}_{(0,1)}, \: \tilde{P}_{(1,2)},\: \tilde{P}_{(2,0)}\},
\end{equation}
is applied to the qutrit at time instances $t_0,t_1,t_2$ $=0,\frac{1}{7}\frac{T}{r},\frac{2}{5}\frac{T}{r},$ where $\frac{T}{r}$ is the total duration of the reference sequence at the $r^{th}$ round of the noise spectroscopy. 
We consider the simpler effective Hamiltonian  given by Eq.\eqref{HQtrit}, that is proven to exist at least for a qutrit system.
The stated switching functions $y_a(t)$ of qutrit correspond to the above reference pulse sequence that are independent of the chosen intervals.
The expectation value of an arbitrary observable of the qutrit is found through Eqs.\eqref{qudit} and \eqref{quditetalambda}, given the qutrit indices of Table \ref{table2}, and variable exchange of $(-a,-b)\rightarrow (a,b)$, as follows:
\begin{align}\label{1.4}
\begin{split}
\langle\langle \hat{O}(t=MT/r) \rangle\rangle^r  & = \eta -\frac{1}{2}\int_0^t\int_0^t dt'dt''\sum_{a,b \in S} \lambda_{ab}
\tilde{y}_{a}(t')\tilde{y}_{b}(t'') \langle\beta_{a}(t')\beta_{b}(t'')\rangle\\
& = \eta -\frac{1}{4\pi}\int_{-\infty}^{\infty}d\omega \sum_{a,b\in S} \lambda_{ab}
F_{a}(\omega,t) F_{-b}^*(\omega,t) S_{ab}(\omega)
\\
& \approx \eta - \frac{M}{2T/r} \sum_{k=-\infty}^{\infty} \sum_{a,b\in S} \lambda_{ab} F_{a}(kr\omega_0,T/r) F_{-b}^*(kr\omega_0,T/r) S_{ab}(kr\omega_0),
\end{split}
\end{align}
where, 
\begin{align}\label{eta_lambda}
\begin{split}
    \eta&=d \sum_{m,n\in S}O_{mn}V_{id-m,id-n} \xi^{m n},\\ 
    \lambda_{ab}&=d \sum_{\substack{m,n,p,q\in S}} O_{mn}V_{pq}\delta_{p,id+a+b-m}\delta_{q,id-n} (1-\xi^{-an})(1-\xi^{-bn}) \xi^{m n},\\
    F_{a}&(\omega,t) = \int_{0}^{t}dt'y_{a}(t')e^{I\omega t'},\:\:\: S_{a,b}(\omega) = \int_{-\infty}^{\infty} dt e^{I\omega t} \langle\beta_a(0)\beta_b(t)\rangle,
\end{split}
\end{align}
and $y_a(t)$ is introduced in Eq.\eqref{HQtrit}. Given the known pulse series, Eqs.\eqref{1.4} and \eqref{eta_lambda} relate the unknown Z-type dephasing noise polyspectra $S_{l,l'}(\omega)$ to the measurement results $\langle\langle\hat{O}(t)\rangle\rangle$. 
We simplify Eq.\eqref{eta_lambda} by assuming that only one term of the Weyl decomposition of the qutrit observable is nonzero and the initial state of the qutrit would have only two non-zero terms:
\begin{align}\label{mnfixedQutrit}
\hat{O}=O_{mn}Z^m X^n,\:\hat{\rho}(0)=V_{00}\mathbbm{1}+V_{p_0,q_0}Z^{p_0} X^{q_0},\:V_{00}=1/3
\end{align}
The density matrix term $V_{00}=1/3$ is assumed to preserve the unit trace condition. The other variables $O_{mn},\:V_{p_0q_0},$ and their fixed indices could be chosen arbitrarily as $m,n\in \{0,\pm 1\}$ and $p_0,q_0\in \{\pm 1\}$. Now the parameters $\eta$ and $\lambda_{ab}$ in Eq.\eqref{eta_lambda} will depend on the fixed indices $(m,n)$ and  $(p,q)$, where $(p,q)\in  S_q = 
\{(0,0)\:\rm and \:(p_0,q_0)\}$ as following: 
\begin{align}\label{etalambdamnfixed-qutrit}
\begin{split}
\eta_{mn}
&=
3O_{mn}V_{id-m,id-n} \xi^{m n},\\
\lambda_{ab}^{mn}
&=
3 \sum_{\substack{p,q\in S_q}} O_{mn}V_{pq}\delta_{p,id+a+b-m}\delta_{q,id-n} (1-\xi^{-an})(1-\xi^{-bn}) \xi^{m n}\\
&=
3O_{mn} (\frac{1}{3})\delta_{n,0}\delta_{a+b-m,0} (1-\xi^{-an})(1-\xi^{-bn}) \xi^{m n}\\
& + 
3O_{mn}V_{id+a+b-m,id-n} (1-\xi^{-an})(1-\xi^{-bn}) \xi^{m n},
\end{split}
\end{align}
where $i\in \mathbbm{Z}$ and $(id+a+b-m),(id-n) \in \{0,\pm 1\}$. 
Using Eq.\eqref{1.4} and appendix \ref{Qutrit-expectation}, we achieve the following:
\begin{align}\label{qutritsimulation} 
\begin{split}
\langle\langle \hat{O}(t_r^A) \rangle\rangle^{r} & = A^r(t_r^A) \approx B^r(t_r^B),\\
A^r(t_r^A) & = \eta_{mn} -\frac{1}{4\pi}\int_{-\Omega}^{\Omega}d\omega \sum_{a,b=\pm1} \lambda_{ab}^{mn}
F_{a}(\omega,t_r^A) F_{-b}^*(\omega,t_r^A) S_{ab}(\omega)\\
& = \eta_{mn} -\frac{1}{4\pi}\int_{\omega_0}^{\Omega}d\omega \sum_{i=0}^3 C_i(m,n,\omega,t_r^A)x_i(\omega),\:\:\: t_r^A=M T/r,\\
B^r(t_r^B) & = \eta_{mn} -
\frac{M}{2T/r} \sum_{k=-N}^N \sum_{a,b=\pm1} \lambda_{ab}^{mn}
F_{a}(rk\omega_0,t_r^B) F_{-b}^*(rk\omega_0,t_r^B) S_{ab}(rk\omega_0)\\
& = \eta_{mn} -
\frac{M}{2T/r} \sum_{k=1}^N \sum_{i=0}^3 C_i(m,n,rk\omega_0,t_r^{B})x_i(rk\omega_0),\:\:\: t_r^B=T/r,
\end{split}
\end{align}
where,\\
\begin{align}\label{Ci-coefficients-qutrit}
\begin{split}
x_0 (\omega) &  =R_1(\omega),\:x_1(\omega)=I_1(\omega),\:x_2(\omega)=E(\omega),\:x_3(\omega)=D(\omega)\\
C_0 (m,& n,\omega,t_r) = 
2\{\lambda_{1,1}^{mn} F_{1}(\omega,t_r) F_{-1}^*(\omega,t_r)
+ \lambda_{-1,-1}^{mn} F_{-1}(\omega,t_r) F_{1}^*(\omega,t_r)\},\\
C_1(m,& n,\omega,t_r) = 
2i\{\lambda_{1,1}^{mn} F_{1}(\omega,t_r) F_{-1}^*(\omega,t_r)
- \lambda_{-1,-1}^{mn} F_{-1}(\omega,t_r) F_{1}^*(\omega,t_r)\},\\
C_2(m,& n,\omega,t_r) = 
2\{\lambda_{1,-1}^{mn} F_{1}(\omega,t_r) F_{1}^*(\omega,t_r)
+ \lambda_{-1,1}^{mn} F_{-1}(\omega,t_r) F_{-1}^*(\omega,t_r)\},\\
C_3(m,& n,\omega,t_r) = 0,\:
N = \lfloor \Omega/\omega_0 \rfloor,%\: \alpha = 2^{(1-\delta_{\omega,0})},
\end{split}
\end{align}
and $\Omega$ is the maximum dephasing noise frequency that we are interested to detect with the qutrit spectator. Note that since $C_3(m,n,\omega,t_r)=0$, there would be only three unknown noise spectra $x_{0,1,2}(\omega)$.

Here we demonstrate how to extract the characteristics of the environmental noise by performing a series of measurements on the qutrit spectator.
We simulate the introduced quantum noise spectroscopy protocol, by first assuming a true spectrum for each of the unknown noise real functions 
$\{x_i(\omega)\}_{i=0}^2$ with a Poissonian form as following:
\begin{align}\label{Poissonianspectra}
\begin{split}
x_0(\omega)&=\omega^2 
e^{-0.18|\omega|}([S])\\ 
x_1(\omega)&=\omega^2 e^{-0.15|\omega|}([S])\\ 
x_2(\omega)&=\omega^2 e^{-0.12|\omega|}([S])
\end{split}
\end{align}

In our calculations the noise functions are defined for the whole frequency range, and the above functions are needed to be even/symmetric with respect to variable $\omega$ (appendix \ref{RealFuncs-qutrit}). So one should assume the similar patterns for the positive and the non-physical negative frequency ranges.
The unit of the noise spectra, or the noise power spectral density, is considered as $[S]$ that could be related to the  units of energy, time, or frequency according to Eqs.\eqref{Sdefinitions} and \eqref{betaEnery}. Moreover, we remind that we have assumed $\hbar =1$ in calculating Eq.\eqref{UnitaryGeneral}. So, all parameters in our simulations should be assumed in the Natural Unit System. 

Now we try to find the noise spectra that the qutrit is experiencing 
at $N$ different frequency points separated by steps $\delta\omega=\omega_0= 2\pi/T$, i.e. $x_i(\omega_0),x_i(2\omega_0),...,x_i(N\omega_0)$, $i=0,1,2$.
We expect that the found noise spectra through our noise spectroscopy protocol would be close to the true spectra of Eq.\eqref{Poissonianspectra}.
These $3N$ unknown variables are found by solving a set of $3N$ linear equations of the form of  Eq.\eqref{qutritsimulation} (the first line).
Each linear equation has $3N$ unknown variables of the form of $x_i(rk\omega_0)$ (the fifth line) where $i=0,1,2$ and $r.k\in\{1,...,N\}$. The functions $x_i(\omega)$ (the third line) are substituted by Eq.\eqref{Poissonianspectra} to find $A^r(t_r^A)$ for each of the linear equations, however,
when we want to use our protocol in an experimental setup, we do not substitute the functions $x_i(\omega)$ by the true spectra to find $A^r(t_r^A)$, but directly measure the observable $\langle\langle \hat{O}(t_r^A) \rangle\rangle^{r} = A^r(t_r^A)$.
Each set of $N$ linear equations correspond to a set of Alvarez-Suter spectroscopy measurements, explained before Eq.\eqref{qudit}. Then the measurements are repeated for three sets, resulting in 
$3N$ linear equations.

The qutrit’s measurable observables and initial states for each of three measurement rounds are assumed to be different as follows:
\begin{align}\label{observablesandinitialrhoQudit}
\begin{split}
\hat{O_0} & =0.3Z^1X^1,   \rho_0(0)=\mathbbm{1}/3+Z^1X^2,\\ 
\hat{O_1} & =0.2Z^{1}X^2, \rho_1(0) = \mathbbm{1}/3+0.7Z^2X^1,\\
\hat{O_2} & =0.4Z^2X^1,   \rho_2(0)=\mathbbm{1}/3+0.6Z^2X^2.
\end{split}
\end{align}
So the fixed indices introduced in  Eq.\eqref{mnfixedQutrit} for each round are defined as above equation.
In our simulations, we ignore the trace-preserving terms of the initial density matrix ($\mathbbm{1}/3$) for simplification purposes.
Note that since all of the $n_i$ powers in Eq.\eqref{observablesandinitialrhoQudit} ($\hat{O_i} =\alpha_i Z^{m_i} X^{n_i}$) are nonzero, the first term of the definition of $\lambda_{ab}^{mn}$ in Eq.\eqref{etalambdamnfixed-qutrit} is eliminated.
Now we put the $3N$ known values of $A^r(t_r^A)$ in the column matrix $\textbf{b}$ of rank $(\rm 3N\times1)$ so that the first/second/or third set of $N$ elements correspond to the first/second/or third set of the Alvarez-Suter experiments and for each set of N experiments we have $r=1,...,N$. We also arrange the unknown noise functions in the column matrix $\tilde{\textbf{x}}$ of rank $(\rm 3N\times1)$, by assigning the three unknown noise functions $x_0(\omega)$, $x_1(\omega)$, and $x_2(\omega)$ subsequently, to the frequencies from the lowest ($\omega_0$) to the highest amount($N\omega_0$). In the following, we convert a set of $3N$ indices to the vectors $\textbf{b}$ and $\tilde{\textbf{x}}$ as described.
\begin{align}\label{Simulation-b}
\begin{split}
&b_{n,1} =A^{n'}(t_{n'}^A),\:\\ &\tilde{x}_{m,1} =
x_{m^{\prime\prime}}(m^{\prime}\omega_0),\:\\
&n , m\in \{0,...,3N-1\},\\
&n' = \mod(n,N)+1 \in \{1,...,N\},\\
&m' = \lfloor m/3\rfloor+1 \in \{1,...,N\},\\ 
&m^{\prime\prime} = \mod(m,3) \in \{0,1,2\}, 
\end{split}
\end{align}
Using Eq.\eqref{qutritsimulation} (the fifth line), we find the coefficient matrix $\textbf{A}$ of rank $\rm 3N\times 3N$, that connects the vectors $\textbf{b}$ and $\tilde{\textbf{x}}$ by the linear matrix equation:
 \begin{align}\label{LinearMatrixEquation}
\begin{split}
 \textbf{A} \tilde{\textbf{x}}=\textbf{b},
\end{split}
\end{align}
where,
\begin{align}\label{Simulation-A}
\begin{split}
&A_{nm} =\delta_{\frac{m'}{n'},\lfloor \frac{m'}{n'}\rfloor} C_{m^{\prime\prime}}(e_{n^{\prime\prime}},f_{n^{\prime\prime}},m'\omega_0,T/n'),\\
&n^{\prime\prime} = \lfloor n/N\rfloor\in\{0,1,2\}\\
& (e_{n^{\prime\prime}},f_{n^{\prime\prime}}) \in \{(m_{i},n_{i})\}_{i=0,1,2}
\end{split}
\end{align}
The factor $m^{\prime}$ multiplying $\omega_0$ was found by corresponding the following two terms of the matrix  $\textbf{A}$ and the vector $\tilde{\textbf{x}}$:
\begin{align}\label{Simulation-C}
\begin{split}
C_{m^{\prime\prime}}(e_{n^{\prime\prime}},f_{n^{\prime\prime}},kn'\omega_0,T/n') &\leftrightarrow 
x_{m^{\prime\prime}}(m'\omega_0),
\end{split}    
\end{align}
which yields $k=m'/n'$, and the conditions $k\in \{1,...,N\}$ and $k\in \mathbbm{Z}$ are satisfied by assuming the factor $\delta_{m'/n',\lfloor m'/n'\rfloor}$ for the elements of the matrix $\textbf{A}$.
Given $\textbf{b}$ that is found by the true spectra (e.g. Eq.\eqref{Poissonianspectra}), and the coefficient matrix $\textbf{A}$ that depends on the filter functions and $\lambda^{mn}_{ab}$, we find the estimated noise spectra $\tilde{\textbf{x}}$ by solving the corresponding linear matrix equation.

\begin{figure}[!t]
\includegraphics[width=1\linewidth]{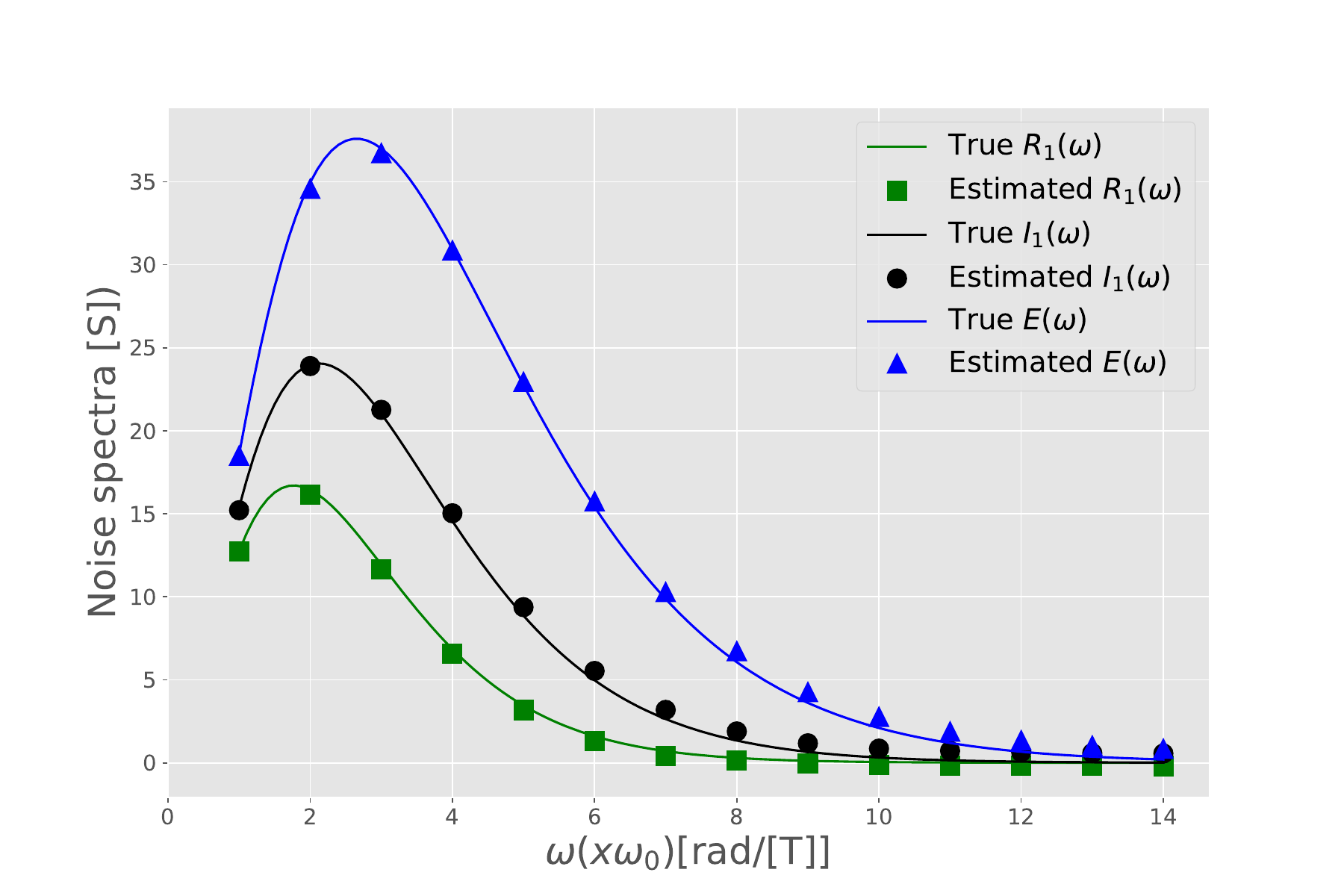}
%\rule{12.8cm}{7.2cm}
\center
\caption{This figure shows the simulation of quantum noise spectroscopy protocol by a qutrit spectator. The true noise spectra are assumed Poissonian as $R_1(\omega)=\omega^2 e^{-0.18\omega}([S]), I_1(\omega)=\omega^2 e^{-0.15\omega}([S])$, and $E(\omega)=\omega^2 e^{-0.12\omega}([S])$, and the estimated noise spectra are found for the $T=1([T])$ duration of the reference pulse sequence, $M=30$ number of repetitions of the refrence pulse sequence, and the maximum detecting frequency of $\Omega=90(rad/[T])$.}
\label{Fig.1}
\end{figure}

Fig.\ref{Fig.1} shows the estimated noise spectra along with the assumed Poissonian true spectra of Eq.\eqref{Poissonianspectra}.
The unit of $\omega_0$ is proportional to the unit of the duration of the reference pulse sequence $([T])$. We assumed the maximum sensible frequency of $\Omega=90(rad/[T])$, and the duration and the number of reference sequence repetitions as $T=1\:([T])$ and $M=30$, respectively. The resolution of the estimated spectra is $\omega_0[rad/[T]]$ that is proportional to the inverse of the chosen parameter $T[T]$. It can be seen that our noise spectroscopy protocol could estimate the true noise spectra with good accuracy.
The simulation of the Poissonian noise type is studied due to its efficient running time with respect to that of an arbitrary non-Poissonian noise function. 
The time-consuming part of the codes correspond to the integral calculations of $A^r(t_r^a)$ which are dismissed in the experimental implementation of our
spectroscopy protocol.

\begin{figure}[!t]
\center
\includegraphics[width=1\linewidth]{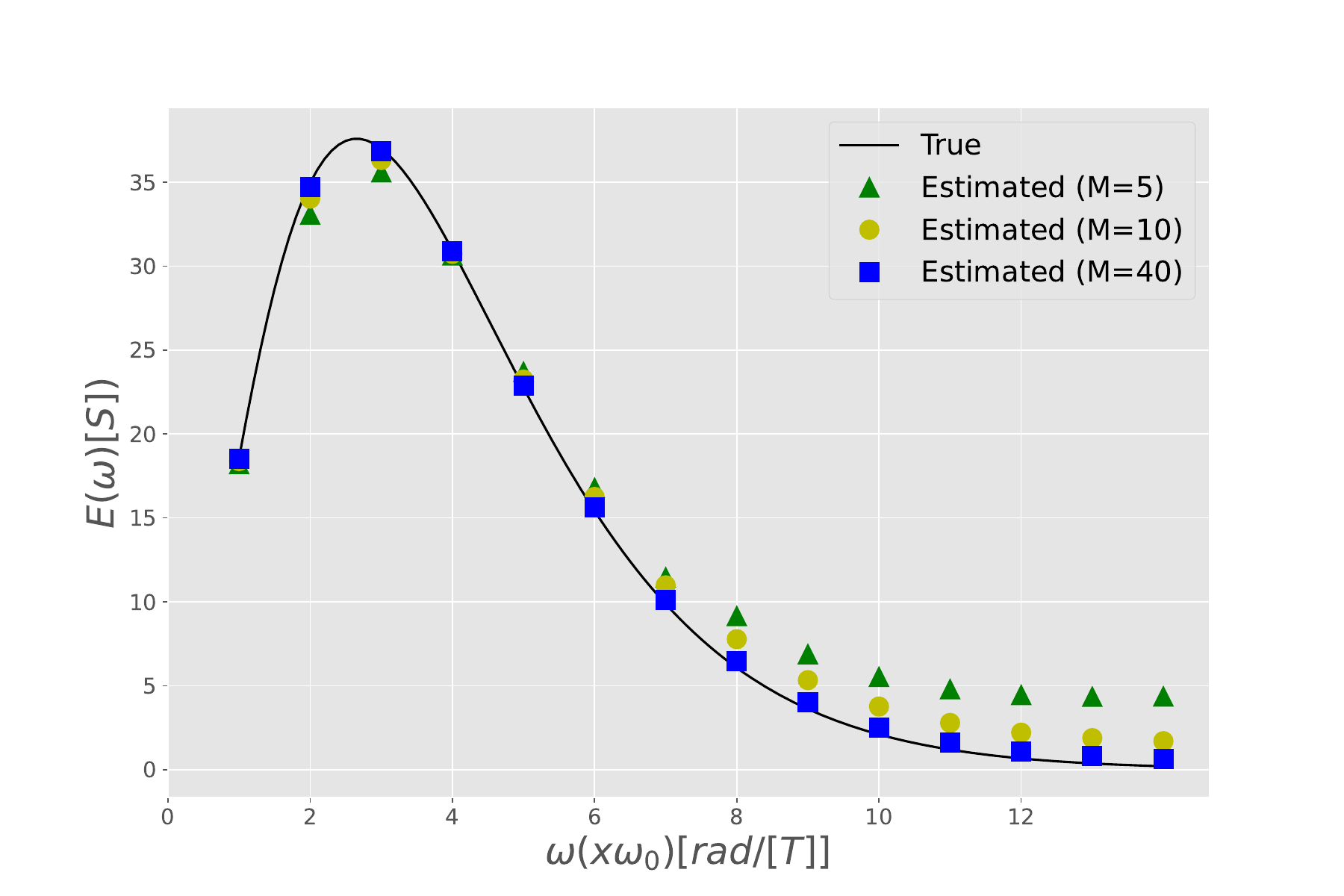}
%\rule{12.8cm}{7.2cm}
\caption{This figure shows how parameter $M$, that is the number of repetitions of the reference pulse sequence, is affecting the accuracy of the noise retrieval by the qutrit noise spectroscopy method. 
A dephasing noise function with true noise spectrum $E(\omega)=\omega^2 e^{-0.12\omega}([S])$ is estimated by a qutrit spectator with our noise spectroscopy protocol. The maximum sensible frequency and the duration of reference pulse sequence are chosen as $\Omega=90(\rm rad/[T])$ and $T=1([T])$, respectively. By increasing the number of repetitions of the reference pulse sequence as $M=5,10,40$, the accuracy of the estimated noise spectrum is increasing.}
\label{Fig.2}
\end{figure}

Next we investigate the effect of the variable $M$, i.e. the number of reference pulse sequence repetitions, on the accuracy of the simulation. Assuming the true Poissonian noise spectrum $E(\omega)=\omega^2 e^{-0.12|\omega|}([S])$, and the parameters 
$T=3([T])$ and $\Omega=100(\rm rad/[T])$, we simulate the noise spectra for three values of $M=5,10,40$. Fig.\ref{Fig.2} shows that by increasing $M$, the accuracy of estimating the true spectra increases. This is due to the fact that in the theoretical framework, we assumed $M \rightarrow \infty$ (appendix \ref{apdx.2.1}, Eq.\eqref{Minfty}). Fig. \ref{Fig.2} also indicates that even for a small value like $M=10$, the estimated spectrum is almost accurate and a large value of $M$ is not required to detect the noise spectra.  
Now we perform the simulation of our noise spectroscopy protocol for non-Poissonian, arbitrary true functions as follows:
\begin{align}\label{Turespectra-arbitrary}
\begin{split}
x_0(\omega)&=
\frac{1}{2}e^{-0.9(\omega-21)^2}+1.5 e^{-0.086(\omega-13)^2}
([S]),\\ 
x_1(\omega)& =e^{-0.16(\omega-3)^2} + e^{-0.286(\omega-9)^2} + e^{-0.03(\omega-14)^2} ([S]),\\
x_2(\omega)&=\frac{1.5}{1+\abs{\omega-\omega_0}}([S]),  
\end{split}   
\end{align}

Note that the above functions are considered for the positive frequency range. Since the above noise spectra are proven to be even, we consider them as a piece-wise function in the full frequency range, so that the same pattern is assumed for the negative frequency range. We also assume a continuous and smooth behaviour for the noise functions around the no-frequency point $\omega=0$.
\begin{figure}[!t]
\center
\includegraphics[width=1\linewidth]{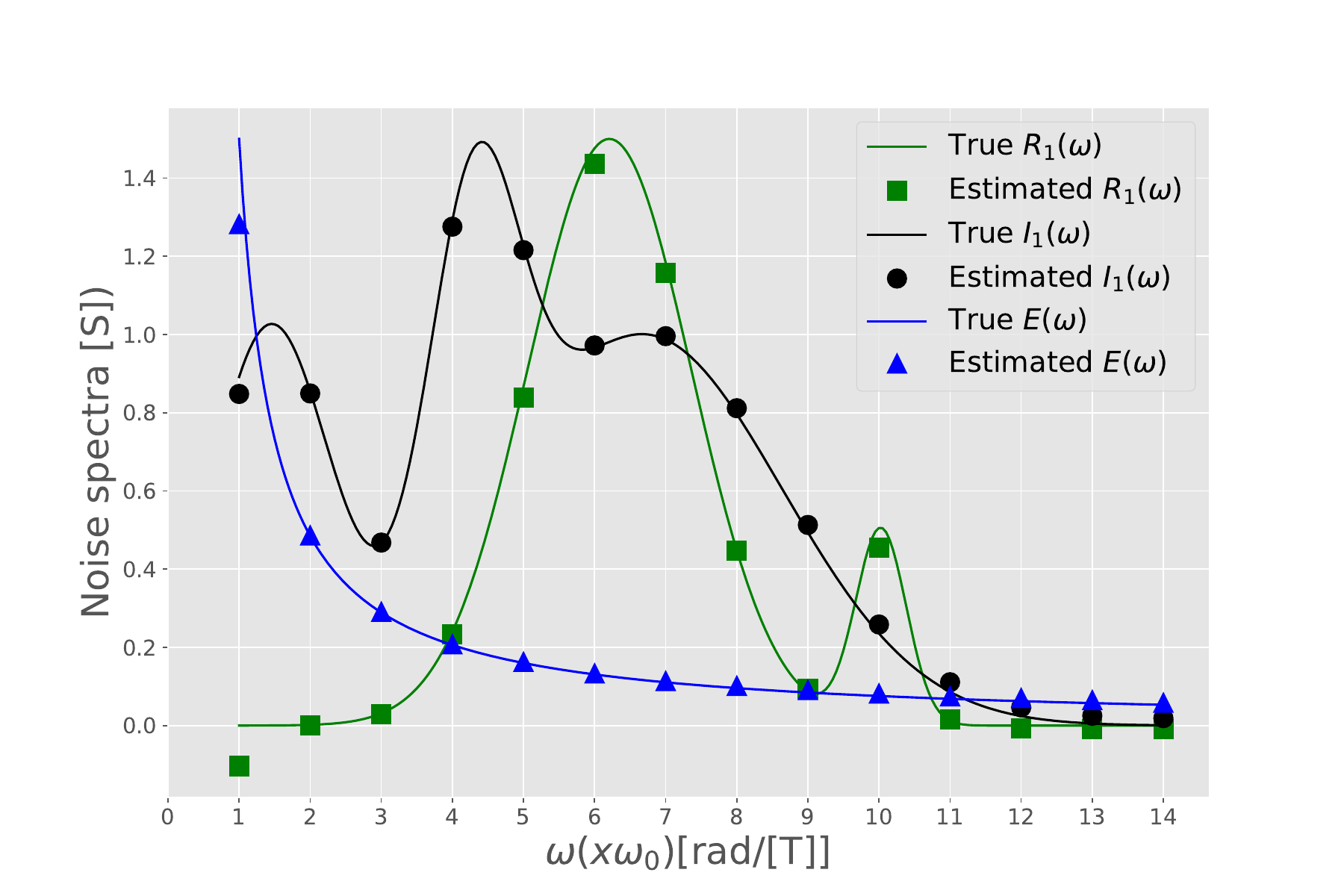}
%\rule{12.8cm}{7.2cm}
\caption{This figure illustrates another simulation of the introduced quantum noise spectroscopy protocol for a qutrit spectator. Here the assumed true noise spectra are non-Poissonian with general forms of $R_1(\omega)=
\frac{1}{2}e^{-0.9(\omega-21)^2}+1.5 e^{-0.086(\omega-13)^2} ([S]), I_1(\omega)=e^{-0.16(\omega-3)^2} + e^{-0.286(\omega-9)^2} + e^{-0.03(\omega-14)^2} ([S])$, and $E(\omega)=\frac{1.5}{1+\abs{\omega-\omega_0}}([S])$. The simulation parameters are as  $\Omega=30(rad/[T])$, the maximum sensible frequency, $T=3([T])$, the duration of reference pulse sequence, and $M=40$, the number of repetitions of the reference sequence in each Alvarez-Suter spectroscopy round. It can be seen that with the chosen set of parameters, our protocol could accurately estimate the general form of noise functions with a qutrit spectator.}
\label{Fig.3}
\end{figure} 
The estimated noise spectra corresponding the above true functions are shown in Fig.\ref{Fig.3} for the fixed parameters $M=40$, $T=3([T])$, and $\Omega=30(rad/[T])$. It can be seen that even for a limited number of reference sequence repetitions ($M=10$), our formalism could find the general and fine patterns of the noise functions with resolution $\omega_0$. The link to all simulation codes is provided in the Code Availability section. 
\section{Simulation: qudit noise spectator with Reduced Weyl Hamiltonian}\label{SecIV}
Here we apply our noise spectroscopy protocol for a spin qudit in the Reduced Weyl basis that is surrounded by Z-type stochastic dephasing noise, as follows:
\begin{align}\label{quditDephasingH}
\begin{split}
H(t) = A(t) I_z, 
\end{split}    
\end{align}
where $I_z$ is the nuclear spin of an ion qudit with the total magnetic moment of $I=(d-1)/2$, and $A(t)$ is the stochastic noisy process.
According to Appendix \ref{AntimonyQuoctHWeyl} (Eqs.\eqref{IcomponentsWeylbasis} and \eqref{I-ComputationalBasis}), the above Hamiltonian can be written in the Reduced Weyl basis as follows:
\begin{align}\label{quditRWH}
\begin{split}
H(t) = \frac{\hbar}{d} A(t) \sum_{a,p \in S} \xi^{-a.p} (I-p) Z^a.
\end{split}
\end{align}
We expose the qudit to the reference pulse sequence of the form of Eq.\eqref{ReferencePulseSequence} applied at time instances $t_i=\frac{i}{d}(T/r),i=0,\dots,d-1$ and the total sequence duration of $T/r$. 
After applying the pulse resonance, the resulted effective Hamiltonian of qudit, noise, and pulses and the corresponding switching function would be as of Eq.\eqref{HRW}. The stochastic dephasing noise coefficient $\beta_a(t)$ in that equation is related to the noisy part of the Hamiltonian of Eq.\eqref{quditRWH} as follows:
\begin{align}\label{betafAt}
\begin{split}
\beta_a(t)= f_a A(t),\: f_a = \frac{\hbar}{d} \sum_{p \in S} \xi^{-a.p} (I-p)
\end{split}
\end{align}
and the corresponding noise spectrum is: 
\begin{align}\label{noisespectrum_quoct_I}
\begin{split}
S_{ab}(\omega)
=& 
\int_{-\infty}^\infty dt \langle
\beta_a(0)
\beta_b(t)\rangle e^{I\omega t} 
=
f_a f_b  \tilde{S}(\omega), \\
\tilde{S}(\omega) =&
\int_{-\infty}^\infty dt \langle
A(0)
A(t)\rangle e^{I\omega t}.
\end{split}
\end{align}
The measured expectation value of the observable $\langle\langle O \rangle\rangle$ of such ion qudit, is found through Eqs.\eqref{qudit} and \eqref{quditetalambda}, by assuming the Reduced Weyl (RW) indexing $(\A_Z, \A_X, \tilde{\A}_Z, \tilde{\A}_X, \tilde{\A}^n_Z, \tilde{\A}^n_X)$ $=(b, 0, \tilde{b}, 0, \tilde{a}, 0)$. 
\begin{align}\label{qudit01}
\begin{split}
\langle\langle  \hat{O}(t=&MT/r) \rangle\rangle^r = \\
&  \eta -\frac{1}{2}\int_0^t\int_0^t dt'dt''\sum_{a,b,\tilde{a},\tilde{b} \in S} \lambda_{b\tilde{b}} \:
y_{a,b}(t')y_{\tilde{a},\tilde{b}}(t'') \langle\beta_a(t')\beta_{\tilde{a}}(t'')\rangle  \\
 = &\:\eta 
-\frac{1}{4\pi}
\int_{-\infty}^{\infty}d\omega
\sum_{a,b,\tilde{a},\tilde{b} \in S} \lambda_{b\tilde{b}} \:
F_{a,b}(\omega,t) F_{-\tilde{a},-\tilde{b}}^*(\omega,t) S_{a{\tilde{a}}}(\omega)\\
 = &\:\eta 
-\frac{1}{4\pi}
\int_{-\infty}^{\infty}d\omega
\sum_{a,b,\tilde{a},\tilde{b} \in S} 
f_a f_{\tilde{a}}\lambda_{b\tilde{b}} \:
F_{a,b}(\omega,t) F_{-\tilde{a},-\tilde{b}}^*(\omega,t) \tilde{S}(\omega)\\
 \approx &\:\eta -
\frac{M}{2T/r} \sum_{k=-\infty}^{\infty} \sum_{a,b,\tilde{a},\tilde{b}\in S} 
f_a f_{\tilde{a}} \lambda_{b\tilde{b}} \: F_{a,b}(rk\omega_0,T/r) F_{-\tilde{a},-\tilde{b}}^*(rk\omega_0,T/r) \tilde{S}(kr\omega_0),
\end{split}
\end{align}
where,
\begin{align}\label{quditetalambda01}
\begin{split}
\eta
& = 
d \sum_{m,n\in S}
O_{mn}\xi^{mn}V_{id-m,id-n}, \\ 
\lambda_{b\tilde{b}}
& = d\sum_{m,n,p \in S} O_{mn} V_{p,-n} \delta_{p,-m-b-\tilde{b}} (1-\xi^{nb}) (1-\xi^{n\tilde{b}}) \xi^{mn},
\end{split}
\end{align} 
We simplify Eqs.\eqref{qudit01} and \eqref{quditetalambda01} by considering that only one term of the Weyl decomposition of the qudit observable is nonzero and the initial state of the quoct would have only two non-zero terms:
\begin{equation}\label{Oprime}
\hat{O^{\prime}}=O_{mn}Z^m X^n,\:\hat{\rho}(0)=\mathbbm{1}/d+V_{p,q}Z^p X^q  
\end{equation}
The term $V_{00}=\mathbbm{1}/d$ is assumed to preserve the trace of the density matrix. However, this term is ignored in our simulation for streamlining purposes. The variables $O_{mn},\:V_{pq}$ and the fixed indices $(m,n,p,q)$ could be chosen arbitrarily while $m,n\in \{0,...,d-1\}$ and $p,q\in \{1,...,d-1\}$. 
Assuming the fixed chosen parameters $(m,n,p,q)$, Eq.\eqref{quditetalambda01} turns to the following:
\begin{align}\label{quditReducedLambdaeta}
\begin{split}
\eta_{mn} = &
d\: O_{mn}\xi^{mn}V_{id-m,id-n}
, \\ 
\lambda^{mnp}_{b\tilde{b}}
& = d \:O_{mn} V_{p,-n} \delta_{p,-m-b-\tilde{b}} (1-\xi^{nb}) (1-\xi^{n\tilde{b}}) \xi^{mn}.
\end{split}
\end{align}
Using Eqs.\eqref{quditI-simulation-01} and \eqref{quditI-simulation-02}, the final form of the expectation value of the observable of Eq.\eqref{Oprime} for a qudit with the Hamiltonian of Eq.\eqref{quditDephasingH} affected by dephasing noise is as follows:
\begin{align}\label{quditI-simulation-03}
\begin{split}
A^r(t_r^A) & = 
\eta_{mn} 
-\frac{1}{4\pi}
\int_{\omega_0}^{\Omega}
d\omega 
C(m,n,p,\omega,t^A_r) S(\omega),\\
B^r(t_r^B) & = \eta_{mn} -
\frac{M}{2T/r} 
\sum_{k=1}^N 
C(m,n,p,rk\omega_0,t^B_r) S(rk\omega_0),\\
C(m,n,p,\omega,t^A_r) = & \:
2\sum_{a,\tilde{a},b,\Tilde{b} \in S}
f_a f_{\tilde{a}} 
\lambda^{mnp}_{b\Tilde{b}} \:
\delta_{p+m,-b-\tilde{b}}
F_{a,b}(\omega,t) 
F_{-\tilde{a},-\tilde{b}}^*(\omega,t).
\end{split}
\end{align}
The other form of above equation is as follows:
\begin{align}\label{quditI simulation}
\begin{split}
& \langle\langle \hat{O^{\prime}}(t = MT/r) \rangle\rangle^r= \\& 
\eta_{mn} 
-\frac{1}{2\pi}
\int_{\omega_0}^{\Omega}
d\omega 
\sum_{a,\tilde{a},b,\Tilde{b} \in S}
f_a f_{\tilde{a}} 
\lambda^{mnp}_{b\Tilde{b}} \:
\delta_{p+m,-b-\tilde{b}}
F_{a,b}(\omega, MT/r) 
F_{-\tilde{a},-\tilde{b}}^*(\omega, MT/r) S(\omega) \\
= & \eta_{mn} -
\frac{M}{T/r} 
\sum_{k=1}^N 
\sum_{a,\tilde{a},b,\Tilde{b} \in S}
f_a f_{\tilde{a}} 
\lambda^{mnp}_{b\Tilde{b}} \:
\delta_{p+m,-b-\tilde{b}}
F_{a,b}(rk\omega_0,T/r) 
F_{-\tilde{a},-\tilde{b}}^*(rk\omega_0,T/r) S(rk\omega_0)
\end{split}
\end{align}
%Measeuring, Assuming a pusle sequence of the form of Eq.\eqref{}, 
The above qudit observable was exposed to to a series of spectroscopy pulses of the form of Eq.\eqref{ReferencePulseSequence1}, resulting in the switching functions $F$. After this exposure, the observable is calculated for a fix $M$ and $N$ values of $r$. 

\begin{figure}[!t]
\center
\includegraphics[width=0.9\linewidth]{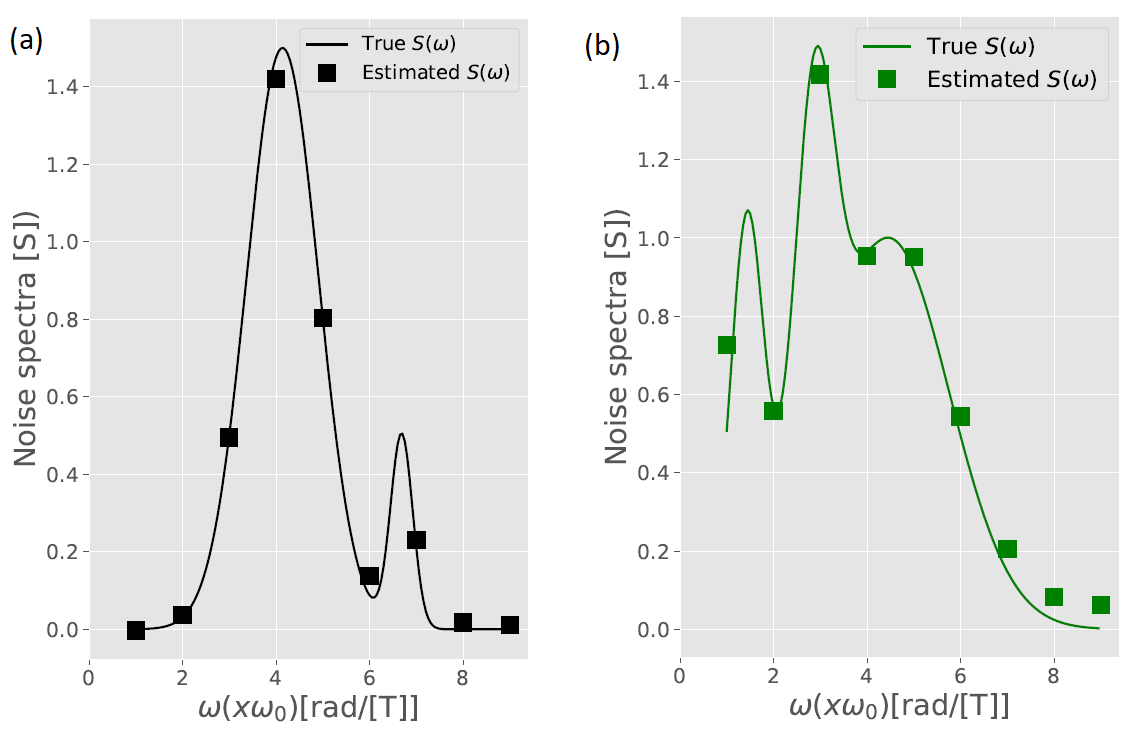}
\caption{This figure shows the simulation of the quantum noise spectroscopy protocol in the Reduced Weyl basis for a ququad and a quoct spectator surrounded by Gaussian Z-type depahsing noise. Part (a) shows the simulation for a ququad (d=4), with initial density matrix of $\hat{\rho}(0)=ZX$. The True noise spectrum is assumed as $S(\omega)=0.5e^{-0.9(\omega-21)^2}+3/2e^{-0.086(\omega-13)^2} ([S])$, and the measured observable is $\hat{O}=Z^3X^3$. The other chosen parameters are the duration and the iterations of the reference pulse sequence considered as $T=2([T])$, and $M=17$, respectively. Part (b) shows the simulation for a quoct (d=8) with initial density matrix of $\hat{\rho}(0)=ZX$. The true spectrum is considered as $S(\omega)=e^{-0.4(\omega-4.5)^2}+e^{-0.28(\omega-9)^2}+e^{-0.03(\omega-14)^2} ([S])$ and the measured observable is $\hat{O}=Z^7X^7$. %and the number of noise spectroscopy iterations to get the estimated spectrum is $M=40$. 
Moreover, the duration and repetitions of reference pulse sequence are chosen as $T=2([T])$, and $M=40$, respectively.
The gauging parameter of $\omega$ in both graphs is $\omega_0=\frac{2\pi}{T}(\frac{\rm rad}{[T]})$.}
\label{Fig4ququadquoct}
\end{figure}

Fig.\ref{Fig4ququadquoct} shows the simulation results for two cases: Fig.\ref{Fig4ququadquoct}(a) presents the simulation of a ququad $(d=4)$. The true noise spectrum is considered non-Poissonian as: 
\begin{equation}\label{ququadSpectrum}
S(\omega) =\frac{1}{2}e^{-0.9(\omega-21)^2}+\frac{1}{3}e^{-0.086(\omega-13)^2} ([S]),
\end{equation}
and the initial density matrix is assumed as $\hat{\rho}(0)=ZX$. The measured observable of the ququad is $O^{\prime}=Z^3X^3$, and the number of spectroscopy sequence iterations to get the estimated noise spectrum is $M=17$.
Fig.\ref{Fig4ququadquoct}(b) shows the simulation of a quoct ($d=8$). The true noise spectrum is assumed to have the following non-Poissonian form:
\begin{equation}\label{quditPoissonianSpectrum}
S(\omega)=e^{-0.4(\omega-4.5)^2}+e^{-0.28(\omega-9)^2}+e^{-0.03(\omega-14)^2} ([S]),    
\end{equation}
and the quoct is assumed to be prepared in the initial state $\hat{\rho}(0)=ZX$. The observable of the quoct that is measured for several times is $O^{\prime}=Z^7X^7$, and the number of iterations for each round of the qudit noise spectroscopy is $M=40$. Fig.\ref{Fig4ququadquoct} indicates that our qudit spectroscopy formalism for the Reduced Weyl case can successfully retrieve the true noise spectra with resolution $\omega_0$.

The running time of our simulations, with a x64-based processor of 12th Gen Intel(R) Core(TM) i5-1245U, 1.60 GHz, and 16.0 GB RAM, is about few hours for the ququad graph, and about 24 hours for the quoct graph. The simulation codes can be found in the Code Availability section. 

We also simulated our qudit noise spectrsocy for other \textit{d} values in this section, and concluded that the increase of \textit{d}, the number of qudit's energy levels, increases the required number of iterations ($M$) to retrieve the true noise spectrum. 

For example, for a fixed true noise spectrum of the Poissonian form of Eq.\eqref{Poissonianspectra}, which leads to a faster simulation process, the required number of iterations for the qudits with $d=3,4,5,6,7,8$, are about $7,15,35,55,70,80$, respectively, which has a linear increasing pattern. We also observed that  for a specific \textit{d}, the required number of $M$ to retrieve the noise spectrum, varies according to the complexity of the true noise form.

\section{Simulation: Antimony quoct noise spectator with Weyl Hamiltonian}\label{SecV}
Now as another example, we simulate our noise spectroscopy protocol for a quoct (an eight level quantum system) in Weyl basis, that is surrounded by XYZ-dephasing noise. The quoct physical system would be the nuclear spin of the Antimony atom $^{123}Sb$ that is doped in silicon. We consider the Hamiltonian of the Antimony quoct as follows:
\begin{align}\label{AntimonyquoctH}
\begin{split}
H(t) = (\gamma_n \tilde{B}_0 \pm \frac{1}{2}A) I_z + Q I_x^2 
%+ \gamma_n \tilde{B}_0 \cos(2\pi f t) I_y, 
\end{split} 
\end{align} 
where the nuclear spin is described by $\textbf{I}=(I_x,I_y,I_z)^T$, $I=\frac{7}{2}$, and $\gamma_n=5.55(\rm MHz/T)$ is the \textit{n}uclear gyromagnetic ratio, $\tilde{B}_0$ is a static magnetic field in the z direction, $Q$ is the effective quadrupole interaction strength, and $A$ is the hyperfine interaction between the electron and nuclear spins of the Antimony atom.
Here we assume that the two parameters $A(t)$ and $Q(t)$ are noisy and stochastic with zero-mean Gaussian probability distributions.

The Hamiltonian of Eq.\eqref{AntimonyquoctH} is part of Eq.(4) of \cite{PhysRevE.98.042206}, and the referrence equation has an extra term $\gamma_n \tilde{B}_0 \cos(2\pi f t) I_y$. The parameter $f$ in this term is the frequency of an oscillating magnetic field $B_1$, equated to the Larmor frequency $\gamma_n \tilde{B}_0$, in the y direction, and is associated to the quantum gate that is coherently controlling the state of the quoct. In our QNS formalism, the Hamiltonian of the incident pulse field is being added to the quoct Hamiltonian in a different process leading to switching functions. This is by applying the gate pulses (Eqs.\eqref{unitarypulse} and \eqref{ReferencePulseSequence1}) to the Hamiltonian of qudit and noises using  Eqs.\eqref{Unitary} and \eqref{TotalU&Hr}, to reach the Hamiltonian of qudit, noises, and pulses that includes switching functions. As a result, we did not include the aforementioned term $\gamma_n \tilde{B}_0 \cos(2\pi f t) I_y$  in the quoct Hamiltonian of Eq.\eqref{AntimonyquoctH}.

Since each of the nuclear spin components $I_{x,y,z}$ could be decomposed in the complete Weyl basis (appendix \ref{AntimonyQuoctHWeyl} Eq.\eqref{IcomponentsWeylbasis}), the Hamiltonian of Eq.\eqref{AntimonyquoctH} which is the  quoct-noise effective Hamiltonian, can be expressed in the complete Weyl basis (W) as follows:
\begin{align}\label{H^W(t)}
\begin{split}
H^W(t) 
=& 
\sum_{ab}\beta_{ab}(t) Z^a X^b 
= \sum_{\substack{i=0,1,2
%,3 
\\ pqq'ab\in S}} 
B_i(t) \tilde{\beta}^{pqq'}_{iab} Z^a X^b,
\end{split}    
\end{align}
where
\begin{align}\label{Bi-beta-vars}
\begin{split}
B_{0}(t),
& B_{1}(t), B_{2}
%(t), B_{3}
=
A(t),Q(t)
%,\rm cos(2 \pi f t)
, 1\\
\tilde{\beta}^{pqq'}_{0ab}=&\pm \frac{1}{2d^3}\xi^{-ap} I_{z,p} \delta_{b,0}\\
\tilde{\beta}^{pqq'}_{1ab}=& \frac{1}{d}\xi^{-ap} 
I_{x,pq}I_{x,qq'}\delta_{p,q+b}\\
%\tilde{\beta}^{pqq'}_{2ab}=& \frac{1}{d^2}\gamma_n \tilde{B}_0  
%\xi^{-ap} I_{y,pq}  \delta_{p,q+b}\\
\tilde{\beta}^{pqq'}_{2ab}=& \frac{1}{d^3} 
\gamma_n \tilde{B}_0 \xi^{-ap} I_{z,p} \delta_{b,0}.
\end{split}
\end{align}
Note that the general Antimony quoct noise function $\beta_{\A_{\beta}}(t)=\beta_{ab}(t)$ consists of two time dependent parts $B_{0,1}(t)$, and some time independent parts, $B_2$ and $\tilde{\beta}^{pqq'}_{iab}$ functions. So the noise spectral functions of the Antimony quoct, i.e. $S_{\A_S} (\omega)=S_{ab}^{\tilde{a}\tilde{b}}(\omega)$, have components that are solely related to the Fourier transform of the time dependent parts as follows:
\begin{align}\label{S_quoct}
\begin{split}
S_{ab}^{\tilde{a}\tilde{b}}(\omega)
=& \int_{-\infty}^\infty dt \langle
\beta_{ab}(0)
\beta_{\tilde{a}\tilde{b}}(t)\rangle e^{I\omega t}
= \sum_{\substack{i,\tilde{i}=0,1,2
%,3 
\\ pqq'\in S\\ \tilde{p}\tilde{q}\tilde{q}'\in S}} \int_{-\infty}^\infty dt 
\langle B_i(0)B_{\tilde{i}}(t)\rangle 
\tilde{\beta}^{pqq'}_{iab}
\tilde{\beta}^{\tilde{p}\tilde{q}\tilde{q}'}_{\tilde{i}\tilde{a}\tilde{b}} 
e^{I\omega t}\\
=& \sum_{\substack{i,\tilde{i}=0,1,2
%,3 
\\ pqq'\in S\\ \tilde{p}\tilde{q}\tilde{q}'\in S}} \tilde{\beta}^{pqq'}_{iab}
\tilde{\beta}^{\tilde{p}\tilde{q}\tilde{q}'}_{\tilde{i}\tilde{a}\tilde{b}}
\tilde{S}_{i\tilde{i}}(\omega),
\end{split}
\end{align}
where,
\begin{align}\label{S_ii}
\begin{split}
 \tilde{S}_{i\tilde{i}}(\omega) 
= \int_{-\infty}^\infty dt 
\langle B_i(0)B_{\tilde{i}}(t)\rangle e^{I\omega t},
\end{split}
\end{align}
where $S_{i\tilde{i}}(\omega)$ are the noise spectra corresponding the self and cross correlations of the dephasing noise processes $A(t)$ and $Q(t)$, as well as correlations between noise processes and the time independent function $B_2$. 
It is discussed and shown in appendices \ref{Antimony-S(w)symproperties} and \ref{RealFuncs-Antimonyquoct} that the unknown noise polyspectra  $\tilde{S}_{i\tilde{i}}(\omega),(i,\tilde{i}=0,1)$ are made of four real functions $R_0(\omega),R'_0(\omega)$, $R_1(\omega)$, and $I_1(\omega)$ as follows:
\begin{align}\label{Antimony-unknownPolyspectra}
\begin{split}
\tilde{S}_{00}(\omega) =  R_0(\omega),&\:\:\:\:
\tilde{S}_{11}(\omega) = R'_0(\omega),\:\:\:\: \\
\tilde{S}_{01}(\omega) = R_1(\omega) + i I_1(\omega),&\:\:\:\:
\tilde{S}_{10}(\omega) = R_1(\omega) - i I_1(\omega).
\end{split}
\end{align}
where $i=\sqrt{-1}$. The first three functions above are even/symmetric, and the last one is odd/anti-symmetric.
Another nonzero correlation term is found in appendix \ref{RealFuncs-Antimonyquoct} (Eq.\eqref{Sii-1}) as follows, where its delta function ensures that it approaches zero for frequency ranges, in our formalism, that are not in the close vicinity of zero.
\begin{align}\label{Antimony-knownPolyspectra}
\begin{split}
%\tilde{S}_{33}(\omega) = 
\tilde{S}_{22}(\omega) = 
2 \pi \delta(\omega) \equiv 0
\end{split}
\end{align}
Now we consider a reference pulse sequence on the Antimony quoct with the form of Eq.\eqref{ReferencePulseSequence1} applied at time instances $t_i=\frac{i}{d}(T/r),i=0,\dots,d-1$ and the total sequence duration of $T/r$. The effective Hamiltonian representing the Antimony quoct, environmental dephasing noise, and the reference pulse sequence is as follows:
\begin{align}\label{}
\begin{split}
H^W(t)
&=
\sum_{\substack{a,b,a',b'  \in S}}\beta_{ab}(t) y^{a b}_{a'b'}(t) Z^{a'}  X^{b^\prime}
=
\sum_{\substack{i=0,1,2 \\ pqq',
aba'b'\in S}} 
B_i(t) \tilde{\beta}^{pqq'}_{iab}  y^{ijab}_{a'b'}(t) Z^{a'} X^{b'},\:\:\:t\in \tau_r
\end{split}    
\end{align}
where the switching functions $y_{a'b'}^{a b}(t)$  are given by  Eq.\eqref{1.0.switchingfunction}.
To find the expectation value of a general observable of the Antimony quoct, we substitute the complete Weyl indices of Table \ref{table2} in Eqs.\eqref{qudit} and \eqref{quditetalambda}, and find the following:
\begin{align}\label{quoct0}
\begin{split}
\langle\langle \hat{O}(t) \rangle\rangle^r
& = \eta -\frac{1}{2}\int_0^t\int_0^t dt'dt''\sum_{\A_{\sum},\tilde{\A}_{\sum} \in S} \lambda_{\A_{\lambda}}
y_{a'b'}^{ab}(t')y_{\tilde{a}'\tilde{b}'}^{\tilde{a}\tilde{b}}(t'') \langle\beta_{ab}(t')\beta_{\tilde{a}\tilde{b}}(t'')\rangle, \\
& = \eta -\frac{1}{4\pi}\int_{-\infty}^{\infty}d\omega \sum_{\A_{\sum},\tilde{\A}_{\sum}\in S} \lambda_{\A_{\lambda}}
F^{ab}_{a'b'}(\omega,t) F^{*^{-\tilde{a},-\tilde{b}}}_{-\tilde{a}',-\tilde{b}'}(\omega,t) S_{ab}^{\tilde{a}\tilde{b}}(\omega),\\
& \approx \eta -
\frac{M}{2T/r} \sum_{k=-\infty}^{\infty} \sum_{\A_{\sum},\tilde{\A}_{\sum}\in S} \lambda_{\A_{\lambda}} F^{ab}_{a'b'}(rk\omega_0,T/r) F^{*^{-\tilde{a},-\tilde{b}}}_{-\tilde{a}',-\tilde{b}'}(rk\omega_0,T/r) S_{ab}^{\tilde{a}\tilde{b}}(kr\omega_0),
\end{split}
\end{align}
where $t=MT/r$, $\A_{\Sigma}=a,b,a',b'$, $\tilde{\A}_{\Sigma}=\tilde{a},\tilde{b},\tilde{a}',\tilde{b}'$, $\A_{Z}=a',\A_{X}=b'$, $\A_{\lambda}= a',b',\tilde{a}',\tilde{b}',\tilde{a},\tilde{b}$, $\omega_0=2\pi/T$, and
\begin{align}\label{etalambda1}
\begin{split}
\eta
 = &
d \sum_{m,n\in S}
O_{mn}\xi^{mn}V_{id-m,id-n}, \\ 
\lambda_{\A_{\lambda}}
= &
_{\tilde{a}\tilde{b}}\lambda^{\tilde{a}'\tilde{b}'}_{a'b'}
=  d\sum_{m,n,p,q \in S} O_{mn}
V_{pq}.
\delta_{p,id-m-(a'+\tilde{a}')}.
\delta_{q,id-n-(b'+\tilde{b}')}.\\
& (1-\xi^{na'-mb'})
(1-\xi^{n\tilde{a}'-m\tilde{b}'})
\xi^{-\tilde{a}' b'
+ m(n+b'+\tilde{b}')
+(a'+\tilde{a}')(b'+\tilde{b}')}
\xi^{\tilde{a}\tilde{b}-\tilde{a}'\tilde{b}'}
\end{split}
\end{align} 
We assume the stationary condition of Eq.\eqref{stationary} for noise functions, and substitute Eqs.\eqref{S_quoct} and \eqref{Antimony-knownPolyspectra} into Eq.\eqref{quoct0} to find the following:
\begin{align}\label{AntimonyQNS}
\begin{split}
\langle\langle &\hat{O}(t=MT/r) \rangle\rangle^r \\
& = \eta -\frac{1}{2}\int_0^t\int_0^t dt'dt''
\sum_{\A_{\sum},\tilde{\A}_{\sum} \in S} \sum_{\substack{i,\tilde{i}=0,1,2
%,3 
\\ pqq'\in S\\ \tilde{p}\tilde{q}\tilde{q}'\in S}}
\tilde{\beta}^{pqq'}_{iab}
\tilde{\beta}^{\tilde{p}\tilde{q}\tilde{q}'}_{\tilde{i}\tilde{a}\tilde{b}}
\lambda_{\A_{\lambda}}
y^{ab}_{a'b'}(t')
y^{\tilde{a}\tilde{b}}_{\tilde{a}'\tilde{b}'}(t'') 
\langle B_i(t')B_{\tilde{i}}(t'')\rangle , \\
& = \eta -\frac{1}{4\pi}\int_{-\infty}^{\infty}d\omega \sum_{\A_{\sum},\tilde{\A}_{\sum}\in S}
\sum_{\substack{i,\tilde{i}=0,1
%,3 
\\ pqq'\in S\\ \tilde{p}\tilde{q}\tilde{q}'\in S}}
\tilde{\beta}^{pqq'}_{iab}
\tilde{\beta}^{\tilde{p}\tilde{q}\tilde{q}'}_{\tilde{i}\tilde{a}\tilde{b}}
\lambda_{\A_{\lambda}}
F^{ab}_{a'b'}(\omega,t) F^{*-\tilde{a},-\tilde{b}}_{-\tilde{a}',-\tilde{b}'}(\omega,t) \tilde{S}_{i\tilde{i}}(\omega),\\
& \approx \eta -
\frac{M}{2T/r} \sum_{k=-\infty}^{\infty} \times \\
& \sum_{\A_{\sum},\tilde{\A}_{\sum}\in S} 
\sum_{\substack{i,\tilde{i}=0,1
%,3 
\\ pqq'\in S\\ \tilde{p}\tilde{q}\tilde{q}'\in S}}
\tilde{\beta}^{pqq'}_{iab}
\tilde{\beta}^{\tilde{p}\tilde{q}\tilde{q}'}_{\tilde{i}\tilde{a}\tilde{b}}
\lambda_{\A_{\lambda}} 
F^{ab}_{a'b'}(rk\omega_0,T/r) F^{*-\tilde{a},-\tilde{b}}_{-\tilde{a}',-\tilde{b}'}(rk\omega_0,T/r) \tilde{S}_{i\tilde{i}}(kr\omega_0).
\end{split}
\end{align}
Here we assume that only one term of the Weyl decomposition of the quoct observable is nonzero and the initial state of the quoct would have only two non-zero terms:
\begin{equation}\label{mnfixed}
\hat{O}=O_{mn}Z^m X^n,\:\hat{\rho}(0)=\mathbbm{1}/8+V_{p_0,q_0}Z^{p_0} X^{q_0}
\end{equation}
We assumed $V_{00}=1/8$ to preserve the trace. The variables $O_{mn},\:V_{pq}$ and the fixed indices could be chosen arbitrarily for each measurement round of the Alvarez-Suter spectroscopy while $m,n\in \{0,\pm 1,\pm 2,\pm3,4\}$ and $p_0,q_0\in \{\pm 1,\pm 2,\pm3,4\}$.
The parameters $\eta$ and $\lambda_{\A_{\lambda}}$ in Eq.\eqref{etalambda1} that depend on these fixed chosen indices will be transformed to: 
\begin{align}\label{etalambdamnfixed-quoct}
\begin{split}
\eta_{mn}
&=
8O_{mn}V_{id-m,id-n} \xi^{m n},\\
\prescript{mnp_0q_0}{\tilde{a}\tilde{b}}{\lambda}^{\tilde{a}'\tilde{b}'}_{a'b'}
& \equiv  \:
\prescript{ind}{\tilde{a}\tilde{b}}{\lambda}^{\tilde{a}'\tilde{b}'}_{a'b'} \\
&= 
8\sum_{m,n,p,q \in S} O_{mn}
V_{pq}.
\delta_{p,id-m-(a'+\tilde{a}')}.
\delta_{q,id-n-(b'+\tilde{b}')}.\\
& (1-\xi^{na'-mb'})
(1-\xi^{n\tilde{a}'-m\tilde{b}'})
\xi^{-\tilde{a}' b'
+ m(n+b'+\tilde{b}')
+(a'+\tilde{a}')(b'+\tilde{b}')}
\xi^{\tilde{a}\tilde{b}-\tilde{a}'\tilde{b}'}\\
& = 
8O_{mn} 
(1-\xi^{na'-mb'})
(1-\xi^{n\tilde{a}'-m\tilde{b}'})
\xi^{-\tilde{a}' b'} \xi^{\tilde{a}\tilde{b}
-\tilde{a}'\tilde{b}'} \times\\
& \Big(
\frac{1}{8}
\delta_{-m,a'+\tilde{a}'}
\delta_{-n,b'+\tilde{b}'} \xi^{mn}
+\\
& V_{p_0,q_0} 
\delta_{-p_0,
+m +a'+\tilde{a}'}
\delta_{-q_0,
+n +b'+\tilde{b}'} \xi^{m(n + b'+\tilde{b}')+
(a'+\tilde{a}')(b'+\tilde{b}')} \Big),
\end{split}
\end{align}
where $ind = (mnp_0q_0)$.
%where $i\in \mathbbm{Z}$ and $(id+a'+b'-m) \in \{0,\pm 1\}$. 
The first term of above equation is related to the first term of the initial density matrix, that we ignore to simplify our simulation.
Note that since $m$, $n$, $p$ and $q$ are fixed parameters, 
there would be no longer summations over these variables in Eq.\eqref{etalambda1}, and the Delta Kronecker term imposes two constraints on varying parameters $a',\tilde{a}',b',\tilde{b}'$ as  $a'+\tilde{a}'=-(q_0+n)$ and $b'+\tilde{b}'=-(p_0+m)$. \\
The unknown complex noise spectra $S_{i\tilde{i}}(\omega)$ in Eq.\eqref{RealFuncs-Antimonyquoct}, are nine different functions Since $i,\tilde{i} \in \{0,\pm 1\}$, and according to appendices \ref{apdx.3.2} and \ref{RealFuncs} they consist of four independent functions introduced in Eq.\eqref{Antimony-unknownPolyspectra}. 
Considering the Alvarez-Suter spectroscopy rounds in our qudit noise spectroscopy formalism and substituting $T$ by $T/r$ and $\omega_0$ by $r\omega_0=2\pi/(T/r)$ in Eq.\eqref{AntimonyQNS}, we find the following equation in which the terms including $S_{2j}(\omega)\:(j=0,1,2)$ or $B_0$ are eliminated as discussed in appendix \ref{RealFuncs-Antimonyquoct}.
%Ignoring the terms $B_{i,\tilde{i}=0}(t)$.
\begin{align}\label{quoct6}
\begin{split}
\langle\langle \hat{O}(t_r^A &) \rangle\rangle^{r}  = A^r(t_r^A) \approx B^r(t_r^B),\:\:\:\:t_r^A=\frac{MT}{r},\:t_r^B=\frac{T}{r}
\end{split}
\end{align}
where,
\begin{align}\label{quoct6-A}
\begin{split}
 A^r(t_r^A)  
 = & \:
\eta_{mn} -\frac{1}{4\pi}
\int_{-\infty}^{\infty}d\omega \sum_{\substack{ab\tilde{a}\tilde{b}\\
a'b'\tilde{a}'\tilde{b}'\\
\in S}}
\sum_{\substack{i,\tilde{i}=0,1
%,2,3 
\\ pqq'\in S\\ \tilde{p}\tilde{q}\tilde{q}'\in S}}
\tilde{\beta}^{pqq'}_{iab}
\tilde{\beta}^{\tilde{p}\tilde{q}\tilde{q}'}_{\tilde{i}\tilde{a}\tilde{b}}
(\prescript{ind}{\tilde{a}\tilde{b}}{\lambda}^{\tilde{a}'\tilde{b}'}_{a'b'})  \\
& \times F^{ab}_{a'b'}(\omega,t_r^A) F^{*-\tilde{a},-\tilde{b}}_{-\tilde{a}',-\tilde{b}'}(\omega,t_r^A) \tilde{S}_{i\tilde{i}}(\omega)\\
 = & \: \eta_{mn} -\frac{1}{4\pi}
\int_{-\infty}^{\infty}d\omega \sum_{i,\tilde{i}=0,1
%,2,3
}
\Xi_{i\tilde{i}}^{mn}(\omega,t_r^A)
\tilde{S}_{i\tilde{i}}(\omega)
\end{split}
\end{align}
and,
\begin{align}\label{quoct6-B}
\begin{split}
B^r(t_r^B) 
& = 
\eta_{mn} -
\frac{M}{2T/r} \sum_{k=-\infty}^{\infty} 
\sum_{\substack{ab\tilde{a}\tilde{b}\\
a'b'\tilde{a}'\tilde{b}'\\
\in S}}
\sum_{\substack{i,\tilde{i}=0,1
%,2,3 
\\ pqq'\in S\\ \tilde{p}\tilde{q}\tilde{q}'\in S}}
\tilde{\beta}^{pqq'}_{iab}
\tilde{\beta}^{\tilde{p}\tilde{q}\tilde{q}'}_{\tilde{i}\tilde{a}\tilde{b}}
(\prescript{ind}{\tilde{a}\tilde{b}}{\lambda}^{\tilde{a}'\tilde{b}'}_{a'b'}) \\
& \times F^{ab}_{a'b'}(rk\omega_0,t_r^B) F^{*-\tilde{a},-\tilde{b}}_{-\tilde{a}',-\tilde{b}'}(rk\omega_0,t_r^B) \tilde{S}_{i\tilde{i}}(kr\omega_0) \\
 = & \: \eta_{mn} -
\frac{M}{2T/r} \sum_{k=-\infty}^{\infty} 
\sum_{i,\tilde{i}=0,1
%,2,3
}
\Xi_{i\tilde{i}}^{mn}(\omega,t_r^B)
\tilde{S}_{i\tilde{i}}(\omega),
\end{split}
\end{align}
considering,
\begin{align}\label{XiAntimony}
\begin{split}
&\Xi_{i\tilde{i}}^{mn}(\omega,t) = 
\sum_{\substack{ab\tilde{a}\tilde{b}\\
a'b'\tilde{a}'\tilde{b}'\\
\in S}}
\sum_{\substack{pqq'\in S\\ \tilde{p}\tilde{q}\tilde{q}'\in S}}
\tilde{\beta}^{pqq'}_{iab}
\tilde{\beta}^{\tilde{p}\tilde{q}\tilde{q}'}_{\tilde{i}\tilde{a}\tilde{b}}
(\prescript{ind}{\tilde{a}\tilde{b}}{\lambda}^{\tilde{a}'\tilde{b}'}_{a'b'})
F^{ab}_{a'b'}(\omega,t) 
F^{*-\tilde{a},-\tilde{b}}_{-\tilde{a}',-\tilde{b}'}(\omega,t).
%& = \eta_{mn} -\frac{1}{4\pi}\int_{0}^{\Omega}d\omega \sum_{i=1}^4 \alpha C_i(m,n,\omega,t_r^A)x_i(\omega),\\
\end{split}
\end{align}
Now using Eq.\eqref{Antimony-unknownPolyspectra}, we assign the unknown real polyspectra to the variables $x_i(\omega)$ as follows:
\begin{equation}\label{x_iVars}
\Big(x_0(\omega),x_1(\omega),x_2(\omega),x_3(\omega)\Big) = \Big( R_0(\omega),R'_0(\omega),R_1(\omega),I_1(\omega)  \Big)
\end{equation}
where the first three variables are even functions and the last one is an odd function. 
Assuming a finite frequency range for the noise spectroscopy protocol as $[-\Omega,\Omega]$, 
the approximation introduced in Eq.\eqref{full-half-freq-range}, and considering the variables in Eq.\eqref{x_iVars}, 
Eq.\eqref{quoct6-A} turns to the following (appendix \ref{Antimony-expectation} Eq.\eqref{Ar-Antimony-2}):
\begin{align}\label{Ar-Antimony-3}
\begin{split}
A^r(t_r^A) &= \eta_{mn} 
%+ \eta_A
-\frac{1}{2\pi}
\int_{\omega_0}^{\Omega}
d\omega 
\sum_{i=0}^2 
C_i^{mn}(\omega,t_r^A)
x_i(\omega),
\end{split}
\end{align}
and Eq.\eqref{quoct6-B} yields (appendix \ref{Antimony-expectation} Eq.\eqref{Br-Antimony}): 
\begin{align}\label{Br-Antimony-2}
\begin{split}
B^r(t_r^{B}) 
&= \eta_{mn} 
%+ \eta_B
-\frac{M}{T/r} \sum_{k=1}^N \sum_{i=0}^2 
C_i^{mn}(rk\omega_0,t_r^{B})
x_i(rk\omega_0)
%& = \eta_{mn} -
%\frac{M}{2t_r^B} \sum_{k=0}^N \sum_{i=1}^4 \alpha C_i(m,n,rk\omega_0,t_r^{B})x_i(rk\omega_0),\:t_r^B=\frac{T}{r}.
\end{split}
\end{align}
where $N=\floor{\Omega/\omega_0}$, and
%and,
\begin{align}\label{Ci-definitions}
\begin{split}
&C_0^{mn}(\omega,t) = 
\Xi_{00}^{mn}(\omega,t),\\
&C_1^{mn}(\omega,t) = 
\Xi_{11}^{mn}(\omega,t),\\
&C_2^{mn}(\omega,t) = 
\Xi_{01}^{mn}(\omega,t) +
\Xi_{10}^{mn}(\omega,t),
\end{split}
\end{align}
Note that the coefficients of variable $x_3(\omega)=I_1(\omega)$ are eliminated.
Now combining Eqs.\eqref{quoct6},\eqref{Ar-Antimony-3}, and \eqref{Br-Antimony-2}, we find:
\begin{align}\label{Antimony-final-0}
\begin{split}
&
%\eta_B
%-
\frac{M}{T/r} \sum_{k=1}^N \sum_{i=0}^2 
C_i^{mn}(rk\omega_0,t_r^{B})
x_i(rk\omega_0) \approx \
%\eta_A
%-
\frac{1}{2\pi} 
\sum_{i=0}^2
\tilde{A}_i^{rmn}
\end{split}
\end{align}
where,
\begin{align}\label{A_i^rmn}
\begin{split}
\tilde{A}_i^{rmn} = \int_{\omega_0}^{\Omega}
d\omega 
x_i(\omega) 
C_i^{mn}(\omega,t_r^{A})
\end{split}
\end{align}
This yields:
\begin{align}\label{Antimony-final-1}
\begin{split}
& 
%2\pi ( \eta_B - \eta_A )
%+ 
\sum_{i=0}^2
\tilde{A}_i^{rmn} \approx
Mr\omega_0 \sum_{k=1}^N \sum_{i=0}^2 
C_i^{mn}(rk\omega_0,t_r^{B})
x_i(rk\omega_0),
\end{split}
\end{align}
where $\omega_0 = \frac {2\pi}{T}$.

Here we simulate the qudit noise spectroscopy protocol for the Antimony quoct system. 
First remind that according to Eq.\eqref{x_iVars}, there are three unknown real noise functions $\{x_i(\omega)\}_{i=0}^2$, since the coefficient of the last one was eliminated in Eqs.\eqref{Ar-Antimony-3} and \eqref{Br-Antimony-2}.
We find and simulate each of these three functions at $N$ different frequencies with step $\delta\omega=\omega_0= 2\pi/T$ i.e.  $x_i(\omega_0),x_i(2\omega_0),...,x_i(N\omega_0)$.
% where $T/r$ is the total duration of the reference pulse sequence. 
The reference pulse sequence is of the form of Eq.\eqref{ReferencePulseSequence1} and is assumed as follows:
\begin{align}\label{Antimony-pulse-sequence}
\begin{split} %[0, 4, 5, 3, 6, 1, 7, 2]
\{ \tilde{P}_{0,4}, \tilde{P}_{4,5}, \tilde{P}_{5,3}, \tilde{P}_{3,6}, \tilde{P}_{6,1}, \tilde{P}_{1,7}, \tilde{P}_{7,2}, \tilde{P}_{2,0} \}
\end{split}
\end{align}
This particular sequence lead to non-zero coefficients for two noise polyspectra $S_{01}(\omega)$ and $S_{10}(\omega)$ in Eqs.\eqref{quoct6-A} and \eqref{quoct6-B}. These functions, according to Eqs.\eqref{Bi-beta-vars}, \eqref{S_quoct}, and \eqref{S_ii}, contain information about the correlations of noise processes $A(t)$ and $Q(t)$, introduced in Eq.\eqref{AntimonyquoctH}.

To explain how this reference pulse sequence leads to non-zero $S_{01}(\omega)$ or $S_{10}(\omega)$, we note that in our simulation, the coefficients of noise spectra $S_{01}(\omega)$ or $S_{10}(\omega)$ (or equivalently $R_1(\omega)$ and $I_1(\omega)$, following Eq.\eqref{Antimony-unknownPolyspectra}) must be non-zero. Given Eq.\eqref{x_iVars}, the coefficients of these noise functions are included in Eqs.\eqref{Ar-Antimony-3} and \eqref{Br-Antimony-2}, and are defined by Eq.\eqref{Ci-definitions}. They consist of $\Xi_{01}^{mn}(\omega,t)$ and $\Xi_{10}^{mn}(\omega,t)$, and to make these coefficients nonzero, according to Eq.\eqref{XiAntimony} and \eqref{Bi-beta-vars}, we need ($\rm b=0$ and $\rm \tilde{b}=\tilde{p}-\tilde{q}$) or ($\rm b=p-q$ and $\rm \tilde{b}=0$). The main challenge is then finding at least one nonzero $F_{a'b'}^{a,0}(\omega,t)$ or $F^{*-\tilde{a},0}_{-\tilde{a}',-\tilde{b}'}(\omega,t)$. Since this filter function is connected to the switching functions or the pulse sequence, we can explore random pulse sequences that yield non-zero $F_{a'b'}^{a,0}(\omega,t)$, and  Eq.\eqref{Antimony-pulse-sequence} serves as one example.

To simulate our quantum noise spectroscopy protocol for the Antimony quoct, since there are three unknown noise parameters $\{ x_i(\omega) \}_{i=0}^2$, we consider three Alvarez-Suter spectroscopy rounds. The quoct's measurable observables $\hat{O}_i$ and the initial states of the quoct $\hat{\rho}_i(t=0)$ for each of these three measurement rounds indexed by $i=0,1,2$ are assumed as follows:
\begin{align}\label{observablesandinitialrho}
\begin{split}
\hat{O_0} & =Z^1X^7,\:\: \hat{\rho}_0(0)=\mathbbm{1}/8+Z^0X^4,\\ 
\hat{O_1} & =Z^{2}X^1,\:\: \hat{\rho}_1(0) = \mathbbm{1}/8+Z^5X^1,\\
\hat{O_2} & =Z^2X^6,\:\: \hat{\rho}_2(0)=\mathbbm{1}/8+Z^1X^0,
\end{split}
\end{align}
To simplify the process of simulating our formalism, the trace preserving terms ($\mathbbm{1}/8$) in the initial density matrices are ignored.
\begin{figure}[!t]
\center
\includegraphics[width=1\linewidth]{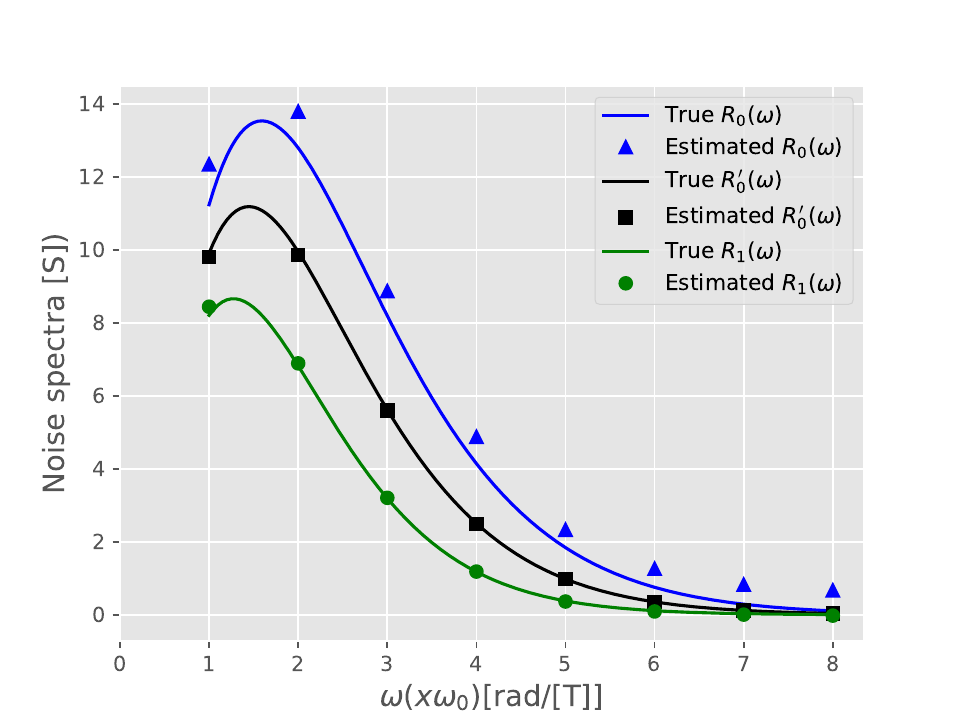}
\caption{This figure shows the simulation of our noise spectroscopy protocol for the Antimony quoct (d=8)  surrounded by XYZ-dephasing noises. The true polyspectra are assumed with Poissonian forms as $R_0(\omega)=\omega^2 e^{-0.2\omega}([S]), R^{\prime}_0(\omega)=\omega^2 e^{-0.22\omega}([S])$, and $R_1(\omega)=\omega^2 e^{-0.25\omega}([S])$. The other simulation parameters are as follows: $\gamma_n$, $\tilde{B}_0$, and $\hbar$ that have no effect on simulation results are assumed to be $1$ with general dimensions. The duration and iterations of reference sequence are $T=1\:[T]$ and $M=800$, and the measured observable of quoct and its initial density matrix are considered as $\hat{O}=ZX^7,\:Z^2X,\:Z^2X^6$ and $\hat{\rho}(0)=X^4,\:Z^5X,\:Z$, for the three measurement rounds, respectively. 
The maximum sensible frequency is $\Omega=55[\rm rad/[T]]$, and the gauging parameter for frequency is $\omega_0 = 2\pi [rad/[T]]$.}
\label{FigAntimonyquoct}
\end{figure}
We also assume the following Poissonian \textit{true} noise functions:
\begin{align}\label{Turespectra-Poissonian-Antimony}
\begin{split}
x_0(\omega) = & \omega^2 e^{-0.20 |\omega|} ([S]),\\ 
x_1(\omega) = & \omega^2 e^{-0.22 |\omega|} ([S]),\\
x_2(\omega) = & \omega^2 e^{-0.25 |\omega|} ([S]),  
\end{split}
\end{align}
Using Eqs.\eqref{Simulation-b}, \eqref{Simulation-A}, and the above assumptions, we find the environmental noise functions around the quoct.

Fig.\ref{FigAntimonyquoct} demonstrates the simulation of our generalized noise spectroscopy protocol developed for the Antimony quoct spectator surrounded by XYZ-dephasing noise. 
The parameters $\gamma_n$ and $\tilde{B}_0$, which are introduced in Eq.\eqref{AntimonyquoctH}, are assumed to be 1. Their units are in the Natural Unit System. This is because we assumed $\hbar =1$ in Eq.\eqref{UnitaryGeneral}, and in our simulations, when using Eq.\eqref{I-ComputationalBasis}. In addition, these parameters are appeared on both sides of Eq.\eqref{quoct6}, and their values or units do not influence the outcome of our simulations.
The other parameters, like
the duration of the reference pulse sequence, the number of its repetitions, and the maximum sensible noise frequency in the simulation are considered as $T= 1([T])$, $M=800$, and $\Omega=55 (rad/[T])$, respectively.

The running time of our simulation was about 10 hours with a x64-based processor of 11th Gen Intel(R) Core(TM) i7-1165G7, 2.80 GHz, and 16.0 GB RAM.
The link to the simulation code is provided in the Code Availability section. 
 
It can be seen from the simulations of Figs.\ref{FigAntimonyquoct} %and \ref{FigAntimonyquoct2} 
that our noise spectroscopy formalism is good to reconstruct the true noise polyspectra surrounded the Antimony quoct, given the chosen reference pulse sequence, and the adequate number of reference sequence repetitions.

In section \ref{SecIII}, we demonstrated the accuracy of our formalism for a qutrit system by examining two distinct types of true polyspectra: Poissonian and general non-Poissonian forms. In section \ref{SecIV}, our focus was exclusively on non-Poissonian forms of true polyspectra to apply our noise spectroscopy formalism to a ququad and a quoct subjected to Z-type dephasing noise. This section (\ref{SecV}) was dedicated to employing Poissonian forms of true polyspectra to simulate our formalism for the Antimony quoct. 
Selecting the Poissonian form of noise polyspectra was sufficient to validate our qudit noise spectroscopy formalism for the following reasons:

First, our noise spectroscopy formalism, represented by the linear matrix Eq.\eqref{LinearMatrixEquation}, was derived under the assumption of an environmental stochastic and stationary XYZ-dephasing noise with zero-mean Gaussian distributions. Given these assumptions, the mathematical structure of Eq.\eqref{LinearMatrixEquation} ensures that our formalism can reconstruct any arbitrary form of noise polyspectra with $\omega_0$ resolution, including those with 
$1/\omega$ dependency, as demonstrated in Section \ref{SecIII} for qutrit systems. Therefore, testing our generalized noise spectroscopy protocol using Poissonian polyspectra is equivalent to that of using non-Poissonian polyspectra.

Second, selecting Poissonian forms simplifies the computation of true noise polyspectra, particularly in the integral process, that is the parameter $A^r(t_r^A)$ in Eqs.\eqref{qutritsimulation} , \eqref{quditI-simulation-03}, and \eqref{Ar-Antimony-3}. Such selection only simplifies the introduced integration processes for extracting equivalent experimental data, to make the numerical calculations more straightforward. However, this does not affect the applicability of our formalism.
For practical applications of our simulation code in experimental scenarios, the integral process is omitted. Instead, the simulation replaces it with actual experimental data derived from measurements.

Thus, employing Poissonian polyspectra ensures the validity of our formalism while simplifying the computation of true noise spectra, making the simulation process more efficient.

\section{Conclusion}
We've established a new theoretical framework for quantum noise spectroscopy using a comb-based approach, incorporating the Alvarez-Suter spectroscopy protocol with resonant pulse sequences. 
A unique aspect of our approach is the utilization of the Weyl basis decomposition and the qudit systems. 
We related any observable property of a qudit with the attributes of the pulses and the unidentified environmental noise functions within the entire Weyl basis. 

Our protocol can identify not just environmental Z-type dephasing noises but also general XYZ-dephasing noises — a feature that many other noise spectroscopy methods lack. 

Moreover, our QNS protocol is designed for qudits of any dimension \textit{d}. When applied to an ion qudit, as opposed to an ion qubit, it interacts across a greater number of energy levels and corresponding coherences with its target environment. This results in an enhanced sensitivity to environmental noise. 
Beyond intensified sensitivity, our method also presents a more compact alternative to employing networks of multiple two-level qubit sensors.

We advanced our QNS framework to address three distinct scenarios. Two of these focused on refining the entire QNS method using two variations of a Reduced Weyl basis, primarily aimed at detecting Z-type dephasing noises. 
The first of these Reduced Weyl cases was tailored for a qutrit system. The second one, while being crafted for a qudit system with a general 'd' energy level, was simulated specifically for a ququad (d=4) and a quoct (d=8) spectator. 

Our final enhancement to the QNS protocol catered to a comprehensive Weyl basis decomposition, making it compatible with all XYZ-dephasing noise types. This version was tested on the Antimony quoct (d=8) spectator experiencing two different noise varieties. 

In the studies detailed here, our simulations were run for authentic Poissonian noise functions and true noise functions of various arbitrary forms. 

During these simulations, we assumed Natural Unit System. In addition, the pulse duration measurements used the arbitrary unit $[T]$, while the extracted noise spectra in the frequency domain utilized the unit $rad/[T]$, mirroring the unit of the pulse sequence. 

The resolution of the achieved noise spectra is represented by $\omega_0=2\pi/T$, where $T$ denotes the length of the reference pulse sequence. This resolution marks the gap between successive frequencies for which the associated noise spectra are gauged using our QNS method. To find the fine details or spur-line of the noise spectra in an experimental application of our QNS protocol, one must enhance the noise estimation resolution, represented by $\omega_0=2\pi/T$, by increasing the duration of the reference pulse sequence, $T$. 

The simulations indicate that, although the theoretical design of the framework suggests a demand for numerous iterations of the reference pulse sequence (i.e., $M \rightarrow \infty$), a substantially high $M$ isn't essential to recognize the actual patterns. The specific parameters of each simulated spectator and the given true spectra determine a threshold $M$, past which the system's precise noise spectra can be captured.

In upcoming work, our QNS framework can be expanded and tested for a broader range of qudit systems, including higher-order natural or synthetic qudits as referenced in \cite{PhysRevLett.119.187702}, and for a sequence of qudits that exhibit spatial inter-qudit dependencies.

\section*{Code Availability}
\addcontentsline{toc}{section}{Code Availability}
The codes demonstrating the simulations of this work are  available at the following url: \url{https://github.com/AUSMURI/Qudit-noise-spectroscopy}.
\begin{acknowledgments}
\addcontentsline{toc}{section}{Acknowledgment}
We acknowledge the valuable instructions, consultations, and reviews provided by A/Professor Gerardo Paz-Silva. The funding for this work was supported by the Australian Government via the AUSMURI grant AUSMURI000002.
\end{acknowledgments}
\bibliographystyle{naturemag}
%\bibliography{apssamp}
\bibliography{Qudit.bib}

\begin{thebibliography}{10}
\expandafter\ifx\csname url\endcsname\relax
  \def\url#1{\texttt{#1}}\fi
\expandafter\ifx\csname urlprefix\endcsname\relax\def\urlprefix{URL }\fi
\providecommand{\bibinfo}[2]{#2}
\providecommand{\eprint}[2][]{\url{#2}}

\bibitem{Plenio_2008}
\bibinfo{author}{Plenio, M.~B.} \& \bibinfo{author}{Huelga, S.~F.}
\newblock \bibinfo{title}{Dephasing-assisted transport: quantum networks and biomolecules}.
\newblock \emph{\bibinfo{journal}{New J. Phys.}} \textbf{\bibinfo{volume}{10}}, \bibinfo{pages}{113019} (\bibinfo{year}{2008}).
\newblock \urlprefix\url{https://doi.org/10.1088/1367-2630/10/11/113019}.

\bibitem{PhysRevA.90.042313}
\bibinfo{author}{Javaherian, C.} \& \bibinfo{author}{Twamley, J.}
\newblock \bibinfo{title}{Robustness of optimal transport in one-dimensional particle quantum networks}.
\newblock \emph{\bibinfo{journal}{Phys. Rev. A}} \textbf{\bibinfo{volume}{90}}, \bibinfo{pages}{042313} (\bibinfo{year}{2014}).
\newblock \urlprefix\url{https://link.aps.org/doi/10.1103/PhysRevA.90.042313}.

\bibitem{PhysRevA.79.032318}
\bibinfo{author}{Ng, H.~K.} \& \bibinfo{author}{Preskill, J.}
\newblock \bibinfo{title}{Fault-tolerant quantum computation versus gaussian noise}.
\newblock \emph{\bibinfo{journal}{Phys. Rev. A}} \textbf{\bibinfo{volume}{79}}, \bibinfo{pages}{032318} (\bibinfo{year}{2009}).
\newblock \urlprefix\url{https://link.aps.org/doi/10.1103/PhysRevA.79.032318}.

\bibitem{PhysRevLett.110.010502}
\bibinfo{author}{Novais, E.} \& \bibinfo{author}{Mucciolo, E.~R.}
\newblock \bibinfo{title}{Surface code threshold in the presence of correlated errors}.
\newblock \emph{\bibinfo{journal}{Phys. Rev. Lett.}} \textbf{\bibinfo{volume}{110}}, \bibinfo{pages}{010502} (\bibinfo{year}{2013}).
\newblock \urlprefix\url{https://link.aps.org/doi/10.1103/PhysRevLett.110.010502}.

\bibitem{Boixo2018}
\bibinfo{author}{Boixo, S.}, \bibinfo{author}{Isakov, S.}, \bibinfo{author}{Smelyanskiy, V.} \& \bibinfo{author}{et~al.}
\newblock \bibinfo{title}{Characterizing quantum supremacy in near-term devices}.
\newblock \emph{\bibinfo{journal}{Nature Phys}} \textbf{\bibinfo{volume}{14}}, \bibinfo{pages}{595–600} (\bibinfo{year}{2018}).
\newblock \urlprefix\url{https://www.nature.com/articles/s41567-018-0124-x#citeas}.

\bibitem{Johnson2017}
\bibinfo{author}{Johnson, P.~D.}, \bibinfo{author}{Romero, J.}, \bibinfo{author}{Olson, J.}, \bibinfo{author}{Cao, Y.} \& \bibinfo{author}{Aspuru-Guzik, A.}
\newblock \bibinfo{title}{Qvector: An algorithm for device-tailored quantum error correction}.
\newblock \emph{\bibinfo{journal}{arXiv:1711.02249}}  (\bibinfo{year}{2017}).
\newblock \urlprefix\url{https://doi.org/10.48550/arXiv.1711.02249}.

\bibitem{PhysRevLett.92.117905}
\bibinfo{author}{Faoro, L.} \& \bibinfo{author}{Viola, L.}
\newblock \bibinfo{title}{Dynamical suppression of $1/f$ noise processes in qubit systems}.
\newblock \emph{\bibinfo{journal}{Phys. Rev. Lett.}} \textbf{\bibinfo{volume}{92}}, \bibinfo{pages}{117905} (\bibinfo{year}{2004}).
\newblock \urlprefix\url{https://link.aps.org/doi/10.1103/PhysRevLett.92.117905}.

\bibitem{PhysRevLett.96.010401}
\bibinfo{author}{Giovannetti, V.}, \bibinfo{author}{Lloyd, S.} \& \bibinfo{author}{Maccone, L.}
\newblock \bibinfo{title}{Quantum metrology}.
\newblock \emph{\bibinfo{journal}{Phys. Rev. Lett.}} \textbf{\bibinfo{volume}{96}}, \bibinfo{pages}{010401} (\bibinfo{year}{2006}).
\newblock \urlprefix\url{https://link.aps.org/doi/10.1103/PhysRevLett.96.010401}.

\bibitem{PhysRevLett.104.133601}
\bibinfo{author}{Wasilewski, W.} \emph{et~al.}
\newblock \bibinfo{title}{Quantum noise limited and entanglement-assisted magnetometry}.
\newblock \emph{\bibinfo{journal}{Phys. Rev. Lett.}} \textbf{\bibinfo{volume}{104}}, \bibinfo{pages}{133601} (\bibinfo{year}{2010}).
\newblock \urlprefix\url{https://link.aps.org/doi/10.1103/PhysRevLett.104.133601}.

\bibitem{PhysRevA.87.022324}
\bibinfo{author}{Wu, R.~B.} \emph{et~al.}
\newblock \bibinfo{title}{Spectral analysis and identification of noises in quantum systems}.
\newblock \emph{\bibinfo{journal}{Phys. Rev. A}} \textbf{\bibinfo{volume}{87}}, \bibinfo{pages}{022324} (\bibinfo{year}{2013}).
\newblock \urlprefix\url{https://link.aps.org/doi/10.1103/PhysRevA.87.022324}.

\bibitem{PhysRevA.92.010302}
\bibinfo{author}{Rossi, M. A.~C.} \& \bibinfo{author}{Paris, M. G.~A.}
\newblock \bibinfo{title}{Entangled quantum probes for dynamical environmental noise}.
\newblock \emph{\bibinfo{journal}{Phys. Rev. A}} \textbf{\bibinfo{volume}{92}}, \bibinfo{pages}{010302} (\bibinfo{year}{2015}).
\newblock \urlprefix\url{https://link.aps.org/doi/10.1103/PhysRevA.92.010302}.

\bibitem{PhysRevApplied.5.014007}
\bibinfo{author}{Zwick, A.}, \bibinfo{author}{\'Alvarez, G.~A.} \& \bibinfo{author}{Kurizki, G.}
\newblock \bibinfo{title}{Maximizing information on the environment by dynamically controlled qubit probes}.
\newblock \emph{\bibinfo{journal}{Phys. Rev. Applied}} \textbf{\bibinfo{volume}{5}}, \bibinfo{pages}{014007} (\bibinfo{year}{2016}).
\newblock \urlprefix\url{https://link.aps.org/doi/10.1103/PhysRevApplied.5.014007}.

\bibitem{PhysRevA.95.022121}
\bibinfo{author}{Paz-Silva, G.~A.}, \bibinfo{author}{Norris, L.~M.} \& \bibinfo{author}{Viola, L.}
\newblock \bibinfo{title}{Multiqubit spectroscopy of gaussian quantum noise}.
\newblock \emph{\bibinfo{journal}{Phys. Rev. A}} \textbf{\bibinfo{volume}{95}}, \bibinfo{pages}{022121} (\bibinfo{year}{2017}).
\newblock \urlprefix\url{https://link.aps.org/doi/10.1103/PhysRevA.95.022121}.

\bibitem{1456701}
\bibinfo{author}{Thomson, D.~J.}
\newblock \bibinfo{title}{Spectrum estimation and harmonic analysis}.
\newblock \emph{\bibinfo{journal}{Proceedings of the IEEE}} \textbf{\bibinfo{volume}{70}}, \bibinfo{pages}{1055--1096} (\bibinfo{year}{1982}).

\bibitem{Fray2017}
\bibinfo{author}{Frey, V.~M.} \emph{et~al.}
\newblock \bibinfo{title}{Application of optimal band-limited control protocols to quantum noise sensing}.
\newblock \emph{\bibinfo{journal}{Nature Communications}} \textbf{\bibinfo{volume}{8}} (\bibinfo{year}{12017}).
\newblock \urlprefix\url{https://www.nature.com/articles/s41467-017-02298-2}.

\bibitem{Fray2020}
\bibinfo{author}{Frey, V.~M.}, \bibinfo{author}{Norris, L.~M.}, \bibinfo{author}{Viola, L.} \& \bibinfo{author}{Biercuk, M.~J.}
\newblock \bibinfo{title}{Application of optimal band-limited control protocols to quantum noise sensing}.
\newblock \emph{\bibinfo{journal}{Phys. Rev. Applied}} \textbf{\bibinfo{volume}{14}}, \bibinfo{pages}{024021} (\bibinfo{year}{2020}).
\newblock \urlprefix\url{https://journals.aps.org/prapplied/abstract/10.1103/PhysRevApplied.14.024021}.

\bibitem{6773659}
\bibinfo{author}{{Slepian}, D.} \& \bibinfo{author}{{Pollak}, H.~O.}
\newblock \bibinfo{title}{Prolate spheroidal wave functions, fourier analysis and uncertainty — i}.
\newblock \emph{\bibinfo{journal}{The Bell System Technical Journal}} \textbf{\bibinfo{volume}{40}}, \bibinfo{pages}{43--63} (\bibinfo{year}{1961}).
\newblock \urlprefix\url{https://ieeexplore.ieee.org/document/6773659}.

\bibitem{6773660}
\bibinfo{author}{{Landau}, H.~J.} \& \bibinfo{author}{{Pollak}, H.~O.}
\newblock \bibinfo{title}{Prolate spheroidal wave functions, fourier analysis and uncertainty — ii}.
\newblock \emph{\bibinfo{journal}{The Bell System Technical Journal}} \textbf{\bibinfo{volume}{40}}, \bibinfo{pages}{65--84} (\bibinfo{year}{1961}).
\newblock \urlprefix\url{https://onlinelibrary.wiley.com/doi/abs/10.1002/j.1538-7305.1961.tb03977.x}.

\bibitem{6773467}
\bibinfo{author}{{Landau}, H.~J.} \& \bibinfo{author}{{Pollak}, H.~O.}
\newblock \bibinfo{title}{Prolate spheroidal wave functions, fourier analysis and uncertainty — iii: The dimension of the space of essentially time- and band-limited signals}.
\newblock \emph{\bibinfo{journal}{The Bell System Technical Journal}} \textbf{\bibinfo{volume}{41}}, \bibinfo{pages}{1295--1336} (\bibinfo{year}{1962}).
\newblock \urlprefix\url{https://ieeexplore.ieee.org/document/6773467}.

\bibitem{6773515}
\bibinfo{author}{{Slepian}, D.}
\newblock \bibinfo{title}{Prolate spheroidal wave functions, fourier analysis and uncertainty — iv: Extensions to many dimensions; generalized prolate spheroidal functions}.
\newblock \emph{\bibinfo{journal}{The Bell System Technical Journal}} \textbf{\bibinfo{volume}{43}}, \bibinfo{pages}{3009--3057} (\bibinfo{year}{1964}).
\newblock \urlprefix\url{https://onlinelibrary.wiley.com/doi/abs/10.1002/j.1538-7305.1964.tb01037.x}.

\bibitem{6771595}
\bibinfo{author}{{Slepian}, D.}
\newblock \bibinfo{title}{Prolate spheroidal wave functions, fourier analysis, and uncertainty — v: the discrete case}.
\newblock \emph{\bibinfo{journal}{The Bell System Technical Journal}} \textbf{\bibinfo{volume}{57}}, \bibinfo{pages}{1371--1430} (\bibinfo{year}{1978}).
\newblock \urlprefix\url{https://ieeexplore.ieee.org/document/6771595}.

\bibitem{1083556}
\bibinfo{author}{{Papoulis}, A.} \& \bibinfo{author}{{Bertran}, M.}
\newblock \bibinfo{title}{Digital filtering and prolate functions}.
\newblock \emph{\bibinfo{journal}{IEEE Transactions on Circuit Theory}} \textbf{\bibinfo{volume}{19}}, \bibinfo{pages}{674--681} (\bibinfo{year}{1972}).
\newblock \urlprefix\url{https://ieeexplore.ieee.org/document/1083556}.

\bibitem{PhysRevLett.107.230501}
\bibinfo{author}{\'Alvarez, G.~A.} \& \bibinfo{author}{Suter, D.}
\newblock \bibinfo{title}{Measuring the spectrum of colored noise by dynamical decoupling}.
\newblock \emph{\bibinfo{journal}{Phys. Rev. Lett.}} \textbf{\bibinfo{volume}{107}}, \bibinfo{pages}{230501} (\bibinfo{year}{2011}).
\newblock \urlprefix\url{https://link.aps.org/doi/10.1103/PhysRevLett.107.230501}.

\bibitem{PhysRevA.98.032315}
\bibinfo{author}{Norris, L.~M.} \emph{et~al.}
\newblock \bibinfo{title}{Optimally band-limited spectroscopy of control noise using a qubit sensor}.
\newblock \emph{\bibinfo{journal}{Phys. Rev. A}} \textbf{\bibinfo{volume}{98}}, \bibinfo{pages}{032315} (\bibinfo{year}{2018}).
\newblock \urlprefix\url{https://link.aps.org/doi/10.1103/PhysRevA.98.032315}.

\bibitem{PhysRevB.89.020503}
\bibinfo{author}{Yoshihara, F.} \emph{et~al.}
\newblock \bibinfo{title}{Flux qubit noise spectroscopy using rabi oscillations under strong driving conditions}.
\newblock \emph{\bibinfo{journal}{Phys. Rev. B}} \textbf{\bibinfo{volume}{89}}, \bibinfo{pages}{020503} (\bibinfo{year}{2014}).
\newblock \urlprefix\url{https://link.aps.org/doi/10.1103/PhysRevB.89.020503}.

\bibitem{PhysRevLett.110.146804}
\bibinfo{author}{Dial, O.~E.} \emph{et~al.}
\newblock \bibinfo{title}{Charge noise spectroscopy using coherent exchange oscillations in a singlet-triplet qubit}.
\newblock \emph{\bibinfo{journal}{Phys. Rev. Lett.}} \textbf{\bibinfo{volume}{110}}, \bibinfo{pages}{146804} (\bibinfo{year}{2013}).
\newblock \urlprefix\url{https://link.aps.org/doi/10.1103/PhysRevLett.110.146804}.

\bibitem{PhysRevA.90.032319}
\bibinfo{author}{Zhao, N.}, \bibinfo{author}{Wrachtrup, J.} \& \bibinfo{author}{Liu, R.-B.}
\newblock \bibinfo{title}{Dynamical decoupling design for identifying weakly coupled nuclear spins in a bath}.
\newblock \emph{\bibinfo{journal}{Phys. Rev. A}} \textbf{\bibinfo{volume}{90}}, \bibinfo{pages}{032319} (\bibinfo{year}{2014}).
\newblock \urlprefix\url{https://link.aps.org/doi/10.1103/PhysRevA.90.032319}.

\bibitem{PhysRevLett.114.017601}
\bibinfo{author}{Romach, Y.} \emph{et~al.}
\newblock \bibinfo{title}{Spectroscopy of surface-induced noise using shallow spins in diamond}.
\newblock \emph{\bibinfo{journal}{Phys. Rev. Lett.}} \textbf{\bibinfo{volume}{114}}, \bibinfo{pages}{017601} (\bibinfo{year}{2015}).
\newblock \urlprefix\url{https://link.aps.org/doi/10.1103/PhysRevLett.114.017601}.

\bibitem{PhysRevLett.116.150503}
\bibinfo{author}{Norris, L.~M.}, \bibinfo{author}{Paz-Silva, G.~A.} \& \bibinfo{author}{Viola, L.}
\newblock \bibinfo{title}{Qubit noise spectroscopy for non-gaussian dephasing environments}.
\newblock \emph{\bibinfo{journal}{Phys. Rev. Lett.}} \textbf{\bibinfo{volume}{116}}, \bibinfo{pages}{150503} (\bibinfo{year}{2016}).
\newblock \urlprefix\url{https://link.aps.org/doi/10.1103/PhysRevLett.116.150503}.

\bibitem{PhysRevA.94.012109}
\bibinfo{author}{Sza\ifmmode~\acute{n}\else \'{n}\fi{}kowski, P.}, \bibinfo{author}{Trippenbach, M.} \& \bibinfo{author}{Cywi\ifmmode~\acute{n}\else \'{n}\fi{}ski, L.}
\newblock \bibinfo{title}{Spectroscopy of cross correlations of environmental noises with two qubits}.
\newblock \emph{\bibinfo{journal}{Phys. Rev. A}} \textbf{\bibinfo{volume}{94}}, \bibinfo{pages}{012109} (\bibinfo{year}{2016}).
\newblock \urlprefix\url{https://link.aps.org/doi/10.1103/PhysRevA.94.012109}.

\bibitem{PhysRevA.97.032101}
\bibinfo{author}{Sza\ifmmode~\acute{n}\else \'{n}\fi{}kowski, P.} \& \bibinfo{author}{Cywi\ifmmode~\acute{n}\else \'{n}\fi{}ski, L.}
\newblock \bibinfo{title}{Accuracy of dynamical-decoupling-based spectroscopy of gaussian noise}.
\newblock \emph{\bibinfo{journal}{Phys. Rev. A}} \textbf{\bibinfo{volume}{97}}, \bibinfo{pages}{032101} (\bibinfo{year}{2018}).
\newblock \urlprefix\url{https://link.aps.org/doi/10.1103/PhysRevA.97.032101}.

\bibitem{PhysRevX.5.021009}
\bibinfo{author}{Loretz, M.} \emph{et~al.}
\newblock \bibinfo{title}{Spurious harmonic response of multipulse quantum sensing sequences}.
\newblock \emph{\bibinfo{journal}{Phys. Rev. X}} \textbf{\bibinfo{volume}{5}}, \bibinfo{pages}{021009} (\bibinfo{year}{2015}).
\newblock \urlprefix\url{https://link.aps.org/doi/10.1103/PhysRevX.5.021009}.

\bibitem{PhysRevE.98.042206}
\bibinfo{author}{Mourik, V.} \emph{et~al.}
\newblock \bibinfo{title}{Exploring quantum chaos with a single nuclear spin}.
\newblock \emph{\bibinfo{journal}{Phys. Rev. E}} \textbf{\bibinfo{volume}{98}}, \bibinfo{pages}{042206} (\bibinfo{year}{2018}).
\newblock \urlprefix\url{https://link.aps.org/doi/10.1103/PhysRevE.98.042206}.

\bibitem{PhysRevLett.119.187702}
\bibinfo{author}{Godfrin, C.} \emph{et~al.}
\newblock \bibinfo{title}{Operating quantum states in single magnetic molecules: Implementation of grover's quantum algorithm}.
\newblock \emph{\bibinfo{journal}{Phys. Rev. Lett.}} \textbf{\bibinfo{volume}{119}}, \bibinfo{pages}{187702} (\bibinfo{year}{2017}).
\newblock \urlprefix\url{https://link.aps.org/doi/10.1103/PhysRevLett.119.187702}.

\end{thebibliography}
\addcontentsline{toc}{section}{References}

\appendix
\section{Switching functions- The effect of spectroscopy pulses on noisy qudits}
\subsection{Weyl decomposed Hamiltonian}\label{apdx.A.1}
In this section, our objective is to determine the effective Hamiltonian of the qudit in Weyl basis, taking into account the XYZ-dephasing noises and spectroscopy pulses. We find this Hamiltonian at arbitrary time instances within one reference pulse sequence. First we assume the effective Hamiltonian of the qudit and noises in Weyl basis as of the second line of Eq.\eqref{HWRW}:
\begin{align}\label{apdxWeylH}
\begin{split}
H  &=\sum_{a,b \in S}\beta_{ab}(t)Z^a X^b
\end{split}
\end{align}
Then the qudit is being exposed to a sequence of resonant pulses where each pulse is represented by Eq.\eqref{unitarypulse}, and the symmetric reference pulse sequence is as Eq.\eqref{symmetricsequence0} or equivalently Eq.\eqref{ReferencePulseSequence1} as follows:
\begin{equation}\label{apdxsymmetricsequence}
\{ ( P^{-1}_{(i_r,j_r)}  ,  P_{(i_r,j_r)}  )  \}_{r=0,...,d-1},
\end{equation} 
with $\tau_r$ as the interval between each pair of pulses, during which the qudit is experiencing the free time evolution with no exposure to any pulses. The unitary operator that applies the whole reference pulse sequence to the qudit initial state  is as follows:
\begin{equation}\label{Unitary}
U = \prod_{r=d-1}^{0} P_{(i_r,j_r)}^{-1}U_{\tau_r}P_{(i_r,j_r)},\:\:\:\:\: U_{\tau_r}= e^{-I/\hbar\: H \: \tau_r}  
\end{equation}
where $U_{\tau_r}$ is the free evolution operator of the qudit affected only by noises during the time interval $\tau_r$. The reverse order of $r$ ensures that the operators of the first pulse are first applied to the qudit's quantum state. %($T=\sum_{i=1}^{N}\tau_i$)
Using the following mathematical expression:
\begin{align}\label{expFormula}
\begin{split}
P^{-1}_{(i,j)}e^{\hat{\Phi}}P_{(i,j)} & = P_{(i,j)}^{-1}(1 + \hat{\Phi} + \frac{1}{2!}\hat{\Phi}^2 +...)P_{(i,j)} \\
 & = 1 + \{P_{(i,j)}^{-1}\hat{\Phi} P_{(i,j)} \} +\frac{1}{2!} \{P_{(i,j)}^{-1}\hat{\Phi} (P_{(i,j)} \} \{P_{i,j}^{-1}) \hat{\Phi} P_{(i,j)} \} +...
= e^{P_{(i,j)}^{-1} \hat{\Phi} P_{(i,j)}}, 
\end{split}
\end{align}
the unitary operator applying the total reference pulse sequence to the qudit is as follows:
\begin{align}\label{TotalU&Hr}
\begin{split}
U    = \prod_{r=d-1}^{0} e^{-I/\hbar P_{(i_r,j_r)}^{-1}H P_{(i_r,j_r)}}   =& \prod_{r=d-1}^{0} e^{-I/\hbar \: H_r  \: \tau_r}  =   e^{-I/\hbar \sum_{r=0}^{d-1}H_r\: \tau_r}, \\
H_r (t \in \tau_r) =&  P_{(i_r,j_r)}^{-1}H P_{(i_r,j_r)},
\end{split}
\end{align} 
$H_r$ is the effective Hamiltonian of the qudit, noises, and the two pulses $\{P_{(i_r,j_r)}^{-1},P_{(i_r,j_r)}\}$ during the interval $\tau_r$. 
So, the effective Hamiltonian $H_r(t)$ is known at any arbitrary time $t \in \tau_r, \:r=0,\ldots,d-1$ during the spectroscopy process. To find $H_r$ we first calculate the term $P_{(i_r,j_r)}^{-1} Z^a X^b P_{(i_r,j_r)}$ as following:
\begin{align}\label{A.1-Weyl}
\begin{split}
& P_{(i,j)}^{-1} Z^a X^b P_{(i,j)}   \\
=&\{ \ket{i}\bra{j} + \ket{j}\bra{i} + \mathbbm{1} - \ket{i}\bra{i} - \ket{j}\bra{j} \}Z^a X^b
 \{\ket{i}\bra{j} + \ket{j}\bra{i} + \mathbbm{1} - \ket{i}\bra{i} - \ket{j}\bra{j} \}\\
= & \{( \ket{i}\bra{j}\xi^{a j} + \ket{j}\bra{i}\xi^{a i} + Z^a - \ket{i}\bra{i}\xi^{a i} - \ket{j}\bra{j}\xi^{a j} \}\\
&\{ \ket{i \oplus b}\bra{j} + \ket{j \oplus b}\bra{i} + X^{b} - \ket{i \oplus b}\bra{i} - \ket{j \oplus b}\bra{j} ) \}\\
=&  \Big( \ket{i}\bra{j}\xi^{a j} \Big) \{ \ket{i \oplus b}\bra{j} + \ket{j \oplus b}\bra{i} + X^{b} - \ket{i \oplus b}\bra{i} - \ket{j \oplus b}\bra{j} ) \}\\
+& \Big( \ket{j}\bra{i}\xi^{a i} \Big) \{ \ket{i \oplus b}\bra{j} + \ket{j \oplus b}\bra{i} + X^{b} - \ket{i \oplus b}\bra{i} - \ket{j \oplus b}\bra{j} ) \}\\
+& \Big( Z^a \Big) \{ \ket{i \oplus b}\bra{j} + \ket{j \oplus b}\bra{i} + X^{b} - \ket{i \oplus b}\bra{i} - \ket{j \oplus b}\bra{j} ) \}\\
+& \Big(  - \ket{i}\bra{i}\xi^{a i} \Big) \{ \ket{i \oplus b}\bra{j} + \ket{j \oplus b}\bra{i} + X^{b} - \ket{i \oplus b}\bra{i} - \ket{j \oplus b}\bra{j} ) \}\\
+& \Big( - \ket{j}\bra{j}\xi^{a j} \Big) \{ \ket{i \oplus b}\bra{j} + \ket{j \oplus b}\bra{i} + X^{b} - \ket{i \oplus b}\bra{i} - \ket{j \oplus b}\bra{j} ) \}\\
= &  \Big( \xi^{a j} \Big) 
\{ 
\delta_{j,i\oplus b} \ket{i}\bra{j} + 
\ket{i}\bra{j \oplus -b} - 
\delta_{j,i \oplus b}\ket{i}\bra{i}) 
\}\\
+ & \Big( \xi^{a i} \Big) 
\{ 
\delta_{i,j \oplus b}\ket{j}\bra{i} + 
\ket{j}\bra{i \oplus -b} - 
\delta_{i,j \oplus b} \ket{j}\bra{j} ) 
\}\\
+ & 
\{ \xi^{a (i \oplus b)}\ket{i \oplus b}\bra{j} + 
\xi^{a (j \oplus b)}\ket{j \oplus b}\bra{i} + 
Z^aX^{b} - 
\xi^{a (i \oplus b)}\ket{i \oplus b}\bra{i} - 
\xi^{a (j \oplus b)}\ket{j \oplus b}\bra{j} ) 
\}\\
+ & \Big(  - \xi^{a i} \Big) 
\{ \delta_{i,j \oplus b}\ket{i}\bra{i} + 
\ket{i}\bra{i \oplus -b} - 
\delta_{i,j \oplus b}\ket{i}\bra{j} ) 
\}\\
+ & \Big( - \xi^{a j} \Big) 
\{ 
\delta_{j,i \oplus b}\ket{j}\bra{j} + 
\ket{j}\bra{j \oplus -b} - 
\delta_{j,i \oplus b}\ket{j}\bra{i} )
\}.\\
\end{split}
\end{align}
We further simplify the above equation as follows:
\begin{align}\label{pijZXpij}
\begin{split}
& P_{(i,j)}^{-1} Z^a X^b P_{(i,j)} =
-  
\Big(\delta_{j,i \oplus b}\xi^{a j}+\delta_{i,j \oplus b}\xi^{a i}\Big)
\Big(\ket{i}\bra{i}+\ket{j}\bra{j}\Big) \\
+ & 
\Big(\delta_{j,i \oplus b} \xi^{a j}+\delta_{i,j \oplus b}\xi^{a i}\Big)\Big(\ket{i}\bra{j} + \ket{j}\bra{i} \Big) \\
+ & 
\Big(   
\xi^{a (i \oplus b)}\ket{i \oplus b}\bra{j} + 
\xi^{a (j \oplus b)}\ket{j \oplus b}\bra{i}  - 
\xi^{a (i \oplus b)}\ket{i \oplus b}\bra{i} - 
\xi^{a (j \oplus b)}\ket{j \oplus b}\bra{j} 
\Big) \\
+ &
\Big( 
\xi^{a j} \ket{i}\bra{j \oplus -b}
+ \xi^{a i} \ket{j}\bra{i \oplus -b} 
- \xi^{a i} \ket{i}\bra{i \oplus -b} 
- \xi^{a j} \ket{j}\bra{j \oplus -b} 
\Big) \\
+ & Z^aX^{b},
\end{split}
\end{align}
where we used Eq.\eqref{generalizedPauliMatrices} and their conjugate form.
Next, we proceed with the decomposition of the term $\ket{i}\bra{j}$ using the Weyl basis. In order to determine the decomposing coefficients in the following, 
we multiply both sides of the expression from right and left sides by $Z^{-k}$ and $X^{-l}$, respectively, where $k,l \in S$ are two arbitrary fixed indices. Then we calculate the trace of both sides of the equation as follows: 
\begin{align}\label{ij-0}
\begin{split}
\ket{i}\bra{j} = &\sum_{a,b\in S} C_{ab} Z^aX^b \\ 
\operatorname{Tr}(Z^{-k}\ket{i}\bra{j}X^{-l}) = & \operatorname{Tr}(\sum_{a,b\in S} C_{ab} Z^{a-k}X^{b-l})\\
\xi^{-ik} \operatorname{Tr}(\ket{i}\bra{j \oplus l}) = & \sum_{a,b\in S} C_{ab} \operatorname{Tr}(Z^{a-k}X^{b-l}) \\
\xi^{-ik}\delta_{i,j \oplus l} = & \sum_{a,b\in S} C_{ab}\delta_{a,k}\delta_{b,l}\operatorname{Tr}(\mathbbm{1})\\
 C_{ab}= & C_{k,l} = \frac{1}{d}\xi^{-ai}\delta_{i,j \oplus b},
\end{split}
\end{align}
To find the fourth line of the above equation, we used the fact that the powers of the generalized Pauli matrices are traceless unless with zero power ($a=k,b=l$) that yields the unit matrix. It could be also calculated as follows:
\begin{align}\label{traceZX}
\begin{split}
\operatorname{Tr}(Z^{\alpha}X^{\beta}) = 
\sum _{l=0}^{d-1}  \bra{l} Z^{\alpha}X^{\beta} \ket{l} =
\sum _{l=0}^{d-1} \xi^{\alpha l}  \bra{l} \ket{l\oplus \beta} = 
\delta_{\beta,0}\sum _{l=0}^{d-1} \xi^{\alpha l} = d. \delta_{\beta,0} \delta_{\alpha,0}.
\end{split}
\end{align}
where we used Eq.\eqref{generalizedPauliMatrices}, $\bra{l}Z^{\alpha} = \xi^{\alpha l} \bra{l}$, and the geometric series 
$\sum_{l=0}^{n-1} ar^{l} = an\:\: (\rm if \:\:r=1)  $ and   
$\sum_{l=0}^{n-1} ar^{l} = a\frac{1-r^n}{1-r},\:\:(\rm otherwise) $.
Now Eq.\eqref{ij-0} turns to the following:
\begin{align}\label{ij}
\begin{split}
\ket{i}\bra{j} = & \frac{1}{d}\sum_{a,b\in S} \xi^{-ai} \delta_{i,j \oplus b} Z^a X^b = 
\frac{1}{d}\sum_{\substack{a\in S\\b=i-j}} \xi^{-ai} Z^a X^b ,
= \frac{1}{d}\sum_{a\in S} \xi^{-a(i+j)}X^{-j} Z^a X^i\\
\hookrightarrow \ket{i}\bra{i} = & \frac{1}{d}\sum_{a\in S} \xi^{-a i} Z^a.
\end{split}
\end{align}
In the first row of the above equation we used $\delta_{i,j \oplus b}=\delta_{i-j, \oplus b}=\delta_{i-j, nd+ b}=\delta_{i-j,b}$, where we assumed $n=0$ for simplifying the notation, and in the second row we used $\ket{i}\bra{j}=(\ket{j}\bra{i})^*$. %($\beta(t)$ is a real function)
Substituting Eq.\eqref{ij} into Eq.\eqref{pijZXpij} we find: 
\begin{align}\label{pijZXpij-0}
\begin{split}
&P_{(i,j)}^{-1} Z^a X^b P_{(i,j)} = 
\frac{1}{d}\sum_{a' \in S} Z^{a'}\Big(
-(\xi^{-a' i}
+\xi^{-a' j})(\delta_{j-i, b}\xi^{a j}+\delta_{i-j, b}\xi^{a i})X^0\\
&+\xi^{-a' i}(\delta_{j-i, b} \xi^{a j}
+\delta_{i-j, b}\xi^{a i})X^{i-j}\\
&+\xi^{-a' j}(\delta_{j-i, b} \xi^{a j}
+\delta_{i-j, b}\xi^{a i})X^{j-i}\\
&+\xi^{-a'(i+b)}\xi^{a(i+b)}X^{(i+b) -j}\\
&+\xi^{-a'(j+b)}\xi^{a(j+b)}X^{(j+ b)-i}\\
&-\xi^{-a'(i+b)}\xi^{a (i+b)}X^{b}\\
&-\xi^{-a'(j+b)}\xi^{a (j+b)}X^{b}\\
&+\xi^{-a' i}\xi^{a j} X^{i-(j-b)}\\
&+\xi^{-a' j}\xi^{a i}X^{j-(i-b)}\\
&-\xi^{-a' i}\xi^{a i}X^{i-(i-b)}\\
&-\xi^{-a' j}\xi^{a j}X^{j-(j-b)}\\
&+\xi^{-a' i}d\delta_{a,a'}X^b\Big)
\end{split}
\end{align}
The above simplification process proceeds as follows:
\begin{align}\label{pijZXpij-1}
\begin{split}
&P_{(i,j)}^{-1} Z^a X^b P_{(i,j)} = 
\frac{1}{d}\sum_{a' \in S} Z^{a'}\Big(
-(\xi^{-a' i}+\xi^{-a' j})(\delta_{j,i +b}\xi^{a j}
+\delta_{i,j+ b}\xi^{a i})X^0\\
&+\xi^{-a' i}(\delta_{j,i+ b} \xi^{a j}
+\delta_{i,j+ b}\xi^{a i})X^{i-j}\\
&+\xi^{-a' j}(\delta_{j,i +b} \xi^{a j}
+\delta_{i,j +b}\xi^{a i})X^{j-i}\\
&+\xi^{-a'i}(\xi^{-a'b+a(i+b)}
+\xi^{aj})X^{i-j+b}\\
&+\xi^{-a'j}(\xi^{-a'b+a(j+b)}
+\xi^{ai})X^{j-i+b}\\
&-(\xi^{-a'i}d\delta_{a,a'}
+\xi^{i(a-a')}+\xi^{j(a-a')}
+\xi^{(a-a')(i+b)}
+\xi^{(a-a')(j+b)})X^b,
\end{split}
\end{align}
that finally yields:
\begin{align}\label{pijZXpij-2}
\begin{split}
&P_{(i,j)}^{-1} Z^a X^b P_{(i,j)} = 
\frac{1}{d}\sum_{a' \in S} Z^{a'}\Big(
-(\xi^{-a' i}
+\xi^{-a' j})(\delta_{j,i +b}\xi^{a j}
+\delta_{i,j+ b}\xi^{a i})X^0\\
&+\xi^{-a' i}(\delta_{j,i+ b} \xi^{a j}
+\delta_{i,j+ b}\xi^{a i})X^{i-j}\\
&+\xi^{-a' j}(\delta_{j,i +b} \xi^{a j}
+\delta_{i,j +b}\xi^{a i})
X^{j-i}\\
&+(\xi^{(a-a')(i+b)}
+\xi^{-a'i+aj})X^{i-j+b}\\
&+(\xi^{(a-a')(j+b)}
+\xi^{-a'j+ai})X^{j-i+b}\\
&+(-(\xi^{(a-a')i}
+\xi^{(a-a')j})(\xi^{(a-a')b} +1)+
\xi^{-a'i}
d\delta_{a,a'})
X^b
\Big).
\end{split}
\end{align}
The above equation can be expressed in the form of the Weyl basis as follows where only specific powers of the generalized Pauli $X$ operator are nonzero:
\begin{align}\label{switchingfunction}
\begin{split}
P_{(i,j)}^{-1} Z^a X^b P_{(i,j)} = & \sum_{a',b'\in S} y_{a' b'}^{i j a b} Z^{a'} X^{b'},\\
y_{a',0}^{i j a b} =& \frac{-1}{d}(\xi^{-a' i}+\xi^{-a' j})(\delta_{j,i+ b}\xi^{a j}+\delta_{i,j+ b}\xi^{a i}),\\
y_{a',i-j}^{i j a b} =& \frac{1}{d}\xi^{-a' i}(\delta_{j,i +b} \xi^{a j}+\delta_{i,j +b}\xi^{a i}),\\
y_{a',j-i}^{i j a b} =& \frac{1}{d}\xi^{-a' j}(\delta_{j,i +b} \xi^{a j}+\delta_{i,j +b}\xi^{a i}),\\
y_{a',i-j+b}^{i j a b} =& \frac{1}{d}(\xi^{(a-a')(i+b)}+\xi^{-a'i+aj}),\\
y_{a',j-i +b}^{i j a b} =& \frac{1}{d}(\xi^{(a-a')(j+b)}+\xi^{-a'j+ai}),\\
y_{a',b}^{i j a b} =& \frac{1}{d}(-(\xi^{(a-a')i}+\xi^{(a-a')j})(\xi^{(a-a')b}+1)+\xi^{-a'i}d \: \delta_{a,a'}),\\
y_{a',b'\notin S'}^{i j a b} =& 0,\:S'=\{ 0,i-j, j-i, i-j+b, j-i+b, b \}\in S.
\end{split}
\end{align}
Using Eqs.\eqref{HWRW}, \eqref{TotalU&Hr}, and \eqref{switchingfunction}, the $H_r$ effective Hamiltonian of the qudit, noise, and pulses during the time interval $\tau_r$ would be as follows:
\begin{align}\label{Heffective-general}
\begin{split}
H_r(t\:\in \tau_r) &= P_{(i_r,j_r)}^{-1} H P_{(i_r,j_r)}  = \sum_{a,b\in S}\beta_{ab}(t)P_{(i_r,j_r)}^{-1} Z^a X^b P_{(i_r,j_r)}
= \sum_{a,b,a',b'\in S}\beta_{ab}(t)y_{a' b'}^{i_r j_r a b} Z^{a'}X^{b'},\\
\hookrightarrow H_r(t) &=\sum_{a',b'\in S}Y_{a'b'}(t)Z^{a'}X^{b'},\:\:\:
Y_{a'b'} = \sum_{a,b\in S}\beta_{ab}(t)y_{a'b'}^{a b} (t),\:t\:\in \tau_r, 
\end{split}
\end{align}
where we assumed $y_{a' b'}^{i_r j_r a b }=y_{a' b'}^{a b} (t\in \tau_r)$, that is defined in Eq.\eqref{switchingfunction}, and is called the switching function. The general Weyl decomposition coefficient $Y_{a' b'}(t)$ in the above equation consists of the switching function associated with spectroscopy pulses and $\beta_{ab}(t)$ function associated with the general XYZ-dephasing noises. 

In our noise spectroscopy protocol, Eq.\eqref{Heffective-general} represents the most comprehensive form of the effective Hamiltonian of a qudit, noises, and two pulses during each pulse interval that is decomposed in the general Weyl basis. 

The unitary operator responsible for the reference sequence is now found by substituting Eq.\eqref{Heffective-general} into Eq.\eqref{TotalU&Hr}:
\begin{align}\label{ReferenceSequenceUnitary}
\begin{split}
U = exp\{-\frac{I}{\hbar} \sum_{r=0}^{d-1} ( \sum_{a,b,a',b'\in S}\beta_{ab}(t) y_{a' b'}^{i_r j_r a b}  Z^{a'}X^{b'})\tau_r\}
\end{split}
\end{align}
\subsection{Reduced Weyl Hamiltonian}\label{apdx.A.2}
Within this section, our goal is to derive the effective Hamiltonian of a qudit, taking into account the effects of noises and two pulses, and perform its decomposition using the Reduced Weyl basis.
To determine the effective Hamiltonian, we employ a similar procedure as in the previous section, leading to Eq.\eqref{TotalU&Hr}. Subsequently, to achieve the equivalent Reduced Weyl decomposition $H_r$, we utilize Eq.\eqref{HWRW} (first line), which represents the effective Hamiltonian of the qudit and noises. We then proceed to calculate the term $P_{(i_r,j_r)}^{-1} Z^a P_{(i_r,j_r)}$ in the following manner:
\begin{align}\label{A.1}
\begin{split}
 P_{(i,j)}^{-1} Z^a P_{(i,j)} = & 
\{ \ket{i}\bra{j} + \ket{j}\bra{i} + \mathbbm{1} - \ket{i}\bra{i} - \ket{j}\bra{j} \}\times \\
& Z^a \{\ket{i}\bra{j} + \ket{j}\bra{i} + \mathbbm{1} - \ket{i}\bra{i} - \ket{j}\bra{j} \}\\
= & \{( \ket{i}\bra{j} + \ket{j}\bra{i} + \mathbbm{1} - \ket{i}\bra{i} - \ket{j}\bra{j} \}\times\\
& \{ \xi^{a i} \ket{i}\bra{j} + \xi^{a j} \ket{j}\bra{i} + Z^{a} - \xi^{a i} \ket{i}\bra{i} - \xi^{a j} \ket{j}\bra{j} ) \}\\
= & (\xi^{a i} - \xi^{a j})(\ket{j}\bra{j} - \ket{i}\bra{i} ) + Z^a,
\end{split}
\end{align}
where we made use of Eq.\eqref{generalizedPauliMatrices}, their conjugates, and also  replaced $-a$ with $a$. 
Now we substitute Eq.\eqref{ij} into Eq.\eqref{A.1} to achieve the following result:
%%($\beta(t)$ is a real function)
\begin{align}\label{A.3}
\begin{split}
P_{(i,j)}^{-1} Z^a P_{(i,j)} & =
 (\xi^{a i} - \xi^{a j})\Big( \sum_{m\in S} \frac{1}{d}\xi^{-mj} Z^m 
 \sum_{n \in S}\frac{1}{d} \xi^{-ni} Z^n \Big) + Z^a  \\
& =  \frac{1}{d}(\xi^{a i} - \xi^{a j}) \sum_{m\in S} (\xi^{-mj} - \xi^{-mi}) Z^m + Z^a  
\end{split}
\end{align}
%Thus, we can achieve the effective Hamiltonian $H_r$ in the following manner:
Consequently, we can derive the effective Hamiltonian $H_r$ by following the subsequent approach:
\begin{flalign}\label{A.4-1-2}
\begin{split}
H_r & = P_{(i_r,j_r)}^{-1} H P_{(i_r,j_r)} = \sum_{a\in S}\beta_{a}(t)P_{(i_r,j_r)}^{-1} Z^aP_{(i_r,j_r)}\\
& = \sum_{a\in S}\beta_{a}(t)\Big(\frac{1}{d}(\xi^{a i} - \xi^{a j}) \sum_{m\in S} (\xi^{-mj} - \xi^{-mi}) Z^m + Z^a\Big)\\
& = \frac{1}{d}\sum_{a\in S} \beta_{a}(t) (\xi^{a i_r} - \xi^{a j_r}) \sum_{m\in S} (\xi^{-mj_r} - \xi^{-mi_r}) Z^m + \sum_{a\in S} \beta_{a}(t) Z^a \\
& =\frac{1}{d}\sum_{a,m\in S} \beta_{a}(t) (\xi^{a i_r} - \xi^{a j_r})(\xi^{-mj_r} - \xi^{-mi_r}) Z^m + \sum_{m\in S} \beta_{m}(t) Z^m \\
& =\frac{1}{d}\sum_{m\in S} \Big(\sum_{a=\in S} \beta_{a}(t) (\xi^{a i_r} - \xi^{a j_r})(\xi^{-mj_r} - \xi^{-mi_r}) + \beta_{m}(t)\Big) Z^m,
\end{split}
\end{flalign}
that is equivalent to the following:
\begin{flalign}\label{Heffective-ReducedWeyl}
\begin{split}
H_r & =\sum_{m\in S}Y_{m}(t)Z^m =\sum_{a,m\in S}\beta_{a}(t) y_{m,a}(t)(t)Z^m,\:\:\:\:\: t\in \tau_r\\
Y_{m}(t) & = \beta_{m}(t) + \frac{1}{d}(\xi^{-mj_r} - \xi^{-mi_r})\sum_{a=\in S} \beta_{a}(t) (\xi^{a i_r} - \xi^{a j_r}) ,\\
y_{m,a}(t) & = \delta_{m,a} +\frac{1}{d}(\xi^{-mj_r} - \xi^{-mi_r})(\xi^{a i_r} - \xi^{a j_r})
\end{split}
\end{flalign}
Eq.\eqref{Heffective-ReducedWeyl} shows the effective Hamiltonian of a qudit, incorporating noises and two pulses within each interval of our noise spectroscopy protocol, which has been decomposed using the Reduced Weyl basis.
\subsection{Reduced Weyl Hamiltonian (Qutrit)}\label{apdx.A.3}
In this section, we demonstrate that the qutrit version of Eq.\eqref{Heffective-ReducedWeyl} has a more abstract structure.
Initially, we note that Eq.\eqref{generalizedPauliMatrices} results in:
%First note that due to the subsequent relations:
\begin{equation}
\begin{split}
Z^d=\mathbbm{1},\: X^d\ket{i}=\ket{i\oplus d}=\mathbbm{1}\ket{i} \\
\hookrightarrow Z^{-a}=Z^{d-a},\: X^{-a}=X^{d-a},
\end{split}    
\end{equation}
So, the indexing of the Weyl decomposition set, denoted by $S=\{0,1,2,...,d-1\}$, has the following alternative form:
\begin{equation}\label{Sequivalents}
\begin{split}
S &= \{0,\pm1,...,\pm\frac{(d-1)}{2}\}\: \rm odd\: d,\\
&= \{0,\pm1,...,\pm\frac{(d-1)}{2},(\frac{d}{2}\:or\:\frac{-d}{2})\}\: \rm even\: d\\
&= \{0\} + S_+ + S_-, 
\end{split}    
\end{equation}
where $S_+$ and $S_-$ has the positive and negative values of $S$, respectively, excluding $0$ term. The above symmetric indexing would be more advantageous within the formalism of this section.
In the subsequent analysis, we utilize Eq.\eqref{Heffective-ReducedWeyl} to compute the effective Hamiltonian of the qutrit influenced by the pulse sequence described in Eq.\eqref{qutritpulse}. This involves determining the function $Y_m(t)$ that corresponds to the following time intervals:
\begin{equation}
\tau_1:\:0\le t < T/7,\:\tau_2:\:T/7\le t < 2T/5,\:\tau_3:\:2T/5\le t < T.
\end{equation}
When applying the pulse indices $(i_1,j_1)=(1,-1)$ for duration $\tau_1$ to Eq.\eqref{Heffective-ReducedWeyl}, we find:
\begin{align}\label{A.5}
\begin{split}
Y_{0}(t)= & \beta_{0}(t)\\
Y_{1}(t)= & \beta_{1}(t) + \frac{1}{3}(\xi^{1} - \xi^{-1})\sum_{a=\in S} \beta_{a}(t) (\xi^{a i_r} - \xi^{a j_r})\\
= & \beta_{1}(t) + \frac{1}{3}(\xi^{1} - \xi^{-1})\Big(\beta_{1}(t) (\xi^{1} - \xi^{-1}) + \beta_{-1}(t) (\xi^{-1} - \xi^{1}) \Big)\\
= &\beta_{1}(t) + \frac{1}{3}(\xi^{1} - \xi^{-1})^2(\beta_{1}(t) - \beta_{-1}(t))
= \beta_{-1}(t)\\
Y_{-1}(t)= &\beta_{-1}(t) + \frac{1}{3}(\xi^{-1} - \xi^{1})\Big( \beta_{1}(t) (\xi^{1} - \xi^{-1}) +\beta_{-1}(t) (\xi^{-1} - \xi^{1}) \Big)\\
= & \beta_{-1}(t) + \frac{1}{3}(\xi^{-1} - \xi^{1})^2(\beta_{-1}(t)-\beta_{1}(t))
= \beta_{1}(t)\\
H_1 = & P_{1,-1}^{-1} H P_{1,-1} = \beta_{0}(t) Z^{0} + \beta_{1}(t) Z^{-1} + \beta_{-1}(t) Z^{1}  \\
= & \sum_{a=\{0,\pm1\}} Y_{a} (t)Z^a = \sum_{a=\{0,\pm1\}}y_{-a}(t)\beta_{-a}(t)Z^a ;\\
y_{0}(t) = &  1 , \: y_{1}(t) = 1,\:y_{-1}(t) = 1
\end{split}
\end{align}
For $t \in \tau_2$ and pulse indices $(i_2,j_2)=(-1,0)$ we have:  
\begin{align}\label{A.6}
\begin{split}
Y_{0}(t) = & \beta_{0}(t),\\
Y_{1}(t) = & \beta_{1}(t) + \frac{1}{3}(\xi^{0} - \xi^{1})\Big(\beta_{1}(t) (\xi^{-1} - \xi^{0}) + \beta_{-1}(t) (\xi^{1} - \xi^{0}) \Big)\\
= & \beta_{1}(t)\Big( 1 + \frac{1}{3}(\xi^{0} - \xi^{1})(\xi^{-1} - \xi^{0})\Big) + \beta_{-1}(t)\Big(\frac{1}{3}(\xi^{0} - \xi^{1})(\xi^{1} - \xi^{0}) \Big)\\
= & \beta_{1}(t)\Big( 1 + \frac{1}{3}(-3)\Big) 
+ \beta_{-1}(t)\Big(-\frac{1}{3}(1+\xi+\xi^2-3\xi^2) \Big) = \xi \beta_{-1}(t)\\
Y_{-1}(t) = & \beta_{-1}(t) + \frac{1}{3}(\xi^{0} - \xi^{-1})\Big(\beta_{1}(t) (\xi^{-1} - \xi^{0}) + \beta_{-1}(t) (\xi^{1} - \xi^{0}) \Big)\\
= & \beta_{-1}(t)\Big( 1 + \frac{1}{3}(\xi^{0} - \xi^{-1})(\xi^{1} - \xi^{0})\Big) + \beta_{1}(t)\Big(\frac{1}{3}(\xi^{0} - \xi^{-1})(\xi^{-1} - \xi^{0}) \Big)\\
= & \beta_{-1}(t)\Big( 1 + \frac{1}{3}(-3)\Big)
+ \beta_{1}(t)\Big(-\frac{1}{3}(1+\xi^2+\xi-3\xi) \Big) = \xi^{-1}\beta_{1}(t)\\
H_2 = & P_{-1,0}^{-1} H P_{-1,0} = \beta_{0}(t) \mathbbm{1} + \beta_{1}(t) \xi^{-1} Z^{-1} + \beta_{-1}(t) \xi^{1} Z^{1} \\
= & \sum_{a=\{0,\pm1\}}Y_{a}(t)Z^a = \sum_{a=\{0,\pm1\}}\beta_{-a}(t)y_{-a}(t)Z^a ;\\
y_{0}(t) = & 1 ,\: y_{-1}(t) = \xi^{1},\:y_{1}(t) = \xi^{-1}
\end{split}
\end{align}
Lastly, for $t$ within the interval $\tau_3$ and with pulse indices $(i_3,j_3)=(-1,0)$, we get:
\begin{align}\label{A.7}
\begin{split}
Y_{0}(t) = &\beta_{0}(t),\\
Y_{1}(t) = &\beta_{1}(t) + \frac{1}{3}(\xi^{-1} - \xi^{0})\Big(\beta_{1}(t) (\xi^{0} - \xi^{1}) + \beta_{-1}(t) (\xi^{0} - \xi^{-1}) \Big)\\
=&\beta_{1}(t)\Big( 1 + \frac{1}{3}(\xi^{-1} - \xi^{0})(\xi^{0} - \xi^{1})\Big) + \beta_{-1}(t)\Big(\frac{1}{3}(\xi^{-1} - \xi^{0})(\xi^{0} - \xi^{-1}) \Big)\\
=&\beta_{1}(t)\Big( 1 + \frac{1}{3}(-3)\Big) 
+ \beta_{-1}(t)\Big(\frac{1}{3}(3\xi^{-1}) \Big) = \xi^{-1} \beta_{-1}(t)\\
Y_{-1}(t) =& \beta_{-1}(t) + \frac{1}{3}(\xi^{1} - \xi^{0})\Big(\beta_{1}(t) (\xi^{0} - \xi^{1}) + \beta_{-1}(t) (\xi^{0} - \xi^{-1}) \Big)\\
=&\beta_{-1}(t)\Big( 1 + \frac{1}{3}(\xi^{1} - \xi^{0})(\xi^{0} - \xi^{-1})\Big) + \beta_{1}(t)\Big(\frac{1}{3}(\xi^{1} - \xi^{0})(\xi^{0} - \xi^{1}) \Big)\\
=&\beta_{-1}(t)\Big( 1 + \frac{1}{3}(-3)\Big)
+ \beta_{1}(t)\Big(\frac{1}{3}(3\xi) \Big) = \xi \beta_{1}(t)\\
H_3 =& P_{0,1}^{-1} H P_{0,1} = \beta_{0}(t) \mathbbm{1} + \beta_{1}(t) \xi^{1} Z^{-1} + \beta_{-1}(t) \xi^{-1} Z^{1}  \\
=& \sum_{a=\{0,\pm1\}}Y_{a}(t)Z^a = \sum_{a=\{0,\pm1\}}\beta_{-a}(t)y_{-a}(t)Z^a ;\\
y_{0}(t) =& 1 ,\: y_{-1}(t) = \xi^{-1},\:y_{1}(t) = \xi^{1}
\end{split}
\end{align}
It could be seen from Eqs.\eqref{A.5}, \eqref{A.6}, and \eqref{A.7} that the effective Hamiltonian of the qutrit, with the given pulse sequence, has a simpler form than that of Eq.\eqref{Heffective-ReducedWeyl} as follows:
\begin{align}\label{qutritHamiltonian}
\begin{split}
H&=\sum_{a=\{0,\pm1\}}\beta_{-a}(t)y_{-a}(t)Z^a,\\
\{y_{a} (t\in \tau_1) &= 1,\:y_{a}(t\in \tau_2)=\xi^{-a},\:y_{a} (t\in \tau_3) = \xi^{a}\},\: \xi=e^{2\pi i/3}
\end{split}
\end{align}
\section{Qudit's expectation values}\label{apdx.2}
 \label{apdx.2.1}
This appendix demonstrates the connection between the expectation value of an arbitrary qudit's observable, the surrounding noise spectra, and the  switching functions (pulse series). In this appendix, $i=\sqrt{-1}$ unless used as a variable index.
\subsection{Weyl decomposed Hamiltonians}
In this subsection, we find the aforementioned relation for the most comprehensive form of the qudit effective Hamiltonian $H_r(t\in\tau_r)$ of Eq.\eqref{Heffective-general}.
The expectation value of the qudit's arbitrary observable $\langle \hat{O} \rangle$ can be decomposed in Weyl basis as Eq.\eqref{Weyldecompositions}. To simplify the calculation of $\langle \hat{O} \rangle$, we find the expectation value of an arbitrary Weyl term $\langle Z^m X^n\rangle$, in the following manner, for a qudit with density matrix $\hat{\rho}(t)$. 
%The notation is simplified considering $H_r(t\in\tau_r) =H(t)$.
\begin{align}\label{start}
\begin{split}
\langle Z^m X^n\rangle 
= &\operatorname{Tr}(\hat{\rho}(t)Z^m X^n) 
= \operatorname{Tr}(U_H\hat{\rho}(0)(Z^m X^nX^{-n}Z^{-m})U^*_H Z^m X^n)\\
= &\operatorname{Tr}(X^{-n}Z^{-m}U^*_H Z^m X^n U_H\hat{\rho}(0)Z^m X^n) \\
= &\operatorname{Tr}(X^{-n}Z^{-m}\mathcal{T}_- e^{i\int_0^t H(t')dt'} Z^m X^n \mathcal{T}_+e^{-i\int_0^t H(t')dt'} \hat{\rho}(0)Z^m X^n) \\
= &\operatorname{Tr}(\mathcal{T}_-e^{i\int_0^t X^{-n}Z^{-m} H(t') Z^m X^n dt'}\mathcal{T}_+e^{-i\int_0^t H(t')dt'} \hat{\rho}(0)Z^m X^n) 
\end{split}
\end{align}
where,
\begin{align}\label{UnitaryGeneral}
\begin{split}
U_H =& \mathcal{T}_+ e^{-i\int_0^t H(t')dt'}
= \mathcal{T}_+ e^{-i( H(0)dt + H(dt)2dt + \dots + H(t-dt)Ndt )}
\\
= & 
e^{-i H(t-dt)Ndt} e^{-i H(t-2dt)(N-1)dt}\dots 
e^{-i H(dt)2dt}e^{-i H(0)dt},\:N \rightarrow \infty
\end{split}
\end{align}
is the unitary operator responsible for the qudit evolution with the following unified effective Hamiltonian (Eq.\eqref{Hgeneral}) at arbitrary time $t$.
\begin{align}\label{apdxHgeneral}
\begin{split}
H(t) =& \sum_{\mathscr{A}_{\sum}\in S} \beta_{\mathscr{A}_{\beta}}(t) y_{\mathscr{A}_y}(t) Z^{\mathscr{A}_z} X^{\mathscr{A}_x},
\end{split}
\end{align}
$U_H$ could be found by applying $\prod_{r=d-1}^{0} \rightarrow   \mathcal{T}_+$, $\sum_{0}^{d-1}\tau_r \rightarrow \int_0^t dt$, and $\hbar=1$ to Eq.\eqref{TotalU&Hr}, where $\mathcal{T}_+$ is the time-ordering operator, and its conjugated form is expressed as follows:
\begin{align}\label{UnitaryGeneralDagger}
\begin{split}
U_H^* =& \mathcal{T}_- e^{i\int_0^t H(t')dt'}
= \mathcal{T}_- e^{i( H(0)dt + H(dt)2dt + \dots + H(t-dt)Ndt )}
\\ 
= & 
e^{i H(0)dt} e^{i H(dt)2dt}\dots 
e^{i H(t-2dt)(N-1)dt}e^{i H(t-dt)Ndt}.
\end{split}
\end{align}
Using Eq.\eqref{Heffective-general}, %that includes the $\A_i$ indices/powers given in table \ref{table2}, 
we can further simplify Eq.\eqref{start} as follows:
\begin{align}\label{B.1.1}
\begin{split}
&\langle \langle Z^m X^n\rangle \rangle
= \\
&\operatorname{Tr}
(\mathcal{T}_-
e^{i\int_0^t \sum_{\A_{\sum} \in S} \langle \beta_{\A_{\beta}}(t^{\prime})\rangle y_{\A_{y}}(t') 
X^{-n}Z^{-m} Z^{\A_Z} X^{\A_X} Z^m X^n dt'}
\mathcal{T}_ +e^{-i\int_0^t \langle H(t')\rangle dt'} \hat{\rho}(0)Z^m X^n) \\
= &
\operatorname{Tr}(\mathcal{T}_-
e^{i\int_0^t\sum_{\A_{\sum} \in S} \langle \beta_{\A_{\beta}}(t^{\prime})\rangle y_{\A_{y}}(t') 
X^{-n}Z^{-m}  Z^{\A_Z} \xi^{-m\A_X} Z^m X^{\A_X}    X^n dt'} \\
&\:\:\:\:\:\:\:\:\:\:\:\:\:\:\: \times \mathcal{T}_+ e^{-i\int_0^t \langle H(t')\rangle dt'} \hat{\rho}(0)Z^m X^n) \\
= &\operatorname{Tr}(\mathcal{T}_-
e^{i\int_0^t\sum_{\A_{\sum} \in S} \langle \beta_{\A_{\beta}}(t^{\prime})\rangle y_{\A_{y}}(t') 
Z^{\A_Z}\xi^{-m\A_X+n\A_Z} X^{-n}  X^{\A_X} X^n dt'}
\\
&\:\:\:\:\:\:\:\:\:\:\:\:\:\:\:
\times \mathcal{T}_ +e^{-i\int_0^t \langle H(t')\rangle dt'} \hat{\rho}(0)Z^m X^n)\\
= &\operatorname{Tr}(\mathcal{T}_-
e^{i\int_0^t\sum_{\A_{\sum} \in S} \langle \beta_{\A_{\beta}}(t^{\prime})\rangle y_{\A_{y}}(t') 
\xi^{-m\A_X+n\A_Z} Z^{\A_Z} X^{\A_X} dt'} 
\mathcal{T}_ +e^{-i\int_0^t \langle H(t')\rangle dt'} \hat{\rho}(0)Z^m X^n) \\
=&\operatorname{Tr}(
\mathcal{T}_-
e^{i\int_0^t \sum_{\A_{\sum}\in S}\langle \beta_{\A_{\beta}}(t^{\prime})\rangle y_{\A_{y}}(t') \xi^{-m\A_X+n\A_Z} Z^{\A_Z} X^{\A_X} dt'}\\
&\mathcal{T}_+
e^{-i\int_0^t \sum_{\A_{\sum} \in S} \langle \beta_{\A_{\beta}}(t^{\prime})\rangle y_{\A_{y}}(t') Z^{\A_Z} X^{\A_X} dt'} 
\hat{\rho}(0)Z^m X^n) \\
= &\operatorname{Tr}(\mathcal{T}_+
e^{-i\int_0^t\sum_{\A_{\sum} \in S} \langle \beta_{\A_{\beta}}(t^{\prime})\rangle y_{\A_{y}}(t') 
(1-\xi^{-m\A_X+n\A_Z}) Z^{\A_Z}X^{\A_X} dt'}
\hat{\rho}(0)Z^m X^n) \\
= &\operatorname{Tr}([1+ [-i\int_0^t\sum_{\A_{\sum} \in S} \langle \beta_{\A_{\beta}}(t^{\prime})\rangle y_{\A_{y}}(t') 
(1-\xi^{-m\A_X+n\A_Z}) Z^{\A_Z}X^{\A_X} dt']\\
& +\frac{1}{2}[-i\int_0^t\sum_{\A_{\sum} \in S} \langle \beta_{\A_{\beta}}(t^{\prime})\rangle y_{\A_{y}}(t') 
(1-\xi^{-m\A_X+n\A_Z}) Z^{\A_Z}X^{\A_X} dt']^2+...]\hat{\rho}(0)Z^m X^n),
\end{split}
\end{align}
where the extra $\langle\rangle$ denotes averaging the expectation values over many realizations of the stochastic noise functions. To derive the above equation, we used Eq.\eqref{expFormula},
the Dyson series, and the assumption of small $t$ values. Furthermore, to arrive at the tenth line of the equation above, we employed Eqs.\eqref{UnitaryGeneral} and \eqref{UnitaryGeneralDagger}. \\
Next note that due to the stochastic noise function $\beta_{a' b'}(t)$ possessing a zero-mean Gaussian distribution and being stationary (Eq.\eqref{stationary}), we have the following:
\begin{align}\label{zeromean}
\begin{split}
&\int_0^t\sum_{\A_{\sum} \in S} \langle\beta_{\A_{\beta}}(t^{\prime})\rangle y_{\A_{y}}(t') 
(1-\xi^{-m\A_X+n\A_Z}) Z^{\A_Z}X^{\A_X} dt'\\
& =\int_0^t dt'\sum_{\A_{\sum} \in S}
\langle \beta_{\A_{\beta}}(t')\rangle\textbf{g}(\A_{y},\A_X,\A_Z,m,n,t')=0,
\end{split}
\end{align}
where $\textbf{g}(\A_{y},\A_X,\A_Z,m,n,t')$ is a function with scalar values multiplied by the powers of generalized Pauli $Z$ and $X$ operators within the specified time intervals. Now Eq.\eqref{B.1.1} can be more simplified as follows:
\begin{align}\label{B.1.1.contineued1}
\begin{split}
&\langle \langle Z^m X^n\rangle \rangle= \operatorname{Tr}([1-\frac{1}{2}\int_0^t dt^{\prime}dt^{\prime\prime} \times \\
&\sum_{\A_{\sum}, \tilde{\A}_{\sum}\in S}
\langle \beta_{\A_{\beta}}(t^{\prime})\beta_{\tilde{\A_{\beta}}}(t^{\prime\prime})\rangle y_{{\A}_y}(t^{\prime})y_{\tilde{\A}_y}(t^{\prime\prime})(1-\xi^{-m\A_X+n\A_Z})(1-\xi^{\tilde{\A}_Z n-\tilde{\A}_X m})\\
& \times \xi^{-\tilde{\A}_Z \A_X} Z^{{\A_Z}+\tilde{\A}_Z}X^{\A_X+\tilde{\A}_X}] \hat{\rho}(0)Z^m X^n)\\
= &\operatorname{Tr}(\hat{\rho}(0)Z^m X^n)
-\operatorname{Tr}(\frac{1}{2}\int_0^t dt^{\prime}dt^{\prime\prime}\times \\
& \sum_{\A_{\sum},\tilde{\A}_{\sum}\in S}\langle \beta_{\A_{\beta}}(t^{\prime})\beta_{\tilde{\A_{\beta}}}(t^{\prime\prime})\rangle
y_{{\A}_y}(t^{\prime}))y_{\tilde{\A}_y}(t^{\prime\prime})
(1-\xi^{-m\A_X+n\A_Z})
(1-\xi^{\tilde{\A}_Z n-\tilde{\A}_X m})\\
& \times 
\xi^{-\tilde{\A}_Z \A_X} Z^{{\A_Z}+\tilde{\A}_Z}X^{\A_X+\tilde{\A}_X} \hat{\rho}(0)Z^m X^n),
\end{split}
\end{align}
Next we assume the Weyl decomposition of an arbitrary initial density matrix operator of the qudit (Eq.\eqref{Weyldecompositions}), and consider the trace of  Weyl operators $\operatorname{Tr}(Z^m X^n)=d\delta_{m=id,n=id}$, where $i\in \mathbbm{Z}$ to achieve the following:
\begin{align}\label{B.1.1.contineued2}
\begin{split}
&\langle \langle Z^m X^n\rangle \rangle= 
\operatorname{Tr}(\sum_{p',q' \in S} V_{p'q'} Z^{p'} X^{q'} Z^m X^n)\\
&-\operatorname{Tr}(\frac{1}{2}\int_0^t  dt^{\prime} dt^{\prime\prime} \sum_{\A_{\sum},\tilde{\A}_{\sum}, p,q \in S}\langle \beta_{\A_{\beta}}(t^{\prime}) \beta_{\tilde{\A_{\beta}}}(t^{\prime\prime})\rangle
y_{{\A}_y}(t^{\prime}))y_{\tilde{\A}_y}(t^{\prime\prime}) \times\\
& (1-\xi^{-m\A_X+n\A_Z})
(1-\xi^{\tilde{\A}_Z n-\tilde{\A}_X m}) 
\xi^{-\tilde{\A}_Z \A_X} Z^{{\A_Z}+\tilde{\A}_Z}
X^{\A_X+\tilde{\A}_X} V_{p,q} Z^p X^q Z^m X^n)\\
=
&\operatorname{Tr}(\sum_{p',q' \in S} V_{p'q'} \xi^{-mq'} Z^{p'+m} X^{q'+n})\\
&-\operatorname{Tr}(\frac{1}{2}\int_0^t dt^{\prime} dt^{\prime\prime} \sum_{\A_{\sum},\tilde{\A}_{\sum}, p,q \in S}\langle \beta_{\A_{\beta}}(t^{\prime}) \beta_{\tilde{\A_{\beta}}}(t^{\prime\prime})\rangle
y_{{\A}_y} (t^{\prime})) y_{\tilde{\A}_y}(t^{\prime\prime}) \times\\
&(1-\xi^{-m\A_X+n\A_Z})
(1-\xi^{\tilde{\A}_Z n-\tilde{\A}_X m}) 
\xi^{-\tilde{\A}_Z \A_X} Z^{{\A_Z}+\tilde{\A}_Z}X^{\A_X+\tilde{\A}_X}  V_{pq} \xi^{-mq} Z^{p+m} X^{q+n})\\
=
& \sum_{p',q' \in S} V_{p',q'} \xi^{-mq'} \delta_{p',-m} \delta_{q',-n} \operatorname{Tr}(1_{d\times d}) \\
&-\operatorname{Tr}(\frac{1}{2}\int_0^t dt^{\prime} dt^{\prime\prime} \sum_{\A_{\sum},\tilde{\A}_{\sum}, p,q \in S}\langle \beta_{\A_{\beta}}(t^{\prime})\beta_{\tilde{\A_{\beta}}}(t^{\prime\prime})\rangle
y_{{\A}_y}(t^{\prime}))y_{\tilde{\A}_y}(t^{\prime\prime}) \times\\
&(1-\xi^{-m\A_X+n\A_Z})
(1-\xi^{\tilde{\A}_Z n-\tilde{\A}_X m}) 
\xi^{-\tilde{\A}_Z \A_X} Z^{{\A_Z}+\tilde{\A}_Z} \times\\&
V_{p,q} \xi^{-mq} 
Z^{p+m} X^{q+n}
X^{\A_X+\tilde{\A}_X} \xi^{-(p+m)(\A_X + \A_{\tilde{X}})})
\end{split}
\end{align}
We continue the simplification as follows:
\begin{align}\label{B.1.1.contineued3}
\begin{split}
&\langle \langle Z^m X^n\rangle \rangle= 
V_{-m,-n} \xi^{mn} . d 
-\operatorname{Tr}(\frac{1}{2}\int_0^t dt^{\prime}dt^{\prime\prime} \times \\&
\sum_{\A_{\sum},\tilde{\A}_{\sum}, p,q \in S}
\langle \beta_{\A_{\beta}}(t^{\prime})\beta_{\tilde{\A_{\beta}}}(t^{\prime\prime})\rangle
y_{{\A}_y}(t^{\prime}))y_{\tilde{\A}_y}(t^{\prime\prime}) 
(1-\xi^{-m\A_X+n\A_Z})
(1-\xi^{\tilde{\A}_Z n-\tilde{\A}_X m})\\
& \times 
V_{pq}
\xi^{-\tilde{\A}_Z \A_X -mq -(p+m)(\A_X + \A_{\tilde{X}})} 
Z^{{\A_Z}+\tilde{\A}_Z + p+m} 
X^{q+n+\A_X+\tilde{\A}_X} )\\
=
& V_{-m,-n} \xi^{mn} . d 
-\frac{1}{2}\int_0^t dt^{\prime}dt^{\prime\prime} \times \\&
\sum_{\A_{\sum},\tilde{\A}_{\sum} \in S}\langle \beta_{\A_{\beta}}(t^{\prime})\beta_{\tilde{\A_{\beta}}}(t^{\prime\prime})\rangle
y_{{\A}_y}(t^{\prime}))y_{\tilde{\A}_y}(t^{\prime\prime})
(1-\xi^{-m\A_X+n\A_Z})
(1-\xi^{\tilde{\A}_Z n-\tilde{\A}_X m})\\
& \times 
\sum_{p,q \in S} V_{pq}
\xi^{-\tilde{\A}_Z \A_X -mq -(p+m)(\A_X + \A_{\tilde{X}})} 
\delta_{{\A_Z}+\tilde{\A}_Z,-p-m} 
\delta_{\A_X+\tilde{\A}_X,-q-n} \operatorname{Tr}( 1_{d \times d})\\
= & V_{-m,-n} \xi^{mn} . d -\frac{1}{2}\int_0^t dt^{\prime}dt^{\prime\prime} \times \\& \sum_{\A_{\sum},\tilde{\A}_{\sum} \in S}\langle \beta_{\A_{\beta}}(t^{\prime})\beta_{\tilde{\A_{\beta}}}(t^{\prime\prime})\rangle
y_{{\A}_y}(t^{\prime}))y_{\tilde{\A}_y}(t^{\prime\prime})
(1-\xi^{-m\A_X+n\A_Z})
(1-\xi^{\tilde{\A}_Z n-\tilde{\A}_X m})\\
& \times d
\sum_{p,q \in S} V_{pq}
\xi^{-\tilde{\A}_Z \A_X -mq +(p+m)(q+n)} 
\delta_{{\A_Z}+\tilde{\A}_Z,-p-m} 
\delta_{\A_X+\tilde{\A}_X,-q-n}
\end{split}
\end{align}
Lastly, the most streamlined form of $\langle \langle Z^m X^n\rangle \rangle$ is found as follows:
\begin{equation}\label{B.1.2}
\begin{split}
\langle\langle Z^m X^n\rangle\rangle= G_{mn} -\frac{1}{2}\int_0^t\int_0^t dt'dt^{\prime\prime}\sum_{\A_{\sum},\tilde{\A}_{\sum}\in S} 
J_{m n \A_J}
\langle
\beta_{\A_{\beta}}(t')
\beta_{\tilde{\A}_{\beta}}(t^{\prime\prime})\rangle 
y_{\A_y}(t')
y_{\tilde{\A}_y}(t^{\prime\prime}),
\end{split}
\end{equation}
where,
\begin{align}\label{Gmn}
G_{mn} & = d \sum_{p,q\in S} \xi^{-mq}V_{p,q}
\delta_{p+m,id}
\delta_{q+n=id}=d\xi^{mn}V_{id-m,id-n},
\end{align}
and
\begin{align}\label{Jmnaa'}
\begin{split}
J_{m n \A_J} 
= & d. \sum_{p,q\in S} 
(1-\xi^{-m\A_X+n\A_Z})
(1-\xi^{\tilde{\A}_Z n-\tilde{\A}_X m})
\xi^{-\tilde{\A}_Z \A_X -mq + 
(p+m)(n+q)}.\\
& V_{p,q}
 \delta_
 {p+m+\A_Z+\tilde{\A}_Z,id} \delta_{q+n+\A_X+\tilde{\A}_X,id}.
\end{split}    
\end{align}
For simplicity of notation, the new indexing variable of $\A_J=\A_Z,\A_X, \tilde{\A}_Z,\tilde{\A}_X$ is considered.
%We assume that the dephasing function $\beta_{\A_{\beta}}(t)$ to be stationary (Eq. \eqref{stationary}). 
Now we define the filter function as the bounded Fourier transform of the switching function in the following:
\begin{align}\label{Fdefinition}
\begin{split}
F_{\A_F}(\omega,t) & =\int_0^t dt' y_{\A_y}(t') e^{i\omega t'},    
\end{split}
\end{align}
where $\A_F=\A_y$. The Hermitian characteristics of the effective Hamiltonian results in Eq.\eqref{yconjugate}, that combined with the above equation yields: 
\begin{align}\label{YFrelation}
\begin{split}
F_{-\A_F}^\ast(\omega,t) =& F_{\A_F}(-\omega,t) \xi^{-\A^n_{Z}\A^n_{X}+\A_{Z}\A_{X}}
\end{split}
\end{align}
On the other hand, the noise spectra and the correlations of the noise functions are the mutual Fourier transforms as follows: 
\begin{align}\label{noise&correlations}
\begin{split}
S_{\A_S,\rightarrow}(\omega) =& S^{\tilde{\A}_{\beta}}_{\A_{\beta}}(\omega) 
=\int_{-\infty}^\infty dt \langle
\beta_{\A_{\beta}}(0)
\beta_{\tilde{\A}_{\beta}}(t)\rangle e^{i\omega t},\\
\langle
\beta_{\A_{\beta}}(t')
\beta_{\tilde{\A}_{\beta}}(t^{\prime\prime})\rangle 
& =\langle
\beta_{\A_{\beta}}(0)
\beta_{\tilde{\A}_{\beta}}(t^{\prime\prime}-t')\rangle
=\frac{1}{2\pi}\int_{-\infty}^\infty d\omega 
S_{\A_S}(\omega) e^{-i\omega (t^{\prime\prime}-t')},
\end{split}
\end{align}
where the new indexing/power variable $\A_S=\A_{\beta},\tilde{\A}_{\beta}$ is defined to simplify the notation.
Using these relations Eq.\eqref{B.1.2} transforms into the following:
\begin{align}\label{B.1.3}
\begin{split}
&\langle\langle Z^m X^n\rangle\rangle 
 = G_{mn}
- \frac{1}{2}
\sum_{\A_{\sum},\tilde{\A}_{\sum}\in S} 
J_{m n \A_J} \Big[\frac{1}{2\pi}\int_{-\infty}^{\infty}d\omega
S_{\A_S}(\omega) e^{-i\omega (t^{\prime\prime}-t')}\Big] \\
& \times \Big[\int_0^tdt'
y_{\A_y}(t')
e^{i\omega t'}\Big]e^{-i\omega t'}
\Big[\int_0^t dt^{\prime\prime}
y^*_{-\tilde{\A}_y}(t^{\prime\prime})\xi^{\tilde{\A}^n_{Z}\tilde{\A}^n_{X}-\tilde{\A}_{Z}\tilde{\A}_{X}}
e^{-i\omega t^{\prime\prime}}\Big]
e^{i\omega t^{\prime\prime}} \\
& = G_{mn} -\frac{1}{4\pi}\int_{-\infty}^{\infty}d\omega 
\sum_{\A_{\sum},\tilde{\A}_{\sum}\in S} 
J_{m n \A_J}
\xi^{\tilde{\A}^n_{Z}\tilde{\A}^n_{X}-\tilde{\A}_{Z}\tilde{\A}_{X}}
S_{\A_S}(\omega) F_{\A_F}(\omega,t) F^*_{-\tilde{\A}_F}(\omega,t)\\
& = G_{mn} -\frac{1}{4\pi}\int_{-\infty}^{\infty}d\omega 
\sum_{\A_{\sum},\tilde{\A}_{\sum}\in S} 
\tilde{J}_{m n \A_J}
S_{\A_S}(\omega) F_{\A_F}(\omega,t) F^*_{-\tilde{\A}_F}(\omega,t),\\
&\tilde{J}_{m n \A_J} = J_{m n \A_J}\xi^{\tilde{\A}^n_{Z}\tilde{\A}^n_{X}-\tilde{\A}_{Z}\tilde{\A}_{X}}.
\end{split}
\end{align}
Now consider that the reference pulse sequence applied within $t\in [0,T]$ is repeated for $M$ times so that the total time of one pulse series consisting of $M$ reference sequences is $t=MT$. The filter function corresponding one pulse series in terms of that of one pulse sequence is as follows:
\begin{align}\label{B.1.4}
\begin{split}
F_{\A_F}(\omega,t)
= & F_{\A_F}(\omega,MT) 
= \int_0^{MT}dt'
y_{\A_y}(t')
e^{i\omega t'}  =\sum_{r=0}^{M-1}\int_{rT}^{(r+1)T}dt'
y_{\A_y}(t')
e^{i\omega t'}\\
= &\sum_{r=0}^{M-1}\int_0^{T}ds 
y_{\A_y}(s)
e^{i\omega s}e^{i\omega rT} = \sum_{r=0}^{M-1}e^{i\omega rT}
F_{\A_F}(\omega,T)= \frac{1-e^{i\omega MT}}{1-e^{i\omega T}}
F_{\A_F}(\omega,T).
\end{split}
\end{align}
In the second row of Eq.\eqref{B.1.4} we applied the variable exchange $t'\rightarrow s+rT$, assumed the switching functions to be identical across different rounds of pulse series (i.e. $y_{\A_y}(t')=y_{\A_y}(s+rT)$), and utilized the \textit{geometric series} relation.
By assuming $\frac{1-e^{i\omega MT}}{1-e^{i\omega T}}=\frac{e^{i M \omega T/2 }}{e^{i\omega T/2}}.\frac{\sin(M \omega T/2)}{\sin(\omega T/2)}$, and substituting Eq.\eqref{B.1.4} into Eq.\eqref{B.1.3}, we find:
\begin{align}\label{B.1.5}
\begin{split}
&\langle\langle Z^m X^n\rangle\rangle= \\
& G_{mn} - 
\frac{1}{4\pi}\int_{-\infty}^{\infty}d\omega \sum_{\A_{\sum},\tilde{\A}_{\sum}\in S} \tilde{J}_{m n \A_J} S_{\A_S}(\omega) F_{\A_F}(\omega,T) F_{-\tilde{\A}_F}^*(\omega,T)  
\frac{\sin^2(M\omega T/2)}{\sin^2(\omega T/2)} \\
& = G_{mn} + \int_{\infty}^{\infty}d\omega \frac{\sin^2(M\omega T/2)}{\sin^2(\omega T/2)} L(\omega);\\
& L(\omega)= -\frac{1}{4\pi} \sum_{\A_{\sum},\tilde{\A}_{\sum}\in S} \tilde{J}_{m n \A_J} S_{\A_S}(\omega) F_{\A_F}(\omega,T) F_{-\tilde{\A}_F}^*(\omega,T)
\end{split}
\end{align}
The last integral term of Eq.\eqref{B.1.5} is simplified by first considering the variable exchange $\omega \rightarrow  \omega'= \omega-m \frac{2\pi}{T}$, next the substitution $\int_{-\infty}^{\infty}d\omega' = \sum_{m=-\infty}^{\infty}\int_{m\frac{2\pi}{T}-\frac{\pi}{T}}^{m\frac{2\pi}{T}+\frac{\pi}{T}}d\omega'$, and finally the variable exchange $\frac{M \omega' T}{2} \rightarrow u$, as following: 
\begin{equation}\label{B.1.6}
\begin{split}
&\int_{-\infty}^{\infty}d\omega \frac{\sin^2(M\omega T/2)}{\sin^2(\omega T/2)} L(\omega) 
= 
\int_{-\infty}^{\infty} d\omega' 
\frac
{\sin^2(
(\omega' + 
m\frac{2\pi}{T}) \frac{MT}{2} 
)}
{\sin^2(
(\omega' + 
m\frac{2\pi}{T}) \frac{T}{2} 
)} 
L(\omega' + m\frac{2\pi}{T} )
)\\
&= \int_{-\infty}^{\infty} d\omega' 
\frac
{\sin^2(
\frac{MT}{2}\omega' + 
mM\pi 
)}
{\sin^2(
\frac{T}{2}\omega' + 
mM\pi
)} 
L(\omega' + m\frac{2\pi}{T} )
) 
= 
\int_{-\infty}^{\infty} d\omega' 
\frac
{\sin^2(
\frac{MT}{2}\omega' 
)}
{\sin^2(
\frac{T}{2}\omega'
)} 
L(\omega' + m\frac{2\pi}{T} )
) \\
&= \sum_{m=-\infty}^{\infty}\int_{m\frac{2\pi}{T}-\frac{\pi}{T}}^{m\frac{2\pi}{T}+\frac{\pi}{T}}
d\omega'
\frac
{\sin^2(
\frac{MT}{2}\omega' 
)}
{\sin^2(
\frac{T}{2}\omega'
)} 
L(\omega' + m\frac{2\pi}{T} )
)\\
&= \frac{2}{MT}
\sum_{m=-\infty}^{\infty}
\int_{mM\pi-\frac{M\pi}{2}}^{mM\pi+\frac{M\pi}{2}}
du 
\frac{\sin^2(u)}{\sin^2(u/M)} L(\frac{2}{M T}u + m \frac{2\pi}{T})
\end{split}
\end{equation}
By making the assumption $M\rightarrow \infty$ we have $\sin(u/M \rightarrow 0)\approx u/M $ and $L(\frac{2}{M T}u + m \frac{2\pi}{T}) \rightarrow L(m \frac{2\pi}{T})$. Consequently, we can find $\int_{mM\pi-M\pi/2}^{mM\pi+M\pi/2}du\frac{\sin^2{u}}{u^2} = \int_{-M\pi/2}^{M\pi/2}du\frac{\sin^2{u}}{u^2} = \int_{-\infty}^{\infty}du\frac{\sin^2{u}}{u^2}\approx \pi$ . Eq.\eqref{B.1.5} can now be expressed in the following form:
\begin{align}\label{Minfty}
\begin{split}
\langle\langle Z^m X^n\rangle\rangle
\approx & G_{mn} + \frac{2\pi}{T}M \sum_{k=-\infty}^{\infty} L(\frac{2\pi}{T} k)\\
\approx & G_{mn} -\frac{M}{2T}\sum_{k=-\infty}^{\infty}\sum_{\A_{\sum},\tilde{\A}_{\sum}\in S} \tilde{J}_{m n \A_J} S_{\A_S}(k\omega_0) F_{\A_F}(k\omega_0,T) F_{-\tilde{\A}_F}^*(k\omega_0,T);
\end{split}
\end{align}
where $\omega_0 = \frac{2\pi}{T}$. Taking the summation of Weyl decomposition coefficients of all parameters of Eq.\eqref{Minfty}, and assuming $T \rightarrow T'=T/r$ and $\omega_0 \rightarrow \omega^\prime_0 = r\omega_0$, we have the following:
\begin{align}\label{mainO}
\begin{split}
&\langle\langle \hat{O}(t=MT/r) \rangle\rangle^r
= \eta -\frac{1}{2}\int_0^t\int_0^t dt'dt''\sum_{\A_{\sum},\tilde{\A}_{\sum} \in S} \lambda_{\A_{\lambda}}
y_{\A_y}(t')y_{\tilde{\A}_y}(t'') \langle\beta_{\A_{\beta}}(t')\beta_{\tilde{\A}_{\beta}}(t'')\rangle  \\
& = \eta -\frac{1}{4\pi}
\int_{-\Omega}^{\Omega}d\omega \sum_{\A_{\sum},\tilde{\A}_{\sum}\in S} \lambda_{\A_{\lambda}}
F_{\A_F}(\omega,t) F_{-\tilde{\A}_F}^*(\omega,t) S_{\A_S}(\omega)\\
& \approx \eta -
\frac{M}{2T/r} \sum_{k=-\Omega}^{\Omega} \sum_{\A_{\sum},\tilde{\A}_{\sum}\in S} \lambda_{\A_{\lambda}} F_{\A_F}(rk\omega_0,T/r) F_{-\tilde{\A}_F}^*(rk\omega_0,T/r) S_{\A_S}(kr\omega_0),
\end{split}
\end{align}
where $\Omega \rightarrow \infty$ is the maximum noise frequency that we are interested to detect with the qudit spectator, $\A_{\lambda}=\A_J$, and,
\begin{align}\label{mainOetalambda1}
\begin{split}
\eta
& = \sum_{m,n\in S} O_{mn} G_{mn} = 
d \sum_{m,n\in S}
O_{mn}\xi^{mn}V_{id-m,id-n}, \\ 
\lambda_{\A_{\lambda}}
& = \sum_{m,n \in S} O_{mn} \tilde{J}_{mn\A_J}\\
& = d\sum_{m,n,p,q \in S} O_{mn} 
 V_{p,q}
 \delta_
 {-p,+m+\A_Z+\tilde{\A}_Z-id}\delta_{-q,+n+\A_X+\tilde{\A}_X-id} \\
& \times (1-\xi^{-m\A_X+n\A_Z})
(1-\xi^{\tilde{\A}_Z n-\tilde{\A}_X m})
\xi^{-\tilde{\A}_Z \A_X
-mq +(m+p)(n+q)} \xi^{\tilde{\A}^n_{Z}\tilde{\A}^n_{X}-\tilde{\A}_{Z}\tilde{\A}_{X}}.
\end{split}
\end{align}
Now, utilizing Eq.\eqref{full-half-freq-range}, the integral and summation in Eq.\eqref{mainO} are transformed to the positive range in the following manner:
\begin{align}\label{mainOprocess}
\begin{split}
&\langle\langle \hat{O}(t=MT/r) \rangle\rangle^r
= \eta -\frac{1}{2}\int_0^t\int_0^t dt'dt''\sum_{\A_{\sum},\tilde{\A}_{\sum} \in S} \lambda_{\A_{\lambda}}
y_{\A_y}(t')y_{\tilde{\A}_y}(t'') \langle\beta_{\A_{\beta}}(t')\beta_{\tilde{\A}_{\beta}}(t'')\rangle  \\
& = \eta -\frac{1}{4\pi} \{
\int_{0^+}^{\Omega}d\omega \sum_{\A_{\sum},\tilde{\A}_{\sum}\in S} 
\lambda_{\A_{\lambda}} \times \\
&\Big(
F_{\A_F}(\omega,MT/r) 
F_{-\tilde{\A}_F}^*(\omega,MT/r) S_{\A_S}(\omega) + 
 F_{\A_F}(-\omega,MT/r) 
F_{-\tilde{\A}_F}^*(-\omega,MT/r) S_{\A_S}(-\omega)
\Big) \\
& + d\omega \times \sum_{\A_{\sum},\tilde{\A}_{\sum}\in S}
\lambda_{\A_{\lambda}}
F_{\A_F}(0,t) 
F_{-\tilde{\A}_F}^*(0,t) 
S_{\A_S}(0) \: \} \\
& \approx \eta -
\frac{M}{2T/r} \{
\sum_{k=1}^{N} 
\sum_{\A_{\sum},\tilde{\A}_{\sum}\in S} \lambda_{\A_{\lambda}} \times \\
&\Big( 
F_{\A_F}(rk\omega_0,T/r) F_{-\tilde{\A}_F}^*(rk\omega_0,T/r) S_{\A_S}(kr\omega_0) \\
& + 
 F_{\A_F}(-rk\omega_0,T/r) F_{-\tilde{\A}_F}^*(-rk\omega_0,T/r) S_{\A_S}(-kr\omega_0)
\Big)  \\
& 
+ \sum_{\A_{\sum},\tilde{\A}_{\sum}\in S} \lambda_{\A_{\lambda}}
\Big( 
F_{\A_F}(0,T/r) F_{-\tilde{\A}_F}^*(0,T/r) S_{\A_S}(0) \Big) \: \}
\end{split}
\end{align}
In the above formula, $S_{\A_S}(0)$ is the noise spectrum at exactly zero frequency. Since $\omega \rightarrow 0$ could be a measurable frequency, the $\omega=0$ means no frequency and so we cannot assign any noise spectrum to $\omega=0$.
Consequently, we ignore the nonphysical terms including $S(0)$ in the integral and summation parts of the above equations, so that we assume:
\begin{align}\label{nonphyscialS(omega)}
\begin{split}
S_{\A_S}(\omega = 0) = 0
\end{split}
\end{align}
To maintain the equality of the integral and summation parts, we start the integration from point $\omega_0$. It is due to the fact that $N=\floor{\Omega/\omega_0}$ and each term of the summation is equivalent to $\omega_0$ portion of the integral range, and if we eliminate the first term of summation ($k=0$), we can start the integration from point $\omega_0$. The equivalence of the integral portions and summation terms can be understood by the following approximation: 
\begin{align}\label{IntegralApprox}
\begin{split}
\int_{0}^{\Omega} f(\omega) d\omega \approx  \sum_{k=0}^{\floor{\Omega/\omega_0}}f(k\omega_0 ) \times \omega_0,\:\:\:\:\:
\omega_0 \rightarrow d\omega\
\end{split}
\end{align}
The above formula in our formalism has been transformed into the less trivial form:
\begin{align}\label{mainOsimplified}
\begin{split}
\int_{0}^{\Omega} f(\frac{MT}{r},\omega) d\omega\approx \sum_{k=0}^{\floor{\Omega/\omega_0}}f(\frac{T}{r},kr\omega_0) \times Mr\omega_0
\end{split}
\end{align}
Using the aforementioned fact of ignoring zero frequency term, we increase the efficiency of
our simulations for retrieving the true spectra. This approximation leads to a significant reduction in the required reference sequence repetitions ($M$)
and yields high resolution for the simulated spectra, i.e. the better overlap of the true and the simulated spectra.
The assumption stated above transforms Eq.\eqref{mainOprocess} into the following form:
\begin{align}\label{mainOfinal}
\begin{split}
&\langle\langle \hat{O}(t=MT/r) \rangle\rangle^r
= \eta 
-\frac{1}{2}
\int_0^t\int_0^t dt'dt''
\sum_{\A_{\sum},\tilde{\A}_{\sum} \in S} \lambda_{\A_{\lambda}}
y_{\A_y}(t')y_{\tilde{\A}_y}(t'') \langle\beta_{\A_{\beta}}(t')\beta_{\tilde{\A}_{\beta}}(t'')\rangle \\
& = \eta -\frac{1}{4\pi}
\int_{\omega_0}^{\Omega}d\omega \sum_{\A_{\sum},\tilde{\A}_{\sum}\in S} 
\lambda_{\A_{\lambda}} \times \\
&\Big(
F_{\A_F}(\omega,MT/r) 
F_{-\tilde{\A}_F}^*(\omega,MT/r) S_{\A_S}(\omega) + 
 F_{\A_F}(-\omega,MT/r) 
F_{-\tilde{\A}_F}^*(-\omega,MT/r) S_{\A_S}(-\omega)
\Big) \\
& \approx \eta -
\frac{M}{2T/r}
\sum_{k=1}^{N} 
\sum_{\A_{\sum},\tilde{\A}_{\sum}\in S} \lambda_{\A_{\lambda}} 
\Big( 
F_{\A_F}(rk\omega_0,T/r) F_{-\tilde{\A}_F}^*(rk\omega_0,T/r) S_{\A_S}(kr\omega_0) \\
&\:\:\:\:\:\:\:\:\:\:\:\:\:\:\:\:\:\:\:\:\:\:\:\:\:\:\:\:\:\:\:\:\:\:\:\:\:\:\:\:\:\:\:\:\:\:\:\:\:\:\:\:\:\:\:\:\:\:\:\:\:+ 
F_{\A_F}(-rk\omega_0,T/r) F_{-\tilde{\A}_F}^*(-rk\omega_0,T/r) S_{\A_S}(-kr\omega_0)
\Big) 
\end{split}
\end{align}
We also realized that in order to have an efficient simulation for point $S_{\A_S}(\omega = \omega_0)$ in Fig.\ref{Fig.1}, 
It is better to start the integration of Eq.\eqref{mainOfinal} from a small value of $\omega$ like
$\omega = 0.001\:[\rm rad/[T]]$, instead of $\omega_0$, while eliminating the corresponding term in the summation part. 
To provide an explanation, we note that in our noise spectroscopy protocol the simulated noise spectra $S_{\A_S}(\omega)$ is determined as follows:
\begin{equation}\label{SOmega0}
S_{\A_S}(\omega) = \G_c \Big(\:\eta,\:F_{\A_F}(\omega,T/r),\:\int_{\omega_0}^\Omega 
d\omega S_{\A_S}(\omega) 
\C 
\Big(\lambda_{\A_{\lambda}},\: F_{\A_F}(\omega,MT/r)\Big) \: \Big)
\end{equation}
where the calligraphic $G_c$ $(\G_c)$ is a function associated with the inverse of the coefficient matrix as defined in our simulations (e.g. Eq.\eqref{Simulation-A}), consisting of multiple variables such as $\eta$ and the filter functions.
In addition, $\G_c$ is a function of the integral mentioned above including the calligraphic $C$ function $(\C)$ that depends on filter functions, $\lambda_{\A_{\lambda}}$, and also the noise polyspectra 
within the frequency range that under our new assumption is $S_{\A_S}(\omega_0 \le \omega \le \Omega)$.
Since $\omega_0$ is on the integration's boundary, to achieve a better estimate of the polyspectra at the exact boundary value $\omega_0$, i.e. $S_{\A_S}(\omega_0)$, it is reasonable to include more polyspetra around $\omega_0$, and not only include $S_{\A_S}(\omega_0^+)$. Therefore, to calculate $S_{\A_S}(\omega_0)$ we should consider the integral limit as $[0^+,\Omega]$ instead of $[\omega_0,\Omega]$.  

Eq.\eqref{mainOfinal} represents the most comprehensive expression found in our noise spectroscopy protocol that relates the expectation value of an arbitrary qudit's observable to the switching functions and the unknown noise polyspectra. 

In the following sections, we simplify this relation for different versions of Reduced Weyl decomposed Hamiltonians of a qutrit and a general qudit, and also the Antimony quoct Hamiltonian in Weyl basis.
\subsection{Qutrit with Reduced Weyl Hamiltonian}\label{Qutrit-expectation}
The aim of this section is to simplify the qutrit form of Eq.\eqref{mainO}, that is Eq.\eqref{1.4}, with the ultimate goal of attaining the qutrit version of Eq.\eqref{mainOfinal}.
Given the fact that in Eq.\eqref{1.4}, $a,b \in \{0,\pm 1\}$, there would be nine unknown complex noise polyspectra $S_{a,b}(\omega)$ that are being detected by the qutrit spectator in our noise spectroscopy protocol. It is shown in appendices \ref{Qutrit-S(w)symproperties} and \ref{RealFuncs-qutrit} that these nine complex variables depend only on four real functions, that one is an asymmetric/odd function, $D(\omega)$, and the other three are symmetric/even functions, $R_1(\omega),I_1(\omega),E(\omega)$.
Assuming these, Eq.\eqref{1.4} transforms to the following:
\begin{align}\label{qutritsimulation-apx}
\begin{split}
\langle\langle \hat{O}(t_r^A) \rangle\rangle^{r} & = A^r(t_r^A) \approx B^r(t_r^B),\\
A^r(t_r^A) & = \eta_{mn} -\frac{1}{4\pi}\int_{-\Omega}^{\Omega}d\omega \sum_{a,b=\pm1} \lambda_{ab}^{mn}
F_{a}(\omega,t_r^A) F_{-b}^*(\omega,t_r^A) S_{ab}(\omega)\\
& = \eta_{mn} -\frac{1}{4\pi}\int_{-\Omega}^{\Omega}d\omega \sum_{i=0}^3 C^\prime_i(m,n,\omega,t_r^A)x_i(\omega),\:\:\: t_r^A=M T/r,\\
B^r(t_r^B) & = \eta_{mn} -
\frac{M}{2T/r} \sum_{k=-N}^N \sum_{a,b=\pm1} \lambda_{ab}^{mn}
F_{a}(rk\omega_0,t_r^B) F_{-b}^*(rk\omega_0,t_r^B) S_{ab}(rk\omega_0)\\
& = \eta_{mn} -
\frac{M}{2T/r} \sum_{k=-N}^N \sum_{i=0}^3 C^\prime_i(m,n,rk\omega_0,t_r^{B})x_i(rk\omega_0),\:\:\: t_r^B=T/r,
\end{split}
\end{align}
where,\\
\begin{align}\label{Ci-coefficients-qutrit-apx}
\begin{split}
x_0 (\omega) &  =R_1(\omega),\:x_1(\omega)=I_1(\omega),\:x_2(\omega)=E(\omega),\:x_3(\omega)=D(\omega)\\
C^\prime_0 (m,& n,\omega,t_r) = 
\{\lambda_{1,1}^{mn} F_{1}(\omega,t_r) F_{-1}^*(\omega,t_r)
+ \lambda_{-1,-1}^{mn} F_{-1}(\omega,t_r) F_{1}^*(\omega,t_r)\},\\
C^\prime_1(m,& n,\omega,t_r) = 
i\{\lambda_{1,1}^{mn} F_{1}(\omega,t_r) F_{-1}^*(\omega,t_r)
- \lambda_{-1,-1}^{mn} F_{-1}(\omega,t_r) F_{1}^*(\omega,t_r)\},\\
C^\prime_2(m,& n,\omega,t_r) = 
\{\lambda_{1,-1}^{mn} F_{1}(\omega,t_r) F_{1}^*(\omega,t_r)
+ \lambda_{-1,1}^{mn} F_{-1}(\omega,t_r) F_{-1}^*(\omega,t_r)\},\\
C^\prime_3(m,& n,\omega,t_r) = 0,\:
N = \lfloor \Omega/\omega_0 \rfloor,%\: \alpha = 2^{(1-\delta_{\omega,0})},
\end{split}
\end{align}
Here we state three points: One is that by substituting $\omega$ with $-\omega$, according to Eq.\eqref{qutrit-variables-apx}, the even functions $x_0(\omega),x_1(\omega),x_3(\omega)$ remain intact, and the odd function $x_3(\omega)$ is multiplied by a minus sign. The second point is that the function $\lambda^{mn}_{a,b}$ in Eq.\eqref{Ci-coefficients-qutrit-apx} is independent of $\omega$. Lastly, according to the qutrit version of Eq.\eqref{YFrelation}, that is $F_{-a}^*(\omega,t) = F_{a}(-\omega,t)$, we have the following relations among the positive and negative frequencies of the filter functions expressed in Eq.\eqref{Ci-coefficients-qutrit-apx}:
\begin{align}\label{FF*qutrit}
\begin{split}
F_{1}(-\omega,t_r) F_{-1}^*(-\omega,t_r) &=F_{1}(\omega,t_r)F_{-1}^*(\omega,t_r),\\ 
F_{-1}(-\omega,t_r)F_{1}^*(-\omega,t_r) &=F_{-1}(\omega,t_r)F_{1}^*(\omega,t_r),\\
F_{1}(-\omega,t_r)F_{1}^*(-\omega,t_r) &=F_{-1}(\omega,t_r)F_{-1}^*(\omega,t_r).    
\end{split}    
\end{align}

Now we map the negative frequency range of the integral and summations in Eq.\eqref{qutritsimulation-apx} to the positive frequency range utilizing Eq.\eqref{full-half-freq-range}, the three aforementioned points, and the approximation introduced in Eq.\eqref{mainOfinal}, to find the following:
\begin{align}\label{qutrit-final-apx}
\begin{split}
A^r(t_r^A) & = \eta_{mn} -\frac{1}{4\pi}\int_{-\Omega}^{\Omega}d\omega \sum_{i=0}^3 C'_i(m,n,\omega,t_r^A)x_i(\omega)\\
& = \eta_{mn} -\frac{1}{4\pi}\int_{\omega_0}^{\Omega}d\omega \sum_{i=0}^3 C_i(m,n,\omega,t_r^A)x_i(\omega), \\
B^r(t_r^B) & = \eta_{mn} -
\frac{M}{2T/r} \{ \sum_{k=-\lfloor\Omega/\omega_0\rfloor}^{\lfloor\Omega/\omega_0\rfloor} \sum_{i=0}^3 C'_i(m,n,rk\omega_0,t_r^{B})x_i(rk\omega_0)\};\\
& = \eta_{mn} -
\frac{M}{2T/r} \{ \sum_{k=1}^{\lfloor\Omega/\omega_0\rfloor} \sum_{i=0}^3 C_i(m,n,rk\omega_0,t_r)x_i(rk\omega_0)\};\\
C_0(m & ,n,\omega,t_r) = C'_0(m,n,\omega,t_r) + C'_0(m,n,-\omega,t_r)
= 2 C'_0(m,n,\omega,t_r),\\
C_1(m & ,n,\omega,t_r)= C'_1(m,n,\omega,t_r) + C'_1(m,n,-\omega,t_r)
= 2 C'_1(m,n,\omega,t_r),\\
C_2(m & ,n,\omega,t_r)= C'_2(m,n,\omega,t_r) + C'_2(m,n,-\omega,t_r)=
2 C'_2(m,n,\omega,t_r)\\
C_3(m & ,n,\omega,t_r) = C'_3(m,n,\omega,t_r) + C'_3(m,n,-\omega,t_r)=
0\\
C_i(m,& n,0,t_r)=C'_i(m,n,0,t_r).
\end{split}
\end{align}
The coefficient $C_3(m ,n,\omega,t_r)$ is zero that corresponds to the odd %(and nonphysical) 
function $x_3(\omega)=D(\omega)$. This implies that only three real noise functions for the specified problem exist and could be achieved through our formalism. 
We ignored the terms corresponding $x_i(\omega=0)$ in both the integral and summation parts, since as discussed in Eq.\eqref{mainOfinal}, $\omega=0$ is not related to a determined wavelength.
\subsection{Qudit with Reduced Weyl Hamiltonian}\label{Qudit-expectation}
In this part, we develop a version of Eq.\eqref{mainO}, dedicated to qudits surrounded by Z-type dephasing noise, that is Eq.\eqref{qudit01}. We first simplify its integral part. Using Eqs.\eqref{full-half-freq-range}, \eqref{quoct-I-noisespectrum}, and the approximation introduced in Eq.\eqref{mainOfinal}, we find the follows:
\begin{align}\label{quditI-simulation-01}
\begin{split}
& A^r(t)  = 
\eta_{mn} 
-\frac{1}{4\pi}
\sum_{a,\tilde{a},b,\Tilde{b} \in S}
f_a f_{\tilde{a}} \lambda^{mnp}_{b\tilde{b}} \delta_{p+m,-b-\tilde{b}} \times \\
&\int_{\omega_0}^{\Omega}
d\omega 
\Big( F_{a,b}(\omega,t) 
F_{-\tilde{a},-\tilde{b}}^*(\omega,t) \tilde{S}(\omega)+
F_{a,b}(-\omega,t) 
F_{-\tilde{a},-\tilde{b}}^*(-\omega,t) \tilde{S}(-\omega) \Big)\\
& =  \eta_{mn} 
-\frac{1}{4\pi}
\sum_{a,\tilde{a},b,\Tilde{b} \in S}
f_a f_{\tilde{a}} \lambda^{mnp}_{b\tilde{b}} \delta_{p+m,-b-\tilde{b}} \times \\
&\int_{\omega_0}^{\Omega}
d\omega 
\Big( F_{a,b}(\omega,t) 
F_{-\tilde{a},-\tilde{b}}^*(\omega,t) (R(\omega)+iI(\omega))+\\
& F_{a,b}(-\omega,t) 
F_{-\tilde{a},-\tilde{b}}^*(-\omega,t) (R(\omega)-iI(\omega)) \Big)\\
& =  \eta_{mn} 
-\frac{1}{4\pi}
\sum_{a,\tilde{a},b,\Tilde{b} \in S}
f_a f_{\tilde{a}} \lambda^{mnp}_{b\tilde{b}} \delta_{p+m,-b-\tilde{b}} \times \\
&\Big(
\int_{\omega_0}^{\Omega}
d\omega 
\Big( F_{a,b}(\omega,t) 
F_{-\tilde{a},-\tilde{b}}^*(\omega,t) +
F_{a,b}(-\omega,t) 
F_{-\tilde{a},-\tilde{b}}^*(-\omega,t) \Big)R(\omega)\\
& +i\Big( F_{a,b}(\omega,t) 
F_{-\tilde{a},-\tilde{b}}^*(\omega,t) -
F_{a,b}(-\omega,t) 
F_{-\tilde{a},-\tilde{b}}^*(-\omega,t) \Big)I(\omega)\Big)\\
& =  \eta_{mn} 
-\frac{1}{4\pi}
\sum_{a,\tilde{a},b,\Tilde{b} \in S}
f_a f_{\tilde{a}} \lambda^{mnp}_{b\tilde{b}} \delta_{p+m,-b-\tilde{b}} \times \\
&\Big(
\int_{\omega_0}^{\Omega}
d\omega 
\Big( F_{a,b}(\omega,t) 
F_{-\tilde{a},-\tilde{b}}^*(\omega,t) +
F_{a,b}(-\omega,t) 
F_{-\tilde{a},-\tilde{b}}^*(-\omega,t) \Big)R(\omega)\\
& +i\Big( F_{a,b}(\omega,t) 
F_{-\tilde{a},-\tilde{b}}^*(\omega,t) -
F^*_{-a,-b}(\omega,t) 
F_{\tilde{a},\tilde{b}}(\omega,t) \Big)I(\omega)\Big)\\
& =  \eta_{mn} 
-\frac{1}{4\pi}
\sum_{a,\tilde{a},b,\Tilde{b} \in S}
f_a f_{\tilde{a}} \lambda^{mnp}_{b\tilde{b}} \delta_{p+m,-b-\tilde{b}} \times \\
&
\int_{\omega_0}^{\Omega}
d\omega 
\Big( 2 F_{a,b}(\omega,t) 
F_{-\tilde{a},-\tilde{b}}^*(\omega,t)  R(\omega)
 +i \times 0 \times I(\omega)\Big),
\end{split}
\end{align}
where $ t = t_r^A=\frac{MT}{r}$. Now we more develop the summation part of Eq.\eqref{qudit01} in the following manner: 
\begin{align}\label{quditI-simulation-02}
\begin{split}
& B^r(t)  = 
\eta_{mn}
-\frac{M}{2T/r}
\sum_{a,\tilde{a},b,\Tilde{b} \in S}
f_a f_{\tilde{a}} \lambda^{mnp}_{b\tilde{b}} \delta_{p+m,-b-\tilde{b}} \times \\
&\sum_{k=1}^{N}
\Big( F_{a,b}(rk\omega_0,t) 
F_{-\tilde{a},-\tilde{b}}^*(rk\omega_0,t) \tilde{S}(rk\omega_0)+\\
&F_{a,b}(-rk\omega_0,t) 
F_{-\tilde{a},-\tilde{b}}^*(-rk\omega_0,t) \tilde{S}(-rk\omega_0) \Big)\\
& =  \eta_{mn} 
-\frac{M}{2T/r}
\sum_{a,\tilde{a},b,\Tilde{b} \in S}
f_a f_{\tilde{a}} \lambda^{mnp}_{b\tilde{b}} \delta_{p+m,-b-\tilde{b}} \times \\
&\sum_{k=1}^{N}
\Big( F_{a,b}(rk\omega_0,t) 
F_{-\tilde{a},-\tilde{b}}^*(rk\omega_0,t) (R(rk\omega_0)+iI(rk\omega_0))+ \\
& F_{a,b}(-rk\omega_0,t) 
F_{-\tilde{a},-\tilde{b}}^*(-rk\omega_0,t) (R(rk\omega_0)-iI(rk\omega_0)) \Big)\\
& =  \eta_{mn}
-\frac{M}{2T/r}
\sum_{a,\tilde{a},b,\Tilde{b} \in S}
f_a f_{\tilde{a}} \lambda^{mnp}_{b\tilde{b}} \delta_{p+m,-b-\tilde{b}} \times \\
&\Big(
\sum_{k=1}^{N}
\Big( F_{a,b}(rk\omega_0,t) 
F_{-\tilde{a},-\tilde{b}}^*(rk\omega_0,t) +
F_{a,b}(-rk\omega_0,t) 
F_{-\tilde{a},-\tilde{b}}^*(-rk\omega_0,t) \Big)R(rk\omega_0)\\
& +i\Big( F_{a,b}(rk\omega_0,t) 
F_{-\tilde{a},-\tilde{b}}^*(rk\omega_0,t) -
F_{a,b}(-rk\omega_0,t) 
F_{-\tilde{a},-\tilde{b}}^*(-rk\omega_0,t) \Big)I(rk\omega_0)\Big) = \\
&  =  \eta_{mn}
-\frac{M}{2T/r}
\sum_{a,\tilde{a},b,\Tilde{b} \in S}
f_a f_{\tilde{a}} \lambda^{mnp}_{b\tilde{b}} \delta_{p+m,-b-\tilde{b}} \times\\
&\Big(
\sum_{k=1}^{N}
\Big( F_{a,b}(rk\omega_0,t) 
F_{-\tilde{a},-\tilde{b}}^*(rk\omega_0,t) +
F_{a,b}(-rk\omega_0,t) 
F_{-\tilde{a},-\tilde{b}}^*(-rk\omega_0,t) \Big)R(rk\omega_0)\\
& +i\Big( F_{a,b}(rk\omega_0,t) 
F_{-\tilde{a},-\tilde{b}}^*(rk\omega_0,t) -
F^*_{-a,-b}(rk\omega_0,t) 
F_{\tilde{a},\tilde{b}}(rk\omega_0,t) \Big)I(rk\omega_0)\Big) = \\
& =  \eta_{mn}
-\frac{M}{2T/r}
\sum_{a,\tilde{a},b,\Tilde{b} \in S}
f_a f_{\tilde{a}} \lambda^{mnp}_{b\tilde{b}} \delta_{p+m,-b-\tilde{b}} \times \\
& \Big(
\sum_{k=1}^{N}
2 F_{a,b}(rk\omega_0,t) 
F_{-\tilde{a},-\tilde{b}}^*(rk \omega_0,t) R(rk\omega_0)
+i\times 0 \times I(rk\omega_0)\Big),
\end{split}
\end{align}
where $ t=t_r^B=\frac{T}{r}$. In the fourth equality of the above equation, we used the Reduced Weyl (RW) version of Eq.\eqref{YFrelation}, that is $F_{a,b}(\omega,t)=F^*_{-a,-b}(-\omega,t)$. In addition, to reach the fifth equality, we applied the variable exchange $a,b \leftrightarrow \tilde{a},\tilde{b}$ and utilized the following relation:
\begin{align}
\begin{split}
\sum_{a,\tilde{a},b,\Tilde{b} \in S}  
F_{a,b}(rk\omega_0,t) 
F_{-\tilde{a},-\tilde{b}}^*(rk\omega_0,t)
= \sum_{\tilde{a},a,\Tilde{b},b \in S}
F^*_{-a,-b}(rk\omega_0,t) 
F_{\tilde{a},\tilde{b}}(rk\omega_0,t)
\end{split}
\end{align}
Eqs.\eqref{quditI-simulation-01} and \eqref{quditI-simulation-02} yield the simplified version of Eq.\eqref{qudit01} as that expressed in Eq.\eqref{quditI-simulation-03}.

\subsection{Antimony quoct with Weyl Hamiltonian}\label{Antimony-expectation}
This section investigates a specific version of Eq.\eqref{mainO} that correspond to the Antimony quoct.
In the following, we find the integral part by substituting Eqs.\eqref{Antimony-unknownPolyspectra}, \eqref{Antimony-knownPolyspectra}, and \eqref{x_iVars} into Eq.\eqref{quoct6-A}: %\eqref{mainOfinal} 
\begin{align}\label{Ar-Antimony}
\begin{split}
A^r(t_r^A) & = \eta_{mn}  -\frac{1}{4\pi}
\int_{-\infty}^{\infty}
d\omega 
\Big(
\Xi_{00}^{mn}(\omega,t_r^A)
x_0(\omega)
+ \Xi_{11}^{mn}(\omega,t_r^A)
x_1(\omega)\\
&+\Xi_{01}^{mn}(\omega,t_r^A)
(x_2(\omega) + Ix_3(\omega))
+ \Xi_{10}^{mn}(\omega,t_r^A) (x_2(\omega)-Ix_3(\omega)) \\
&+
%(
\Xi_{22}^{mn}(\omega,t_r^A)
%+ 
%\Xi_{32}^{mn}(\omega,t_r^A)) 
\pi \delta(\omega) 
%+ (
%\Xi_{33}^{mn}(\omega,t_r^A)+ 
%\Xi_{23}^{mn}(\omega,t_r^A)
%)
%\pi \delta(\omega)
\Big)\\
& = \eta_{mn}  
-\frac{1}{4\pi}
\int_{-\infty}^{\infty}
d\omega \Big(
\Xi_{00}^{mn}(\omega,t_r^A)
x_0(\omega)
+ \Xi_{11}^{mn}(\omega,t_r^A)
x_1(\omega)\\
&+(
\Xi_{01}^{mn}(\omega,t_r^A)+
\Xi_{10}^{mn}(\omega,t_r^A))
x_2(\omega) 
+ I(
\Xi_{01}^{mn}(\omega,t_r^A)-\Xi_{10}^{mn}(\omega,t_r^A))
x_3(\omega) +\eta^\prime_A(\omega)\Big), 
\end{split}
\end{align}
where $I = \sqrt{-1}$, and,
\begin{align}\label{etaPrimeA}
\begin{split}
\eta^\prime_A(\omega) &= 
%(
%\Xi_{22}^{mn}(\omega,t_r^A)
%+ 
%\Xi_{32}^{mn}(\omega,t_r^A)) \pi \delta(\omega \pm \omega^\prime_0) 
%+ (
%\Xi_{33}^{mn}(\omega,t_r^A)+ 
%\Xi_{23}^{mn}(\omega,t_r^A)
%)
\Xi_{22}^{mn}(\omega,t_r^A)
\tilde{S}_{22}(\omega) =
\Xi_{22}^{mn}(\omega,t_r^A)
\pi \delta(\omega) 
\end{split}
\end{align}
Using the Dirac delta function property $\int_{-\infty}^{\infty} d\omega f(\omega-a)\delta(\omega) =  f(a)$ and Eq.\eqref{Xi-negative}, we define and calculate the following parameter:
\begin{align}\label{etaA}
\begin{split}
\eta_A &= -\frac{1}{4\pi}
\int_{-\infty}^{\infty}  \eta^\prime_A(\omega) d\omega 
=
-\frac{1}{4\pi}
\int_{-\infty}^{\infty}  \Xi_{22}^{mn}(\omega,t_r^A) \pi \delta(\omega) d\omega \\
&=
-\frac{1}{4}
\Xi_{22}^{mn}(0,t_r^A) \equiv 0  \\
%&= 
%-\frac{1}{4}
%\Big(
%\Xi_{22}^{mn}(\pm \omega^\prime_0,t_r^A)
%+ 
%\Xi_{32}^{mn}(\pm \omega^\prime_0,t_r^A) 
%+ 
%\Xi_{33}^{mn}(0,t_r^A)+ 
%\Xi_{23}^{mn}(0,t_r^A)
%\Big)\\
%&= 
%-\frac{1}{4}
%\Big(2
%\Xi_{22}^{mn}( \omega^\prime_0,t_r^A)
%+ 
%2\Xi_{32}^{mn}(\omega^\prime_0,t_r^A) 
%+ 
%\Xi_{33}^{mn}(0,t_r^A)+ 
%\Xi_{23}^{mn}(0,t_r^A)
%\Big)\\
%&= 
%-\frac{1}{2}
%\Big(
%\Xi_{22}^{mn}( \omega^\prime_0,t_r^A)
%+ 
%\Xi_{32}^{mn}(\omega^\prime_0,t_r^A) 
%\Big),
\end{split}
\end{align}
Please note that the above parameter is calculated before applying the frequency range separation of the integral demonstrated in Eq.\eqref{full-half-freq-range}.
%and has not yet exclude the zero frequency polyspectra.
% Here we find:
% \begin{align}\label{Ar-Antimony-2-0}
% \begin{split}
% \eta_A = 0
% \end{split}
% \end{align}
The reason for $\eta_A$ being zero originates from the approximation made in Eq.\eqref{mainOfinal}, which results in Eq.\eqref{Antimony-knownPolyspectra}. Consequently, this leads to a zero value for $\eta'_A$ in Eq.\eqref{etaPrimeA}, and consequently $\eta_A$ becomes zero in Eq.\eqref{etaA}.

Now we simplify $A^r(t_r^A)$ in Eq.\eqref{Ar-Antimony} in terms of unknown noise functions. We use the finite frequency range for the noise spectroscopy protocol
($\int_{-\infty}^{\infty} \rightarrow$
$\int_{-\Omega}^{\Omega}$), utilize Eq.\eqref{Xi-negative}, the approximation introduced in Eq.\eqref{mainOfinal}, and substitute Eqs.\eqref{AntimonyRealFunctions} and \eqref{etaA} 
%or \eqref{etaA-2}) 
into Eq.\eqref{Ar-Antimony}, to achieve the following:
\begin{align}\label{Ar-Antimony-2}
\begin{split}
A^r(t_r^A) 
=& \eta_{mn}  - \frac{1}{4\pi} \int_{\omega_0}^{\Omega} d\omega \{ \\
&
\Big(\Xi_{00}^{mn}(\omega,t_r^A) + 
\Xi_{00}^{mn}(-\omega,t_r^A)\Big)
x_0(\omega)
+
\Big(\Xi_{11}^{mn}(\omega,t_r^A) + 
\Xi_{11}^{mn}(-\omega,t_r^A)\Big)
x_1(\omega)+\\
&
\Big(\Xi_{01}^{mn}(\omega,t_r^A) + 
\Xi_{10}^{mn}(\omega,t_r^A) +
\Xi_{01}^{mn}(-\omega,t_r^A) +
\Xi_{10}^{mn}(-\omega,t_r^A)\Big)
x_2(\omega)+\\
&
I\Big(\Xi_{01}^{mn}(\omega,t_r^A) - 
\Xi_{10}^{mn}(\omega,t_r^A) -
\Xi_{01}^{mn}(-\omega,t_r^A) +
\Xi_{10}^{mn}(-\omega,t_r^A)\Big)
x_3(\omega)
\} \\
=& \eta_{mn} + \eta_A - \frac{1}{2\pi} \times \\ 
& \int_{\omega_0}^{\Omega} d\omega \{ 
\Xi_{00}^{mn}(\omega,t_r^A) 
x_0(\omega)
+
\Xi_{11}^{mn}(\omega,t_r^A) 
x_1(\omega)+
\Big(\Xi_{01}^{mn}(\omega,t_r^A) + 
\Xi_{10}^{mn}(\omega,t_r^A)\Big)
x_2(\omega)
\}.
\end{split}
\end{align}
Next, we simplify the summation part of the quoct's expectation value. We substitute Eq.\eqref{Antimony-unknownPolyspectra}
%and \eqref{Antimony-knownPolyspectra},
into Eq.\eqref{quoct6-B},
and consider the variables name of Eq.\eqref{x_iVars}, 
to achieve the following:
\begin{align}\label{Br-Antimony0}
\begin{split}
B^r(t_r^B) & = \eta_{mn}  -\frac{M}{2T/r}
\sum_{k=-\infty}^{\infty} 
\Xi_{i\tilde{i}}^{mn}(rk\omega_0,t_r^B)
\tilde{S}_{i\tilde{i}}(rk\omega_0) \\
=&  \eta_{mn}  -\frac{M}{2T/r}
\sum_{k=-\infty}^{\infty} \\
\Big(&
\Xi_{00}^{mn}(rk\omega_0,T/r)
x_{0}(rk\omega_0)+
\Xi_{11}^{mn}(rk\omega_0,T/r)
x_1(rk\omega_0)\\
+&\Xi_{01}^{mn}(rk\omega_0,T/r)
(x_2(rk\omega_0) + ix_3(rk\omega_0)) \\
+& \Xi_{10}^{mn}(rk\omega_0,T/r) (x_2(rk\omega_0)-ix_3(rk\omega_0)) \\
%+&(
%\Xi_{22}^{mn}(rk\omega_0,T/r)
%+ 
%\Xi_{32}^{mn}(rk\omega_0,T/r)) \pi \delta(\omega \pm \omega^\prime_0) \\
+& 
%(\Xi_{33}^{mn}(rk\omega_0,T/r)+ 
%\Xi_{22}^{mn}(rk\omega_0,T/r)
%%)
%\pi \delta(\omega)
\eta^\prime_B(rk\omega_0)
\Big)\\
%=& \eta_{mn}  -\frac{M}{2T/r}
%\sum_{k=-\infty}^{\infty} \\
%& \Big(
%\Xi_{00}^{mn}(rk\omega_0,T/r)
%x_0(\omega)
%+ \Xi_{11}^{mn}(rk\omega_0,T/r)
%x_1(\omega)\\
%+&(
%\Xi_{01}^{mn}(rk\omega_0,T/r)+
%\Xi_{10}^{mn}(rk\omega_0,T/r))
%x_2(\omega) \\
%+& i(
%\Xi_{01}^{mn}(rk\omega_0,T/r)-\Xi_{10}^{mn}(rk\omega_0,T/r))
%x_3(\omega) +\eta^\prime_B(\omega)\Big),
\end{split}
\end{align}
where
\begin{align}\label{etaPrimeB}
\begin{split}
\eta^\prime_B(rk\omega_0) &= 
\Xi_{22}^{mn}(rk\omega_0,T/r)
\pi \delta(rk\omega_0)
%(
%\Xi_{22}^{mn}(rk\omega_0,T/r)
%+ 
%\Xi_{32}^{mn}(rk\omega_0,T/r)) \pi \delta(rk\omega_0 \pm \omega^\prime_0) \\
%&+ (
%\Xi_{33}^{mn}(rk\omega_0,T/r)+ 
%\Xi_{23}^{mn}(rk\omega_0,T/r)
%)
%\pi \delta(rk\omega_0)
\end{split}
\end{align}
%%For the Kronecker delta function we have $\sum_{i=-\infty}^{\infty} f_i\delta_{ai} =  f_a$. 
%%Due to the summation in front of $\delta(rk\omega_0 \pm \omega^\prime_0)$ in Eq.\eqref{Br-Antimony}, the Dirac delta function turns to the Kronecker delta function as $\delta_{rk\omega_0, \pm \omega^\prime_0}$, and by assuming $ \Xi^{mn}_{i\tilde{i}}(ak, T/r) = {\Xi^{\prime}}_{k}^{a}$ we have the following:
There is a summation in Eq.\eqref{Br-Antimony0} that precedes the 
%$\delta(rk\omega_0 \pm \omega^\prime_0)$
$\delta(r\omega_0\times k - 0)$
of Eq.\eqref{etaPrimeB}. This summation transforms the continuous Dirac delta function to its discrete alternative, i.e. Kronecker delta function of 
%$\delta_{rk\omega_0, \pm \omega^\prime_0}$. 
$\delta_{r\omega_0.k,\:0}$. Utilizing the property of the Kronecker delta function $\sum_{k=-\infty}^{\infty} f(\alpha k)\delta_{\alpha k,\:0} = f(0)$, 
%and considering a simplified and index notation of the function $ \Xi^{mn}_{22}(ak, T/r) \equiv {\Xi^{\prime}}_{k}^{a}$,
we find the following:
\begin{align}\label{etaB-0}
\begin{split}
\sum_{k=-\infty}^{\infty}
\Xi^{mn}_{22}(rk\omega_0, T/r) \delta(rk\omega_0 )
=
\sum_{k=-\infty}^{\infty}
\Xi^{mn}_{22}(rk\omega_0, T/r) \delta_{rk\omega_0,0}
%&\equiv 
%\sum_{k=-\infty}^{\infty} %{\Xi^{\prime}}_{k}^{(r\omega_0)}
%\delta_{k,0} 
%= 
% {\Xi^{\prime}}_
% {0}^{(r\omega_0)} \\
% & \equiv  
=
\Xi^{mn}_{22}(0, T/r ) 
\end{split}
\end{align}
Here we assumed $\delta_{r\omega_0k,0}\equiv \delta_{k,0}$, since the parameters $r$ and $\omega_0$ are nonzero.
This yields: 
\begin{align}\label{etaB}
\begin{split}
\eta_B 
=& 
-\frac{M}{2T/r} \sum_{k=-\infty}^{\infty}  \eta^\prime_B(rk\omega_0) \\
=& 
-\frac{M\pi}{2T/r} \sum_{k=-\infty}^{\infty}  
\Xi^{mn}_{22}(rk\omega_0, T/r) \delta(rk\omega_0 )  \\
=&
-\frac{M\pi}{2T/r} \Xi^{mn}_{22}(0, T/r ) 
\equiv 0
\end{split}
\end{align}
% where,
% \begin{align}\label{Br-Antimony-0}
% \begin{split}
% \eta_B = 0
% \end{split}
% \end{align}
The above zero equivalence is because of the fact that the approximation of Eq.\eqref{mainOfinal} yields Eq.\eqref{Antimony-knownPolyspectra}, leading to zero $\eta'_B$ in Eq.\eqref{etaPrimeB}, and zero $\eta_B$ in Eq.\eqref{etaB}.
Now Eq.\eqref{Br-Antimony0} turns to the following:
\begin{align}\label{Br-Antimony1}
\begin{split}
B^r(t_r^B) &= \eta_{mn} 
 - \frac{M}{2T/r}
\sum_{k=-\infty}^{\infty} \\
\Big( &
\Xi_{00}^{mn}(rk\omega_0,T/r)
x_0(rk\omega_0)
+ \Xi_{11}^{mn}(rk\omega_0,T/r)
x_1(rk\omega_0)\\
+&(
\Xi_{01}^{mn}(rk\omega_0,T/r)+
\Xi_{10}^{mn}(rk\omega_0,T/r))
x_2(rk\omega_0) \\
+& i(
\Xi_{01}^{mn}(rk\omega_0,T/r)-\Xi_{10}^{mn}(rk\omega_0,T/r))
x_3(rk\omega_0) \Big),
\end{split}
\end{align}
Using the finite frequency range for the noise spectroscopy protocol
($\Sigma_{k=-\infty}^{\infty} \rightarrow$
$\Sigma_{k=-N}^{N},\:N=\floor{\Omega}/{\omega_0}$), utilizing Eq.\eqref{Xi-negative}, and the approximation introduced in Eq.\eqref{mainOfinal},
%, and assuming Eqs.\eqref{AntimonyRealFunctions} and (\eqref{etaB} or \eqref{etaB-2}) in Eq.\eqref{Br-Antimony}, 
we achieve the following:
\begin{align}\label{Br-Antimony}
\begin{split}
B^r(t_r^B) &= \eta_{mn} 
 - \frac{M}{2T/r}
\sum_{k=1}^{N} \\
& \Big(
2\Xi_{00}^{mn}(rk\omega_0,T/r)
x_0(rk\omega_0)
+ 2\Xi_{11}^{mn}(rk\omega_0,T/r)
x_1(rk\omega_0)\\
+&2(
\Xi_{01}^{mn}(rk\omega_0,T/r)+
\Xi_{10}^{mn}(rk\omega_0,T/r))
x_2(rk\omega_0) + i \times 0 \times
x_3(rk\omega_0) \Big) \\
&= \eta_{mn} 
+ \eta_B - \frac{M}{T/r} 
\sum_{k=1}^{N} \\
& \Big(
\Xi_{00}^{mn}(rk\omega_0,T/r)
x_0(rk\omega_0)
+ \Xi_{11}^{mn}(rk\omega_0,T/r)
x_1(rk\omega_0)\\
+&(
\Xi_{01}^{mn}(rk\omega_0,T/r)+
\Xi_{10}^{mn}(rk\omega_0,T/r))
x_2(rk\omega_0) 
\Big)
\end{split}
\end{align}
Next, we further simplify the integral part of the expectation value of the Antimony quoct of Eq.\eqref{Ar-Antimony-3} as follows:
\begin{align}\label{Ar-Antimony-4}
\begin{split}
A^r(MT/r) 
&= \eta_{mn}
%+ \eta_A
-\frac{1}{2\pi} 
\sum_{i=0}^2
\tilde{A}_i^{rmn},
\end{split}
\end{align}
where,
\begin{align}\label{Atilde}
\begin{split}
& \tilde{A}_i^{rmn} = \int_{\omega_0}^{\Omega}
d\omega 
x_i(\omega) 
C_i^{mn}(\omega,MT/r)\\
&\tilde{A}_0^{rmn} = 
\int_{\omega_0}^{\Omega}
d\omega 
x_0(\omega) 
\Xi_{00}^{mn}(\omega,\frac{MT}{r}),\\
&\tilde{A}_1^{rmn} = 
\int_{\omega_0}^{\Omega}
d\omega 
x_1(\omega) 
\Xi_{11}^{mn}
(\omega,\frac{MT}{r}),\\
&\tilde{A}_2^{rmn} = 
\int_{\omega_0}^{\Omega}
d\omega 
x_2(\omega) 
(\Xi_{01}^{mn}(\omega,\frac{MT}{r})+
\Xi_{10}^{mn}(\omega,\frac{MT}{r})).
\end{split}
\end{align}
and,
\begin{align}\label{Atilde2}
\begin{split}
&\tilde{A}_0^{rmn} = 
A^{rmn}_{0,00},\:\:\: 
\tilde{A}_1^{rmn} = 
A^{rmn}_{1,11},\:\:\: 
\tilde{A}_2^{rmn} = 
A^{rmn}_{2,01}+
A^{rmn}_{2,10},\\
&A^{rmn}_{i,l\tilde{l}}=
\int_{\omega_0}^{\Omega}  d\omega 
x_i(\omega) \Xi_{l\tilde{l}}^{mn}(\omega,\frac{MT}{r})
\end{split}
\end{align}
Combining Eqs.\eqref{Ar-Antimony-4} and \eqref{FsimplifiedIntegral} results in the following:%a formula consists of terms as follows: 
\begin{align}\label{IntegralAntimony0}
\begin{split} 
&A^{rmn}_{i,l\tilde{l}}=
\int_{\omega_0}^{\Omega}  d\omega 
x_i(\omega) \Xi_{l\tilde{l}}^{mn}(\omega,\frac{MT}{r}) 
= \\
&\sum_{\substack{ab\tilde{a}\tilde{b}\\
a'b'\tilde{a}'\tilde{b}'\\
\in S}}
\sum_{\substack{pqq'\in S\\ \tilde{p}\tilde{q}\tilde{q}'\in S}}
\tilde{\beta}^{pqq'}_{lab}
\tilde{\beta}^{\tilde{p}\tilde{q}\tilde{q}'}_{\tilde{l}\tilde{a}\tilde{b}}
(\prescript{ind}{\tilde{a}\tilde{b}}{\lambda}^{\tilde{a}'\tilde{b}'}_{a'b'}) \times  \int_{\omega_0}^{\Omega}  d\omega x_i(\omega)
F^{ab}_{a'b'}(\omega,\frac{MT}{r}) 
F^{*-\tilde{a},-\tilde{b}}_{-\tilde{a}',-\tilde{b}'}(\omega,\frac{MT}{r}) = \\
&
\sum_{\substack{ab\tilde{a}\tilde{b}\\
a'b'\tilde{a}'\tilde{b}'\\
\in S}}
\sum_{\substack{pqq'\in S\\ \tilde{p}\tilde{q}\tilde{q}'\in S}}
\tilde{\beta}^{pqq'}_{lab}
\tilde{\beta}^{\tilde{p}\tilde{q}\tilde{q}'}_{\tilde{l}\tilde{a}\tilde{b}}
(\prescript{ind}{\tilde{a}\tilde{b}}{\lambda}^{\tilde{a}'\tilde{b}'}_{a'b'}) \int_{\omega_0}^{\Omega}  d\omega 
\frac{x_i(\omega)}{\omega^2} \times\\
& \sum_{n_1,h_1=0}^{M-1,d-1}  \sum_{n_2,h_2=0}^{M-1,d-1}
y^{i_{h_1},j_{h_1}}_{\A_y} y^{*i_{h_2},j_{h_2}}
_{-\tilde{\A}_y} 
(e^{It^{h_1+1}_r\omega} - 
e^{It^{h_1}_r\omega}  )
(e^{-It^{h_2+1}_r\omega} - 
e^{-It^{h_2}_r\omega}  )
e^{Id(t^{n_1}_r-t^{n_2}_r)\omega}
= \\
&
\sum_{\substack{ab\tilde{a}\tilde{b}\\
a'b'\tilde{a}'\tilde{b}'\\
\in S}}
\sum_{\substack{pqq'\in S\\ \tilde{p}\tilde{q}\tilde{q}'\in S}}
\tilde{\beta}^{pqq'}_{lab}
\tilde{\beta}^{\tilde{p}\tilde{q}\tilde{q}'}_{\tilde{l}\tilde{a}\tilde{b}}
(\prescript{ind}{\tilde{a}\tilde{b}}{\lambda}^{\tilde{a}'\tilde{b}'}_{a'b'}) 
\sum_{n_1,h_1=0}^{M-1,d-1}  \sum_{n_2,h_2=0}^{M-1,d-1}
y^{i_{h_1},j_{h_1}}_{\A_y} y^{*i_{h_2},j_{h_2}}_{-\tilde{\A}_y} \times\\
& 
\int_{\omega_0}^{\Omega}  d\omega 
\frac{x_i(\omega)}{\omega^2} 
(e^{It^{h_1+1}_r\omega} - 
e^{It^{h_1}_r\omega}  )
(e^{-It^{h_2+1}_r\omega} - 
e^{-It^{h_2}_r\omega}  )
e^{Id(t^{n_1}_r-t^{n_2}_r)\omega},
\end{split}
\end{align}
where $\A_y=a,b,a',b'$, and $\tilde{\A}_y=\Tilde{a},\Tilde{b},\Tilde{a}',\Tilde{b}'$.
\subsubsection{Poissonian noise spectra}\label{AppendixPoissonianNoiseSpectra}
To simulate our noise spectroscopy protocol for the Antimony quoct, we assume the true Poissonian function $x_i(\omega) = \omega^2 e^{g_i|\omega|}$ for the noise polyspectra. 
Therefore, Eq.\eqref{IntegralAntimony0} transforms into the following equation:
\begin{align}\label{Antimony5-1}
\begin{split}
&A^{rmn}_{i,l\tilde{l}}=
\sum_{\substack{ab\tilde{a}\tilde{b}\\
a'b'\tilde{a}'\tilde{b}'\\
\in S}}
\sum_{\substack{pqq'\in S\\ \tilde{p}\tilde{q}\tilde{q}'\in S}}
\tilde{\beta}^{pqq'}_{lab}
\tilde{\beta}^{\tilde{p}\tilde{q}\tilde{q}'}_{\tilde{l}\tilde{a}\tilde{b}}
(\prescript{ind}{\tilde{a}\tilde{b}}{\lambda}^{\tilde{a}'\tilde{b}'}_{a'b'}) 
\sum_{n_1,h_1=0}^{M-1,d-1}  \sum_{n_2,h_2=0}^{M-1,d-1}
y^{i_{h_1},j_{h_1}}_{\A_y} y^{*i_{h_2},j_{h_2}}
_{-\tilde{\A}_y} 
\times\\
& 
\int_{\omega_0}^{\Omega}  d\omega 
e^{-g_i\omega}
(e^{It^{h_1+1}_r\omega} - 
e^{It^{h_1}_r\omega}  )
(e^{-It^{h_2+1}_r\omega} - 
e^{-It^{h_2}_r\omega}  )
e^{Id(t^{n_1}_r-t^{n_2}_r)\omega} = \\
& 
\sum_{\substack{ab\tilde{a}\tilde{b}\\
a'b'\tilde{a}'\tilde{b}'\\
\in S}}
\sum_{\substack{pqq'\in S\\ \tilde{p}\tilde{q}\tilde{q}'\in S}}
\tilde{\beta}^{pqq'}_{lab}
\tilde{\beta}^{\tilde{p}\tilde{q}\tilde{q}'}_{\tilde{l}\tilde{a}\tilde{b}}
(\prescript{ind}{\tilde{a}\tilde{b}}{\lambda}^{\tilde{a}'\tilde{b}'}_{a'b'}) 
\sum_{n_1,n_2=0}^{M-1} \sum_{h_1,h_2=0}^{d-1}
y^{i_{h_1},j_{h_1}}_{\A_y} y^{*i_{h_2},j_{h_2}}
_{-\tilde{\A}_y}  \times \\
& \int_{\omega_0}^{\Omega} d\omega 
\sum_{j=0}^3 (-1)^{f_j}e^{G_j(g_i,n_1,h_1,n_2,h_2)\omega} = \\
&
\sum_{\substack{ab\tilde{a}\tilde{b}\\
a'b'\tilde{a}'\tilde{b}'\\
\in S}}
\sum_{\substack{pqq'\in S\\ \tilde{p}\tilde{q}\tilde{q}'\in S}}
\tilde{\beta}^{pqq'}_{lab}
\tilde{\beta}^{\tilde{p}\tilde{q}\tilde{q}'}_{\tilde{l}\tilde{a}\tilde{b}}
(\prescript{ind}{\tilde{a}\tilde{b}}{\lambda}^{\tilde{a}'\tilde{b}'}_{a'b'}) 
\sum_{n_1,n_2=0}^{M-1}  \sum_{h_1,h_2=0}^{d-1}
y^{i_{h_1},j_{h_1}}_{\A_y} y^{*i_{h_2},j_{h_2}}
_{-\tilde{\A}_y} \times \\
& \sum_{j=0}^3 (-1)^{f_j} 
D_j(g_i,n_1,h_1,n_2,h_2) = \\
& 
\sum_{\substack{ab\tilde{a}\tilde{b}\\
a'b'\tilde{a}'\tilde{b}'\\
\in S}}
\sum_{\substack{pqq'\in S\\ \tilde{p}\tilde{q}\tilde{q}'\in S}}
\tilde{\beta}^{pqq'}_{lab}
\tilde{\beta}^{\tilde{p}\tilde{q}\tilde{q}'}_{\tilde{l}\tilde{a}\tilde{b}}
(\prescript{ind}{\tilde{a}\tilde{b}}{\lambda}^{\tilde{a}'\tilde{b}'}_{a'b'}) 
\sum_{n_1,n_2=0}^{M-1}  \sum_{h_1,h_2=0}^{d-1}
y^{i_{h_1},j_{h_1}}_{\A_y} y^{*i_{h_2},j_{h_2}}
_{-\tilde{\A}_y}   
\Tilde{D}(g_i,h_1,h_2)\\
\end{split}
\end{align}
where,
\begin{align}\label{Antimony5Variables}
\begin{split}
& G_0(g_i,n_1,h_1,n_2,h_2) = -g_i + Id(t_r^{n_1} - t_r^{n_2}) +
I(t^{h_1+1}_r - t^{h_2+1}_r), \\
& G_1(g_i,n_1,h_1,n_2,h_2) = -g_i + Id(t_r^{n_1} - t_r^{n_2}) +
I(t^{h_1+1}_r - t^{h_2}_r), \\
& G_2(g_i,n_1,h_1,n_2,h_2) = -g_i + Id(t_r^{n_1} - t_r^{n_2}) +
I(t^{h_1}_r - t^{h_2+1}_r), \\
& G_3(g_i,n_1,h_1,n_2,h_2) = -g_i + Id(t_r^{n_1} - t_r^{n_2}) +
I(t^{h_1}_r - t^{h_2}_r), \\
& f_0 = f_3 = 0, \:
f_1 = f_2 = 1,\: 
I = \sqrt{-1},\\
&D_j(g_i,n_1,h_1,n_2,h_2) = 
1/G_j(g_i,n_1,h_1,n_2,h_2)
\Big(
e^{\Omega G_j(g_i,n_1,h_1,n_2,h_2)} - 
e^{\omega_0 G_j(g_i,n_1,h_1,n_2,h_2)}
\Big),\\
& \tilde{D}(g_i,h_1,h_2) = 
\sum_{n_1,n_2=0}^{M-1}
\sum_{j=0}^{3}
(-1)^{f_j}D_j(g_i,n_1,h_1,n_2,h_2),
\end{split}
\end{align}
and $\Omega$ is the maximum noise frequency that the quantum system senses.
This yields:
\begin{align}\label{Antimony5-0}
\begin{split}
&A^{rmn}_{i,l\tilde{l}}=
\sum_{\substack{\tilde{a}\tilde{b}\tilde{a}'\tilde{b}',\\
aba'b',\\
h_1h_2
\in S}}
\tilde{\beta}^{1}_{lab}
\tilde{\beta}^{1}_{\tilde{l}\tilde{a}\tilde{b}}
(\prescript{ind}{\tilde{a}\tilde{b}}{\lambda}^{\tilde{a}'\tilde{b}'}_{a'b'}) 
y^{i_{h_1},j_{h_1}}_{aba'b'} y^{*i_{h_2},j_{h_2}}
_{-\tilde{a},-\tilde{b},-\tilde{a}',-\tilde{b}'}    
\tilde{D}(g_i,h_1,h_2),
\end{split}
\end{align}
where,
\begin{align}\label{Antimony5-0-variables}
\begin{split}
\tilde{\beta}^{1}_{lab}
= \sum_{pqq'\in S}\tilde{\beta}^{pqq'}_{lab}
\end{split}
\end{align}
\section{Proofs}
This appendix presents the detailed proofs supporting the results discussed in various sections, providing  derivations essential to our formalism. Throughout this appendix, \textit{i} denotes $\sqrt{-1}$ unless explicitly used as a variable index.
\subsection{Fluctuating qudit energies and noise functions in the Weyl basis}\label{apdx.3.1} 
In this section, we find the function mapping the environmental noises ($\beta_{ab}(t)$) to the eigen-energies and the interaction energies (coherences) of a qudit ($\varepsilon_{ij}(t)$) surrounded by noises. To achieve this, we equate the effective Hamiltonian of the qudit and noise in the Weyl basis (Eq.\eqref{HWRW}), to that of in computational basis (Eq.\eqref{Hcomputationalbasis}). Notably, the complex nature of the time-dependent noise coefficients ($\beta_{ab}(t)$) results in the qudit's eigen-energies and interaction energies (coherences) in Eq.\eqref{Hcomputationalbasis} becoming both time-dependent and complex, indicating the presence of all XYZ type of noises. In the following, we detail the derivation of the mapping function for the general case of the complete Weyl Hamiltonians. Here we equate the second line of Eq.\eqref{HWRW} to Eq.\eqref{Hcomputationalbasis} and utilize Eq.\eqref{ij} as follows:
\begin{align}\label{C1-betaEpsilonMain}
\begin{split}
& H = \sum_{i,j \in S} \varepsilon_{ij}(t) \ket{i}\bra{j} = \sum_{i,j \in S} \varepsilon_{ij}(t) \Big(\frac{1}{d}\sum_{a \in S}\xi^{-a i} Z^a X^{i-j}\Big) \equiv \sum_{a,b \in S} \beta_{ab}(t) Z^aX^b \\
&\implies \beta_{ab}(t)=\frac{1}{d}\sum_{i,j \in S} \xi^{-a i} \delta_{b,i-j} \varepsilon_{ij}(t)\\
\end{split}
\end{align}
By considering both time-dependent and time-independent components of the noise function and energies, the following results are found:
\begin{align}\label{C1-betaEpsilon-01}
\begin{split}
&\beta_{ab}(t) = \beta_{ab}^0 + 
\beta^1_{ab}(t) = \frac{1}{d} \sum_{i,j \in S} 
\xi^{-a i} \delta_{b,i-j} 
(\varepsilon^0_{ij} + \delta\varepsilon_{ij}(t)) \\
& \rightarrow \beta_{ab}^0 = \frac{1}{d}\sum_{i,j \in S}\xi^{-a i}\delta_{b,i-j} \varepsilon^0_{ij},\:\:\: \beta^1_{ab}(t) = \frac{1}{d}\sum_{i,j \in S}\xi^{-a i}\delta_{b,i-j} \delta\varepsilon_{ij}(t)
\end{split}
\end{align}
Next, we express energies in terms of noise functions. Initially, we discover the following relationship:
\begin{align}\label{C1-ZaXbKetBraNM}
\begin{split}
Z^aX^b =& \sum_{n,m \in S} 
R_{nm} 
\ket{n}\bra{m},\\
\bra{l}Z^aX^b\ket{l'} = &
\sum_{n,m \in S} 
R_{nm} 
\bra{l}\ket{n}
\bra{m}\ket{l'} \\
\delta_{l,l' \oplus b}
\xi^{al} =& \sum_{n,m}R_{nm}\delta_{l,n}
\delta_{l',m},\\
\xi^{al} \delta_{l,l' \oplus b} =& R_{ll'} \\
%\xi^{al} =& R_{l,l \oplus b} \\
\hookrightarrow Z^aX^b =& \sum_{n,m \in S} 
\xi^{an} \delta_{n,m \oplus b}
\ket{n}\bra{m},\\
\end{split}
\end{align}
Using the above equation we have:
\begin{align}\label{C1-EpsilonBetaMain}
\begin{split}
&H = \sum_{a,b \in S} \beta_{ab}(t) Z^aX^b = 
\sum_{a,b \in S} \beta_{ab}(t) [\sum_{n,m \in S} 
\xi^{an} \delta_{n,m \oplus b}
\ket{n}\bra{m}] 
\equiv 
\sum_{n,m \in S} \varepsilon_{nm}(t) \ket{n}\bra{m} \\
&\implies 
\varepsilon_{nm}(t) = 
\sum_{a,b \in S} \xi^{an}  \delta_{n,m \oplus b} \beta_{ab}(t)
\end{split}
\end{align}
Now assuming the time-dependent and time-independent terms we have:
\begin{align}\label{C1-EpsilonBeta-01}
\begin{split}
&\varepsilon_{nm}(t) = \varepsilon^0_{nm}+\delta\varepsilon_{nm}(t) = \sum_{a,b \in S} \xi^{an} \delta_{n,m \oplus b} 
(\beta_{ab}^0+ \beta^1_{ab}(t)) \\
&\rightarrow 
\varepsilon^0_{nm} = \sum_{a,b \in S} \xi^{an} \delta_{n,m \oplus b} \beta_{ab}^0, \:\:\:
\delta\varepsilon_{nm}(t) = \sum_{a \in S} \xi^{an} \delta_{n,m \oplus b} \beta^1_{ab}(t)
\end{split}
\end{align}
Eqs.\eqref{C1-betaEpsilonMain} and \eqref{C1-EpsilonBetaMain} demonstrate a linear relationship between the variations of the qudit's eigen-energies or interaction energies and those of the XYZ-dephasing noise functions.
For non-dissipative noise types like dephasing noises, the Hermitian characteristics of the effective Hamiltonian of Eq.\eqref{C1-EpsilonBetaMain}, i.e. $\varepsilon_{nm}(t) = \varepsilon^*_{mn}(t)$ $\implies$ $\varepsilon_{nn}(t) \in \mathbbm{R}$,  imposes a constraint on the noise functions as follows: 
\begin{align}
\begin{split}
\varepsilon_{nn}(t) & =  \sum_{a,b \in S} \xi^{an}  \delta_{n,n \oplus b} \beta_{ab}(t) = 
\sum_{a \in S}\xi^{a n}\beta_{a0}(t) \\
& = f + \Big(\sum_{a \in S_+}
\xi^{a n}\beta_{a0}(t) + 
\sum_{a \in S_-}
\xi^{a n}\beta_{a0}(t)
\Big)
\\
& = f + \Big(\sum_{a \in S_+}\xi^{a n}\beta_{a0}(t) + \sum_{a \in S_+}
\xi^{-a n}\beta_{-a,0}(t)\xi^{-a \times 0}\Big)
\\
& = f + \Big(\sum_{a \in S_+}\xi^{a n}\beta_{a0}(t) + \sum_{a \in S_+}
(\xi^{a n}\beta_{a0}(t))^*\Big),\\
f &= \beta_{00}(t) + \beta_{d/2,0}(t),\:\rm d \:even\\
f &= \beta_{00}(t),\:\rm d \:odd
\end{split}
\end{align}
As the terms within the parentheses in the above equation are each other's conjugates, the entire expression inside the parentheses becomes a real number. Considering that $\varepsilon_{nn}(t) \in \mathbbm{R}$, the parameter $f$ needs to be real, which is not a trivial result for even d.

\subsection{Noise power spectral density in Weyl and computational bases}\label{energy-correlations}
To gain a better understanding of the noise polyspectra, we investigate the connecting function of the qudit's energies and the statistical average of the correlations of the noise functions, denoted as $\langle\beta_{ab}(0)\beta_{a'b'}(t)\rangle$. In particular, we derive this relationship for the most general case, which is the complete Weyl case, utilizing Eq.\eqref{C1-betaEpsilonMain}. The achieved relation is as follows:
\begin{align}\label{C2-BetaCorrelations-EnergyCorrelations}
\begin{split}
&\langle\beta_{ab}(0)\beta_{a'b'}(t)\rangle
= 
\frac{1}{d^2}\sum_{m n m' n' \in S}
\xi^{-(a m  + a' m' )}
\delta_{b,m-n}\delta_{b',m'-n'}
\langle\varepsilon_{mn}(0)  \varepsilon_{m'n'}(t)\rangle \\
&= 
\frac{1}{d^2}\sum_{m m' \in S}
\xi^{-(a m  + a' m' )}
\langle\varepsilon_{m,m-b}(0)  \varepsilon_{m',m'-b'}(t)\rangle \\
&=
\frac{1}{d^2}\sum_{m m'\in S}
\xi^{-(a m  + a' m' )}
\varepsilon^0_{mn} \varepsilon^0_{m'n'}
+
\frac{1}{d^2}\sum_{m m'\in S}
\xi^{-(a m  + a' m' )}
\langle\delta \varepsilon_{m,m-b}(0) 
\delta \varepsilon_{m',m'-b'}(t)\rangle,
\end{split}
\end{align}
where we used the following equation:
\begin{align}\label{C2-EnergyCorrelationsExpectations}
\begin{split}
& \langle\varepsilon_{mn}(0)  \varepsilon_{m'n'}(t)\rangle
=
\langle 
\Big(  \varepsilon^0_{mn}  + \delta \varepsilon_{mn}(0)  \Big)
\Big(  \varepsilon^0_{m'n'}+ \delta \varepsilon_{m'n'}(t)   \Big) 
\rangle\\
&= 
\Big(\langle\varepsilon^0_{mn} \varepsilon^0_{m'n'}\rangle
+
\langle\varepsilon^0_{mn} 
\delta \varepsilon_{m'n'}(t)\rangle
+
\langle\delta \varepsilon_{mn}(0) \varepsilon^0_{m'n'}\rangle
+
\langle\delta \varepsilon_{mn}(0) 
\delta \varepsilon_{m'n'}(t)\rangle\Big)\\
&=
\varepsilon^0_{mn} \varepsilon^0_{m'n'}
+
\langle\delta \varepsilon_{mn}(0) 
\delta \varepsilon_{m'n'}(t)\rangle
\end{split}
\end{align}
Here we used the fact that the energies are uniformly oscillating around their time-independent values, i.e. possessing a zero-mean Gaussian distribution, and so the average of the energy variation of a single energy level as well as that of an interaction energy is zero. Eq.\eqref{C2-BetaCorrelations-EnergyCorrelations} indicates that the correlation of the noise functions 
($\langle\beta_{ab}(0)\beta_{a'b'}(t)\rangle,\:(a,b,\Tilde{a},\Tilde{b}\in S)$) 
is proportional to the sum of the correlations of the variations of qudit's energies ($\langle\delta \varepsilon_{mn}(0) \delta \varepsilon_{m'n'}(t)\rangle$) with the following conditions:
\begin{align}\label{energy-noise-condition}
\begin{split}
b=n-m,\:\:\:b'=n'-m'
\end{split}
\end{align}
Now we find the reverse relation of Eq.\eqref{C2-BetaCorrelations-EnergyCorrelations}.
\begin{align}\label{C2-EnergyCorrelations-BetaCorrelations}
\begin{split}
 \langle      \varepsilon_{mn}(0)  \varepsilon_{m'n'}(t)   \rangle
&= 
\sum_{a b a' b' \in S}
\xi^{a m  + a' m' }
\delta_{n,m+b}\delta_{n',m'+b'}
\langle    \beta_{ab}(0)\beta_{a'b'}(t)  \rangle \\
&= 
\sum_{a a' \in S}
\xi^{a m  + a' m' }
\langle    \beta_{a, n-m}(0)\beta_{a', n'-m'}(t)  \rangle 
\end{split}
\end{align}

Combining Eq.\eqref{C2-EnergyCorrelationsExpectations} and \eqref{C2-EnergyCorrelations-BetaCorrelations}, we find the following:
\begin{align}\label{C2-NoiseCorrelationsExpectations-betafunctions-final}
\begin{split}
&  \langle\delta \varepsilon_{mn}(0)  \delta \varepsilon_{m'n'}(t)\rangle = 
-  \varepsilon^0_{mn} \varepsilon^0_{m'n'}  
+ \sum_{a a' \in S}
\xi^{a m  + a' m'}
\langle  \beta_{a,n-m}(0)  \beta_{a',n'-m'}(t)   \rangle,
\end{split}
\end{align}
where the first term of the right side is time independent. 
In our noise spectroscopy formalism, we find noise power spectral densities in the Weyl basis that are the Fourier transform of the noise correlations:
\begin{align}\label{NoisePolyspectra-NoiseCorrelation}
\begin{split}
S_{ab}^{a'b'}(\omega) = \F\Big( \langle  \beta_{ab}(0)  \beta_{a'b'}(t)   \rangle \Big).
\end{split}
\end{align}
Using the linearity of the Fourier transform, and the fact that the Fourier transform of a constant expression is $\F[C] = 2\pi C \delta(\omega)$, Eq.\eqref{C2-NoiseCorrelationsExpectations-betafunctions-final} yields: 
\begin{align}\label{C2-EnergyCorrelations-NoiseCorrelations-FourierTransform}
\begin{split}
 \F \Big( \langle\delta \varepsilon_{mn}(0)  \delta \varepsilon_{m'n'}(t)\rangle  \Big) = &
\sum_{a a' \in S}
\xi^{a m  + a' m'}
\F \Big( \langle  \beta_{a,n-m}(0)  \beta_{a',n'-m'}(t)   \rangle  \Big)  \\
= &
\sum_{a a' \in S}
\xi^{a m  + a' m'}
S_{a,n-m}^{a',n'-m'}(\omega)
\end{split}
\end{align}
where we ignored the first term including $\delta(\omega\rightarrow 0)=1$, since our noise spectroscopy formalism studies frequencies larger than $\omega_0=2\pi /T$.
Now We remind that $\delta \varepsilon_{nn}(t)$ denotes the fluctuation of the $n^{th}$ energy level, while $\delta \varepsilon_{nm}(t)$ represents the variation of the interaction energy or coherences. The latter can be envisioned as the energy variation associated with an electron in an entangled state $\frac{1}{\sqrt{2}}(|m\rangle + |n\rangle)$. These fluctuations in energy arise due to the stochastic energy exchanges with different quanta of energy in the qudit's surroundings. Such exchanges take place through the absorption or emission of energy carriers with frequency $\omega=\delta \varepsilon_{nm}(t)/\hbar$.
Consequently, the noise energy emitted or absorbed by the qudit is the sum of all energy variations of the qudit, i.e. $\Big(\sum_{mn}\delta\varepsilon_{nm}(t)\Big)$.

In our analysis, we assumed that the environmental noises affecting the qudit's energy levels are stochastic with \textbf{\textit{zero-mean}} Gaussian distributions and are \textbf{\textit{stationary}} as described in Eq.\eqref{stationary}. By making this assumption, we should calculate the noise energy not by summing over the energy variations, rather by summing over the average of many realizations of the \textbf{auto and cross correlations} of the energy variations, that is $\sum_{mnm'n'} \langle\delta \varepsilon_{mn}(0) \delta \varepsilon_{m'n'}(t)\rangle$. 

The Fourier transform of this summation yields the corresponding quantity in the frequency domain, that is the total noise spectral density at  frequency $\omega$ affected the qudit:
\begin{align}\label{FourierTransform-EnergyCorrelations}
\begin{split}
S'(\omega) =  \sum_{mnm'n'} \F \Big(  \langle\delta \varepsilon_{mn}(0)  \delta \varepsilon_{m'n'}(t)\rangle  \Big) =  \sum_{mnm'n'} \bar{S}_{mn}^{m'n'}(\omega),
\end{split}
\end{align}
%is the noise spectrum detected by the qudit at frequency $\omega$. 
where,
\begin{align}\label{NoisePower-WeyAndComputationalBases}
\begin{split}
\bar{S}_{mn}^{m'n'}(\omega) =&  \F \Big(  \langle\delta \varepsilon_{mn}(0)  \delta \varepsilon_{m'n'}(t)\rangle \Big).
\end{split}
\end{align}
is defined as the noise power spectral density in the computational basis at frequency $\omega$, associated to energy levels $(m,n,m',n')$. 
Combining Eqs.\eqref{C2-EnergyCorrelations-NoiseCorrelations-FourierTransform} and \eqref{FourierTransform-EnergyCorrelations}, we find:
\begin{align}\label{NoiseSpectrum-DifferentRepresentations}
\begin{split}
S'(\omega) = 
 \sum_{mnm'n' \in S}
 \sum_{a a' \in S}
\xi^{a m  + a' m'}
S_{a,n-m}^{a',n'-m'}(\omega),
\end{split}
\end{align}
that relates the noise polyspectra in the Weyl basis respresenation $S_{ab}^{a'b'}(\omega)$,  to the total noise spectral density $S'(\omega)$ at frequency $\omega$ that is detected by the qudit. 

The equations of sections \ref{apdx.3.1} and \ref{energy-correlations} are derived for the general Weyl basis representation. To find these equations for the Reduced Weyl basis, given Eq.\eqref{HWRW}, we consider $b=0$ and $b'=0$ or equivalently $m=n$ and $m'=n'$ in each equation.  Here we present Eq.\eqref{NoiseSpectrum-DifferentRepresentations} for the Reduced Weyl case as follows:
\begin{align}\label{NoiseSpectrum-DifferentRepresentations-reducedWeyl}
\begin{split}
S'(\omega) = 
\sum_{mm'a a' \in S}
\xi^{a m  + a' m'}
S_{a,a'}(\omega),
\end{split}
\end{align}
where we ignored the known zero indices in $S_{a,0}^{a',0}(\omega)$, and transformed it to the usual notation of the Reduced Weyl sections, that is $S_{a,a'}(\omega)$.

\subsection{Symmetric and anti-symmetric properties of noise spectra}\label{apdx.3.2} 
In our qudit noise spectroscopy formalism, when considering the complete Weyl basis in the general case, the symmetries of noise polyspectra are found as follows:
\begin{align}\label{general_S(w,t)}
\begin{split}
S^{\tilde{A}_{\beta}}_{\A_{\beta}}(\omega) = S^{\A_{\beta}}_{\tilde{\A}_{\beta}}(-\omega)
= S^{*-\tilde{\A}_{\beta}}_{-\A_{\beta}}(-\omega)\xi^{\A^n_z \A^n_x + \tilde{\A}^n_z \tilde{\A}^n_x},     
\end{split}    
\end{align}
The proof of the above statement is as follows:
\begin{align}\label{C.2-0}
\begin{split}
S_{\A_S,\rightarrow}(\omega) & = S^{\tilde{\A}_{\beta}}_{\A_{\beta}}(\omega) = \int_{-\infty}^{\infty} dt e^{i\omega t} \langle\beta_{\A_{\beta}}(0)\beta_{\tilde{\A}_{\beta}}(t)\rangle,\\
S_{\A_S,\leftarrow}(-\omega) & = S^{\A_{\beta}}_{\tilde{\A}_{\beta}}(-\omega)  =\int_{-\infty}^{+\infty} dt e^{i(-\omega) t}\langle \beta_{\tilde{\A}_{\beta}}(0)\beta_{\A_{\beta}}(t)\rangle= \int_{-\infty}^{+\infty} dt e^{i\omega t} \langle \beta_{\tilde{\A}_{\beta}}(t)\beta_{\A_{\beta}}(0)\rangle\\
S^{*-\tilde{\A}_{\beta}}_{-\A_{\beta}}(-\omega) & = \int_{-\infty}^{+\infty} dt \{e^{i(-\omega) t}\}^* \langle \beta_{-\A_{\beta}}^*(0)\beta_{-\tilde{\A}_{\beta}}^*(t)\rangle \\
&=\int_{-\infty}^{+\infty} dt e^{i\omega t} 
\xi^{\A^n_z \A^n_x + \tilde{\A}^n_z \tilde{\A}^n_x}
\langle \beta_{\A_{\beta}}(0)\beta_{\tilde{\A}_{\beta}}(t)\rangle 
\end{split}
\end{align}
The final expression in the above equation was derived using Eq.\eqref{beta_condition}.
\subsubsection{Qutrit with Reduced Weyl Hamiltonian}\label{Qutrit-S(w)symproperties}
In this section, we demonstrate the Qutrit version of Eq.\eqref{general_S(w,t)}, which is as follows:
%and to prove it we used Eqs.\eqref{H^W(t)} and \eqref{beta_condition}.
\begin{align}
\begin{split}
S_{a,b}(\omega)=& S_{-a,-b}^*(-\omega)= S_{b,a}(-\omega).
\end{split}    
\end{align}
We prove the above statement as follows: %We used Eqs.\eqref{H^W(t)} and \eqref{beta_condition}.
\begin{align}\label{C.2}
\begin{split}
S_{a,b}(\omega) & = \int_{-\infty}^{\infty} dt e^{i\omega t} \langle\beta_a(0)\beta_b(t)\rangle,\\
S_{-a,-b}^*(-\omega) & = \int_{-\infty}^{+\infty} dt \{e^{i(-\omega) t}\}^* \langle \beta_{-a}^*(0)\beta_{-b}^*(t)\rangle=\int_{-\infty}^{+\infty} dt e^{i\omega t} \langle \beta_{a}(0)\beta_{b}(t)\rangle,\\
S_{b,a}(-\omega) & =\int_{-\infty}^{+\infty} dt e^{i(-\omega) t}\langle \beta_{b}(0)\beta_{a}(t)\rangle
=\int_{+\infty}^{-\infty}d(-t) e^{i(-\omega)(-t)}\langle \beta_{b}(0)\beta_{a}(-t)\rangle
\\
=& \int_{-\infty}^{+\infty} dte^{i\omega t} \langle \beta_{b}(t)\beta_{a}(0)\rangle
= S_{a,b}(\omega).
\end{split}
\end{align}
In the second line of the above equation, we utilized $\beta_a(t)=\beta_{-a}(t)^*$, which is the qutrit equivalent of Eq.\eqref{beta_condition}. Additionally, in the third line, we substituted $t\rightarrow -t$, applied the property $\int_{a}^{b}f(x)dx=\int_{-a}^{-b}f(-x)d(-x)=\int_{-b}^{-a}f(-x)dx$, and took into account the stationary condition for the dephasing noise functions as follows:
\begin{align}\label{Stationary2}
\begin{split}
\langle\beta_a(0)\beta_b(-t)\rangle =\langle\beta_a(\tau)\beta_b(-t+\tau)\rangle= \langle\beta_a(t)\beta_b(0)\rangle.
\end{split}
\end{align}
It is evident that the right-hand sides of all three functions in Eq.\eqref{C.2} are identical.
\subsubsection{Qudit with Reduced Weyl Hamiltonian}\label{Qudit-s(w)symproperties}
In this part, we find the qudit version of Eq.\eqref{general_S(w,t)} specifically for the Reduced Weyl case. To demonstrate this, we consider Eq.\eqref{noisespectrum_quoct_I} and assume the complex noise polyspectra as $\tilde{S}(\omega)= R(\omega)+iI(\omega)$. As a result, we arrive at the following:
\begin{align}\label{quoct-I-noisespectrum}
\begin{split}
\tilde{S}(\omega)= \int_{-\infty}^{\infty} dt e^{i\omega t}  
\langle A(0)A(t)\rangle
= \Big(
\int_{-\infty}^{\infty} dt e^{i(-\omega) t}  
\langle A(0)A(t)\rangle \Big)^*
= \tilde{S}^*(-\omega)
\end{split}
\end{align}
\subsubsection{Antimony quoct with Weyl Hamiltonian}\label{Antimony-S(w)symproperties}
In this section, we derive a particular case of Eq.\eqref{general_S(w,t)} applicable to a quoct with general noise. The specific equation takes the following form:
\begin{align}
\begin{split}
S_{ab}^{a'b'}(\omega) =& S^{\ast-\tilde{a},-\tilde{b}}_{-a,-b}(-\omega)= S^{\tilde{\A}_{\beta}}_{\A_{\beta}}(-\omega)\xi^{ab + \tilde{a}\tilde{b}}.\\
\end{split}   
\end{align}
The undetermined noise polyspectra for the Antimony quoct are denoted as $\tilde{S}_{i\tilde{i}}(\omega)$, and they are connected to the primary quoct noise polyspectra in the following manner:
\begin{align}\label{Antimony_noise_spectra}
\begin{split}
S_{ab}^{a'b'}(\omega) =& \sum_{\substack{i,\tilde{i}=0,1,2 \\ pqq'\in S\\ \tilde{p}\tilde{q}\tilde{q}'\in S}} \tilde{\beta}^{pqq'}_{iab}
\tilde{\beta}^{\tilde{p}\tilde{q}\tilde{q}'}_{\tilde{i}\tilde{a}\tilde{b}} \tilde{S}_{i\tilde{i}}(\omega),\:\:\:\:\:\:
\tilde{S}_{i\tilde{i}}(\omega)
= \int_{-\infty}^\infty dt 
\langle B_i(0)B_{\tilde{i}}(t)\rangle e^{i\omega t}.
\end{split}    
\end{align}
The following relationship can be derived for the Antimony quoct noise polyspectra:
\begin{align}\label{Sym2}
\begin{split}
\tilde{S}_{i\tilde{i}}(-\omega)=\tilde{S}_{\tilde{i}i}(\omega),
\end{split}    
\end{align}
and the proof is as follows:
\begin{align}
\begin{split}
\tilde{S}_{i\tilde{i}}(-\omega)=&
\int_{-\infty}^\infty dt 
\langle B_i(0)B_{\tilde{i}}(t)\rangle e^{i(-\omega) t}
=
\int_{+\infty}^{-\infty} d(-t) 
\langle B_i(0)B_{\tilde{i}}(-t)\rangle e^{i(-\omega)(-t)}\\
=& \int_{-\infty}^{+\infty} dt 
\langle B_i(t)B_{\tilde{i}}(0)\rangle e^{i\omega t}
= \tilde{S}_{\tilde{i}i}(\omega),
\end{split}    
\end{align}
To find the above expression, we employed the transformation $t\rightarrow -t$ and considered the stationary condition (Eq.\eqref{stationary}) for the $B_i(t)$ dephasing noise function.
Considering that the noise polyspectra $B_i(t)$ for the Antimony quoct are real functions (Eq.\eqref{H^W(t)}), the following result can be deduced:
\begin{align}\label{SrealSym}
\begin{split}
\tilde{S}^*_{i\tilde{i}}(\omega)=
\tilde{S}_{i\tilde{i}}(-\omega),
\end{split}    
\end{align}
and the proof is as follows:
\begin{align}
\begin{split}
\tilde{S}^*_{i\tilde{i}}(\omega)=
\int_{-\infty}^{+\infty} dt 
\langle B^*_i(0)B^*_{\tilde{i}}(t)\rangle e^{-i\omega t}
=& \int_{-\infty}^{+\infty} dt 
\langle B_i(0)B_{\tilde{i}}(t)\rangle e^{i(-\omega) t}
= \tilde{S}_{i\tilde{i}}(-\omega),
\end{split}    
\end{align}
Please note that Eq.\eqref{Sym2} is only valid for $i,\:\tilde{i}=0,1$, since for the other indices, the corresponding $B_i(t)$ are known functions of time and do not possess any stochastic and stationary distribution. 

\subsection{Real compositions of noise polyspectra}\label{RealFuncs}
\subsubsection{Qutrit with Reduced Weyl Hamiltonian}\label{RealFuncs-qutrit}
In this section, we demonstrate that the nine complex noise spectra of the qutrit, denoted as $S_{a,b}(\omega)$ with $a, b \in {0, \pm 1}$, can be expressed using only four real functions, which are even or odd with respect to the variable $\omega$.\\
Firstly, it is worth noting that the noise spectra $S_{a,b}(\omega)$ in Eq.\eqref{1.4} for the indices $a=0$ or $b=0$ can be assumed to be zero. This is attributed to the fact that they correspond to the Fourier transforms of the terms $\langle \beta_{0}(0)\beta_{0}(t)\rangle$, $\langle \beta_{0}(0)\beta_{1,2}(t)\rangle$, or $\langle \beta_{1,2}(0)\beta_{0}(t)\rangle$. For a qutrit, the term involving $\beta_{0}(t)$ in the noise-qudit effective Hamiltonian of Eq.\eqref{HWRW} is $\beta_{0}(t)Z^0$. As $Z^0=I$, this term uniformly shifts all energy eigenvalues of the noisy qutrit by $\beta_{0}(t)$ at any given time. Consequently, its average value ($\langle \beta_{0}(0)\beta_{0}(t)\rangle$) appears to have no influence on the qutrit dynamics, rendering it undetectable during qutrit's noise spectroscopy.
Therefore, the nonzero qutrit noise spectra are given as follows:
\begin{align}
\begin{split}
S_{1,1}(\omega) &= R_1(\omega) + i I_1(\omega),\:
S_{-1,-1}(\omega) = R'_1(\omega) + i I'_1(\omega),\\
S_{1,-1}(\omega) &= R_2(\omega) + i I_2(\omega),\: S_{-1,1}(\omega) = R'_2(\omega) + i I'_2(\omega).
\end{split}
\end{align}
Referring to appendix \ref{apdx.3.2}, we have $S_{a,b}(\omega)=S_{b,a}(-\omega)$, leading to $S_{1,1}(\omega) = S_{1,1}(-\omega)$. Consequently, the following functions exhibit symmetry/evenness with respect to $\omega$:
\begin{equation}\label{even}
R_1(\omega)=R_1(-\omega),\:I_1(\omega)=I_1(-\omega)
\end{equation}
Furthermore, the relation $S_{a,b}(\omega)=S_{-a,-b}(-\omega)$ from appendix \ref{apdx.3.2} implies that $S_{-1,-1}(\omega)=S_{1,1}(-\omega)$, which can be expressed as:
\begin{equation}
R'_1(\omega)=R_1(-\omega), \: I'_1(\omega)= - I_1(-\omega).
\end{equation}
Now using Eq.\eqref{even} we find: 
\begin{equation}
R'_1(\omega)= R_1(\omega),\: I'_1(\omega)= -I_1(\omega).
\end{equation}
On the other hand, the appendix \ref{apdx.3.2} yields $S_{1,-1}(\omega) = S_{-1,1}(-\omega) = S_{-1,1}^*(-\omega)$ that results in the following: 
\begin{align}
\begin{split}
R_2(\omega) + i I_2(\omega) = R'_2(-\omega) + i I'_2(-\omega) = R'_2(-\omega) - i I'_2(-\omega)\\
\hookrightarrow 
R_2(\omega) = R'_2(-\omega),\: I'_2(-\omega) = -I'_2(-\omega) = 0.\\
\end{split}    
\end{align}
Next, we decompose the arbitrary $R_2(\omega)$ function in terms of a symmetric and asymmetric function as following:
\begin{align}
\begin{split}
R_2(\omega) &= E(\omega) + D(\omega),\\
E(\omega) &= \frac{1}{2}(R_2(\omega) + R_2(-\omega)),\\
D(\omega) &= \frac{1}{2}(R_2(\omega) - R_2(-\omega)). 
\end{split}
\end{align}
As a result, all complex noise spectra $S_{a,b}(\omega)$ would depend on four real functions $R_1(\omega),I_1(\omega),E(\omega)$, and $D(\omega)$ where all are symmetric except $D(\omega)$. They could be noted as following:
\begin{align}\label{qutrit-variables-apx}
\begin{split}
x_0 (\omega) &= R_1(\omega) = x_0 (-\omega),\\
x_1(\omega)  &= I_1(\omega) = x_1 (-\omega),\\
x_2(\omega)  &= E(\omega)   = x_2 (-\omega),\\
x_3(\omega)  &= D(\omega)   = -x_3 (-\omega),\\
\end{split}
\end{align}
Here is the qutrit complex noise spectra in terms of these four even or odd functions:
\begin{align}
\begin{split}
S_{1,1}(\omega)  &= R_1(\omega)+i I_1(\omega),\\
S_{-1,-1}(\omega) &= R_1(\omega)-i I_1(\omega),\\ 
S_{1,-1}(\omega) &= E(\omega) + D(\omega),\\ 
S_{-1,1}(\omega) &= E(\omega) - D(\omega).
\end{split}
\end{align}
%%%%%%%%%%%%%%%%%%%%%%%%%%%
\subsubsection{Qudit with Reduced Weyl Hamiltonian}\label{RealFuncs-qudit}
Here we show that the complex noise spectra surrounding a qudit in Reduced Weyl basis rely on two real functions.
Based on Eq.\eqref{noisespectrum_quoct_I}, the Z-type dephasing noise spectrum of a qudit is denoted as $\tilde{S}(\omega)$. By employing Eq.\eqref{quoct-I-noisespectrum} and assuming the real and imaginary parts of the noise spectrum, the following result is found:
\begin{align}
\begin{split}
\tilde{S}(\omega) =& R(\omega)+iI(\omega)
= \tilde{S}^*(-\omega)
= R(-\omega) - iI(-\omega)\\
\hookrightarrow & 
R(\omega) = R(-\omega),\:
I(\omega) = - I(-\omega).
\end{split}
\end{align}
So the independent real and imaginary parts of the noise spectrum, i.e. $R(\omega)$ and $I(\omega)$, are even and odd functions, respectively.
%%%%%%%%%%%%%%%%%%%%%%%%%%%
\subsubsection{Antimony Quoct with Weyl Hamiltonian}\label{RealFuncs-Antimonyquoct}
In this section, we demonstrate that the complex noise spectra of the Antimony quoct, denoted as $\tilde{S}_{i,\tilde{i}}(\omega)$, can be expressed using four real functions that are symmetric/even or asymmetric/odd with respect to the variable $\omega$.
In the following we find all possible self and cross correlations in terms of $\tilde{S}_{i,\tilde{i}}(\omega)$. First, we find the self correlation related to the indices $i,\tilde{i}=2$:
\begin{align}\label{Sii-1}
\begin{split}
%\tilde{S}_{22}(\omega) &= \int_{-\infty}^{+\infty} dt e^{I\omega t}\langle \cos(\omega_0^\prime \times 0)\cos(\omega_0^\prime t) \rangle = 
%\int_{-\infty}^{+\infty} dt e^{I\omega t} \cos(\omega_0^\prime t) =  
%\pi \delta(\omega \pm \omega_0^\prime)
%,\\
\tilde{S}_{22}(\omega) &= \int_{-\infty}^{+\infty} dt e^{I\omega t}\langle 1 \rangle = 
2\pi \delta(\omega) \equiv 0
%,\\
%\tilde{S}_{23}(\omega) &= \int_{-\infty}^{+\infty} dt e^{I\omega t}\langle \cos(\omega_0^\prime \times 0)\times 1 \rangle = 
%2\pi \delta(\omega)
%,\\
%\tilde{S}_{32}(\omega) &= \int_{-\infty}^{+\infty} dt e^{I\omega t}\langle 1 \times \cos(\omega_0^\prime t) \rangle = 
%\pi \delta(\omega \pm \omega'_0)
%,
\end{split}
\end{align}
Please note that we will ignore the above term by equating it to zero. This is because the delta function is nonzero only in the very close vicinity of the zero frequency, and in our simulations, we will not consider frequencies that are extremely close to zero.\\
The cross correlations with indices $i$ or $\Tilde{i}=2$ 
%in $\{2\}$ 
that are related to the known function $B_2=1$, and indices $i$ or $\Tilde{i} \in \{0,1\}$, that are related to the unknown noise functions $B_{j}(t)=A(t),Q(t);\:j=0,1$ are as follows:
\begin{align}\label{Sii-2}
\begin{split}
\tilde{S}_{j2}(\omega) = \tilde{S}_{2j}(\omega) &= \int_{-\infty}^{+\infty} dt e^{I\omega t}\langle B_{j}(0) \times 1 \rangle = 
\int_{-\infty}^{+\infty} dt e^{I\omega t}\langle 1\times B_{j}(t) \rangle = 0,
\end{split}
\end{align}
In the above equation, we utilized $\langle B_j(t)\rangle=\langle B_j(0)\rangle=0$, which is due to assuming a zero mean Gaussian distribution for these stochastic noisy variables. Eqs.\eqref{Sii-1} and \eqref{Sii-2} show that
some correlations that are related to  $i,\tilde{i}=2
%,3
$, 
%where 2 in this notation could be symbolically substituted by -1
are known or zero. This is due to the fact that these correlations involve Fourier transforms of constant functions yielding delta function centered at zero frequency, or incorporate zero terms such as 
%$B_0\langle A(t)\rangle$ or $B_0\langle Q(t)\rangle$, 
$\langle A(0)\rangle$ or $\langle Q(t)\rangle$,
which is a result of assuming zero-mean Gaussian distribution for these noisy variables. \\
Next, we list the rest of unknown and nonzero Antimony quoct noise polyspectra ($\tilde{S}_{i,\tilde{i}}(\omega),\:i,\tilde{i}=0,1$) in terms of their Fourier Transforms (F.T.) and their real and imaginary components as follows:
\begin{align}
\begin{split}
\tilde{S}_{00}(\omega) &= F.T.(\langle A(0)A(t) \rangle) =  
R_0(\omega) + i I_0(\omega),\\
\tilde{S}_{11}(\omega) &= 
F.T.(\langle Q(0)Q(t) \rangle) =
R'_0(\omega) + i I'_0(\omega),\\
\tilde{S}_{01}(\omega) &= 
F.T.(\langle A(0)Q(t) \rangle) = 
R_1(\omega) + i I_1(\omega),\\ \tilde{S}_{10}(\omega) &= 
F.T.(\langle Q(0)A(t) \rangle) =
R'_1(\omega) + i I'_1(\omega),
\end{split}
\end{align}
Initially, as expressed above, we encounter eight unknown functions representing the real and imaginary parts of the noise spectra. 
For the specific case of the Antimony quoct, 
%we first assess $\tilde{S}_{00}(\omega)$ that corresponds to the Fourier transform of $\langle A(0)A(t) \rangle$, and $\tilde{S}_{11}(\omega)$ that corresponds to the Fourier transform of $\langle Q(0)Q(t) \rangle$. 
we utilize Eqs.\eqref{Sym2}, and deduce that $\tilde{S}_{00}(\omega) = \tilde{S}_{00}(-\omega)$,\: $\tilde{S}_{11}(\omega) = \tilde{S}_{11}(-\omega)$, and $\tilde{S}_{01}(\omega) = \tilde{S}_{10}(-\omega)$, leading to:
\begin{align}\label{even-2}
\begin{split}
R_0(\omega)=R_0(-\omega),\:I_0(\omega)=I_0(-\omega),\\
R'_0(\omega)=R'_0(-\omega),\:I'_0(\omega)=I'_0(-\omega)\\
R'_1(\omega)= R_1(-\omega),\: I'_1(\omega)= I_1(-\omega).
\end{split}
\end{align}
According to the above equations, the functions $R_0(\omega), I_0(\omega),R'_0(\omega)$ and $I'_0(\omega)$ are symmetric/even with respect to $\omega$. In addition, the two functions $R'_1(\omega)$ and $I'_1(\omega)$ depend on the functions $R_1(\omega)$ and $I_1(\omega)$.
From Eq.\eqref{SrealSym} we find $S^*_{00}(\omega)=S_{00}(-\omega)$ , $S^*_{11}(\omega)=S_{11}(-\omega)$, and $S^*_{01}(\omega)=S_{10}(-\omega)$, resulting in the following:
\begin{align}\label{R1I1}
\begin{split}
R_0(\omega)-i I_0(\omega)=
R_0(-\omega)+i I_0(-\omega),\\
R'_0(\omega)-i I'_0(\omega)=
R'_0(-\omega)+i I'_0(-\omega),\\
R_1(\omega)-i I_1(\omega)=
R_1(-\omega)+i I_1(-\omega),\\
\end{split}
\end{align}
Upon combining Eqs.\eqref{even-2} and \eqref{R1I1}, we deduce that $I_0(\omega) = I'_0(\omega) = 0$. Furthermore, it becomes apparent that the functions $R_1(\omega)$ and $I_1(\omega)$ are even and odd, respectively. Consequently, all the unknown complex noise polyspectra of the Antimony quoct, denoted as $\tilde{S}_{i,\tilde{i}}(\omega)$, depend on four real functions as follows:
\begin{align}\label{AntimonyRealFunctions}
\begin{split}
R_0(\omega)=R_0(-\omega),\:\:R'_0(\omega)=
R'_0(-\omega),\\
R_1(\omega)=R_1(-\omega),\:\: I_1(\omega)=
-I_1(-\omega)    
\end{split}
\end{align}
where the first three variables are symmetric/even and the last function is anti-symmetric/odd. %(Eq.\eqref{Antimony-unknownPolyspectra}).
%%%%%%%%%%%
\subsection{Map of frequency range to positive frequencies}\label{apdx.3.4}
In our noise spectroscopy formalism, we need to map the whole frequency range to the positive frequency range. To reach this goal, we use the following formula:
\begin{align}\label{Integrals}
\begin{split}
\int_a^bf(x)dx=\int_{-a}^{-b}f(-x)(-dx)=\int_{-b}^{-a}f(-x)dx, 
\end{split}
\end{align}
That yields: 
\begin{align}\label{full-half-freq-range}
\begin{split}
\int_{-\Omega}^{\Omega}g(\omega)d\omega =&
\int_{-\Omega}^{0^-}
g(\omega)d(\omega)
+\int_{0}^{\Omega}g(\omega)
d\omega
=\int_{\Omega}^{0^+}
g(-\omega)d(-\omega)
+\int_{0}^{\Omega}g(\omega)
d\omega \\
& =\int_{0^+}^{\Omega}
g(-\omega)d\omega + \int_{0}^{\Omega}g(\omega)d\omega = 
g(0)d\omega + 
\int_{0^+}^{\Omega}
(g(-\omega)+g(\omega))
d\omega 
\\
\sum_{k=-N}^{N}g(rk\omega_0)    
& =\sum_{k=-N}^{-1}g(rk\omega_0) 
+g(0)
+\sum_{k=1}^{N}g(rk\omega_0) 
= g(0) + \sum_{k=1}^{N}(g(-rk\omega_0)+g(rk\omega_0))
\end{split}
\end{align}
where $N=\lfloor\Omega/\omega_0\rfloor$. In the second line of the equation above, when $d\omega$ approaches zero, if $g(0)\rightarrow \infty$, the term $g(0)d\omega$ converges to a finite value and cannot be disregarded. However, in the typical scenario where $g(0)$ remains finite, the term $g(0)d\omega$ can be neglected. These expressions present the symbolic representation of the primary integral and summation components of Eq.\eqref{qudit}.

\subsection{Antimony quoct Hamiltonian in Weyl basis}\label{AntimonyQuoctHWeyl}
This section derives the Weyl basis representation of the effective Hamiltonian of the Antimony quoct introduced in Eq.\eqref{AntimonyquoctH}. First, we find the spin magnetic moment matrices of a qudit ($I_{x/y/z}$) in the computational basis as follows:
\begin{align}\label{I-ComputationalBasis}
\begin{split}
I_z =&\sum_{p=1}^{d} I'_{z,p} \ket{p}\bra{p},\: 
I'_{z,p} = \hbar (I+1-p)\\
=& \sum_{p=0}^{d-1} I_{z,p} \ket{p}\bra{p},\: 
I_{z,p} = \hbar (I-p)\\
I_{x}=&\sum_{p,q=1}^{d} I'_{x,pq} \ket{p}\bra{q},\:I'_{x,pq} = \frac{\hbar}{2} (\delta_{p,q+1}+\delta_{p+1,q})\sqrt{(I+1)(p+q-1)-pq}\\
=& \sum_{p,q=0}^{d-1} I_{x,pq} \ket{p}\bra{q},\:I_{x,pq} = \frac{\hbar}{2} (\delta_{p,q+1}+\delta_{p+1,q})\sqrt{(I+1)(p+q+1)-(p+1)(q+1)}\\
I_{y}=&\sum_{p,q=1}^{d} I'_{y,pq} \ket{p}\bra{q},\:I'_{y,pq}=\frac{i\hbar}{2} (\delta_{p,q+1}-\delta_{p+1,q})\sqrt{(I+1)(p+q-1)-pq}\\
=&\sum_{p,q=0}^{d-1} I_{y,pq} \ket{p}\bra{q},\:I_{y,pq}=\frac{i\hbar}{2} (\delta_{p,q+1}-\delta_{p+1,q})\sqrt{(I+1)(p+q+1)-(p+1)(q+1)}
\end{split}    
\end{align}
where $I=\frac{d-1}{2}$ for a spin qudit and $I=\frac{7}{2}$ for the Antimony quoct with $d=8$.
In the above transformation, after shifting the indices by $1$, we considered the new state numbers intrinsically shifted by $-1$ as $\{\ket{p},p=0,...,d-1\}$.
Using Eq.\eqref{ij} we have:
\begin{align}\label{IcomponentsWeylbasis}
\begin{split}
I_z =& \sum_{p=0}^{d-1} I_{z,p} \ket{p}\bra{p} = \frac{1}{d}\sum_{ap} \xi^{-ap}I_{z,p}Z^a,\\
I_{y}=& \sum_{pq=0}^{d-1} I_{y,pq} \ket{p}\bra{q} =\frac{1}{d}\sum_{pqab} \xi^{-ap}\delta_{p,q+b}I_{y,pq}Z^aX^b
= \frac{1}{d}\sum_{pqab} \xi^{-ap}I_{y,pq}Z^aX^{p-q},\\
I_{x}=& \sum_{pq=0}^{d-1} I_{x,pq} \ket{p}\bra{q} =\frac{1}{d}\sum_{pqab} \xi^{-ap}\delta_{p,q+b}I_{x,pq}Z^aX^b
= \frac{1}{d}\sum_{pqab} \xi^{-ap}I_{x,pq}Z^aX^{p-q} \\
I_{x}^2 =& 
\sum_{pqp'q'=0}^{d-1} I_{x,pq}I_{x,p'q'} \ket{p}\bra{q}\ket{p'}\bra{q'}
= \sum_{pqp'q'=0}^{d-1} I_{x,pq}I_{x,p'q'} \delta_{q,p'}\ket{p}\bra{q'} \\
=& \sum_{pqq'=0}^{d-1} I_{x,pq} I_{x,qq'} \ket{p}\bra{q'} 
= \frac{1}{d}\sum_{pqq'ab} \xi^{-ap}\delta_{p,q'+b}I_{x,pq}I_{x,qq'}Z^aX^b \\
=& \frac{1}{d}\sum_{pqq'a} \xi^{-ap}I_{x,pq}I_{x,qq'}Z^a X^{p-q'},
\end{split}    
\end{align}
where $I_{z,p},\:I_{y,pq},\:I_{x,pq}\:$ are defined in Eq.\eqref{I-ComputationalBasis}. Now the Antimony Hamiltonian of Eq.\eqref{AntimonyquoctH} could be found in the complete Weyl basis as follows:
\begin{align}\label{H(t)1}
\begin{split}
&H(t) 
=(\gamma_n \tilde{B}_0 \pm \frac{1}{2}A(t)) I_z + Q(t) I_x^2 \\
%+ \gamma_n \tilde{B}_0 \cos(2\pi f t) I_y\\
=& (\gamma_n \tilde{B}_0 \pm \frac{1}{2}A(t))
\Big(\frac{1}{d}\sum_{abp} \xi^{-ap}I_{z,p}\delta_{0,b}Z^aX^b\Big)\\
+& Q(t) 
\Big(\frac{1}{d}\sum_{pqq'ab} \xi^{-ap}\delta_{p,q'+b}I_{x,pq}I_{x,qq'}Z^aX^b\Big) \\ 
%+ \gamma_n \tilde{B}_0 \cos(2\pi f t) 
%\Big(\frac{1}{d}\sum_{pqab} \xi^{-ap}\delta_{p,q+b}I_{y,pq}Z^aX^b\Big) \\
=& \sum_{ab}\Big(\Big[
\frac{1}{d}\gamma_n \tilde{B}_0\sum_{p} \xi^{-ap}I_{z,p}\delta_{b0}
%+
%\frac{1}{d}\gamma_n \tilde{B}_0 \cos(2\pi f t)
%\sum_{pq} \xi^{-ap}\delta_{p,q+b}I_{y,pq}
\Big]\\
+&\Big[
\pm \frac{1}{2d}\sum_{p} \xi^{-ap}I_{z,p}\delta_{0,b}
\Big]A(t)
+\Big[
\frac{1}{d}\sum_{pqq'} \xi^{-ap}\delta_{p,q'+b}I_{x,pq}I_{x,qq'}
\Big]Q(t) \Big) Z^a X^b\\ 
=& \sum_{ab}
\Big(
\Big[
\pm \frac{1}{2d^3}\sum_{pqq'} \xi^{-ap}I_{z,p}\delta_{0,b}
\Big]A(t)
+\Big[
\frac{1}{d}\sum_{pqq'} \xi^{-ap}\delta_{p,q'+b}I_{x,pq}I_{x,qq'}
\Big]Q(t) \\
%+& \Big[
%\frac{1}{d^2}\gamma_n \tilde{B}_0
%\sum_{pqq'} \xi^{-ap}\delta_{p,q+b}I_{y,pq}
%\Big]  \cos(2\pi f t) 
+& \Big[
\frac{1}{d^3}\gamma_n \tilde{B}_0\sum_{pqq'} \xi^{-ap}I_{z,p}\delta_{b0}
\Big]\times 1
\Big) Z^a X^b\\ 
=& \sum_{pqq'ab}\Big(B_0 \tilde{\beta}^{pqq'}_{0ab}
+ B_{1}(t) \tilde{\beta}^{pqq'}_{1ab} 
+ B_{2}(t) \tilde{\beta}^{pqq'}_{2ab} 
%+ B_{3}(t) \tilde{\beta}^{pqq'}_{3ab}
\Big) Z^a X^b \\
&= \sum_{\substack{i=0,1,2
%,3 
\\ pqq'ab\in S}} 
B_i(t)\tilde{\beta}^{pqq'}_{iab} Z^a X^b
= \sum_{ab}\beta_{ab}(t) Z^a X^b,
\end{split}    
\end{align}
where, 
\begin{align}\label{H(t)2}
\begin{split}
\beta_{ab}(t)&= \sum_{\substack{i=0,1,2
%,3 
\\ pqq'\in S}}B_i(t) \tilde{\beta}^{pqq'}_{iab}\\
B_{0}(t),
& B_{1}(t), B_{2}
%(t), B_{3}
=
A(t),Q(t)
%, \rm cos(2 \pi f t)
, 1\\
\tilde{\beta}^{pqq'}_{0ab}=&\pm \frac{1}{2d^3}\xi^{-ap} I_{z,p} \delta_{b,0}\\
\tilde{\beta}^{pqq'}_{1ab}=& \frac{1}{d}\xi^{-ap} 
I_{x,pq}I_{x,qq'}\delta_{p,q+b}\\
%\tilde{\beta}^{pqq'}_{2ab}=& \frac{1}{d^2}\gamma_n \tilde{B}_0  
%\xi^{-ap} I_{y,pq}  \delta_{p,q+b}\\
\tilde{\beta}^{pqq'}_{2ab}=& \frac{1}{d^3} 
\gamma_n \tilde{B}_0 \xi^{-ap} I_{z,p} \delta_{b,0} 
\end{split}    
\end{align}
\subsection{Characteristics of Weyl coefficient function}
In this section, we demonstrate that the relation $\Xi_{i,\tilde{i}}(m,n,\omega,t_r) = \Xi_{i,\tilde{i}}(m,n,-\omega,t_r)$ holds true for this Weyl coefficient function, as explained below:
\begin{align}\label{Xi-negative}
\begin{split}
%\Xi_{2,1}(m,n,-\omega,t_r^A) = &
&\Xi_{i\tilde{i}}^{mn}
(-\omega,t_r) = \\
&\sum_{\substack{ab\tilde{a}\tilde{b}\\
a'b'\tilde{a}'\tilde{b}'\\
\in S}}
\sum_{\substack{pqq'\in S\\ \tilde{p}\tilde{q}\tilde{q}'\in S}}
\tilde{\beta}^{pqq'}_{iab}
\tilde{\beta}^{\tilde{p}\tilde{q}\tilde{q}'}_{\tilde{i}\tilde{a}\tilde{b}}
(\prescript{ind}{\tilde{a}\tilde{b}}{\lambda}^{\tilde{a}'\tilde{b}'}_{a'b'})
F^{ab}_{a'b'}(-\omega,t_r) F^{*-\tilde{a},-\tilde{b}}_{-\tilde{a}',-\tilde{b}'}(-\omega,t_r)\\
&=\sum_{\substack{ab\tilde{a}\tilde{b}\\
a'b'\tilde{a}'\tilde{b}'\\
\in S}}
\sum_{\substack{pqq'\in S\\ \tilde{p}\tilde{q}\tilde{q}'\in S}}
\tilde{\beta}^{pqq'}_{iab}
\tilde{\beta}^{\tilde{p}\tilde{q}\tilde{q}'}_{\tilde{i}\tilde{a}\tilde{b}}
(\prescript{ind}{\tilde{a}\tilde{b}}{\lambda}^{\tilde{a}'\tilde{b}'}_{a'b'})
F^{*-a,-b}_{-a',-b'}(\omega,t_r) \xi^{ab-a'b'} F^{\tilde{a},\tilde{b}}_{\tilde{a}',\tilde{b}'}(\omega,t_r) \xi^{-\tilde{a}\tilde{b}+\tilde{a}'\tilde{b}'}\\
&= \sum_{\substack{
a'b'\tilde{a}'\tilde{b}'\\
ab\tilde{a}\tilde{b}\\
\in S}}
\sum_{\substack{pqq'\in S\\ \tilde{p}\tilde{q}\tilde{q}'\in S}}
\tilde{\beta}^{pqq'}_{i\tilde{a}\tilde{b}}
\tilde{\beta}
^{\tilde{p}\tilde{q}\tilde{q}'}
_{\tilde{i}ab}
(\prescript{ind}{\tilde{a}\tilde{b}}{\lambda}^{\tilde{a}'\tilde{b}'}_{a'b'})
F^{*-a,-b}_{-a',-b'}(\omega,t_r) F^{\tilde{a},\tilde{b}}_{\tilde{a}',\tilde{b}'}(\omega,t_r).
= \Xi_{i\tilde{i}}^{mn}
(\omega,t_r) 
\end{split}
\end{align}
To arrive at the final line of the above equation, we used Eq.\eqref{etalambdamnfixed-quoct} in the following manner:
\begin{align}\label{XiPMomega}
\begin{split}
\prescript{ind}{\tilde{a}\tilde{b}}{\lambda}^{\tilde{a}'\tilde{b}'}_{a'b'}
&\equiv
\prescript{mnp_0q_0}{\tilde{a}\tilde{b}}{\lambda}^{\tilde{a}'\tilde{b}'}_{a'b'} \\
&= 
8O_{mn} V_{p_0,q_0}  
\delta_{-p_0-m,
a'+\tilde{a}'}
\delta_{-q_0-n,
b'+\tilde{b}'} \times
\\
&
(1-\xi^{na'-mb'})
(1-\xi^{n\tilde{a}'-m\tilde{b}'})
\xi^{m(n + b'+\tilde{b}')+
(a'+\tilde{a}')(b'+\tilde{b}')}\xi^{-\tilde{a}' b'} \xi^{\tilde{a}\tilde{b}
-\tilde{a}'\tilde{b}'} ,\\
\prescript{mnp_0q_0}{\tilde{a}\tilde{b}}{\lambda}^{\tilde{a}'\tilde{b}'}_{a'b'} \:
&\xi^{ab-a'b'} 
\xi^{-\tilde{a}\tilde{b}+\tilde{a}'\tilde{b}'} =\\
& 
8O_{mn} V_{p_0,q_0}  
\delta_{-p_0-m,
a'+\tilde{a}'}
\delta_{-q_0-n,
b'+\tilde{b}'} \times
\\
&
(1-\xi^{na'-mb'})
(1-\xi^{n\tilde{a}'-m\tilde{b}'})
\xi^{m(n + b'+\tilde{b}')+
(a'+\tilde{a}')(b'+\tilde{b}')}\xi^{-\tilde{a}' b'} \xi^{\tilde{a}\tilde{b}
-\tilde{a}'\tilde{b}'}\times\\
& \xi^{ab-a'b'} 
\xi^{-\tilde{a}\tilde{b}+\tilde{a}'\tilde{b}'}\\
&= 
8O_{mn} V_{p_0,q_0}  
\delta_{-p_0-m,
a'+\tilde{a}'}
\delta_{-q_0-n,
b'+\tilde{b}'} \times
\\
&
(1-\xi^{na'-mb'})
(1-\xi^{n\tilde{a}'-m\tilde{b}'})
\xi^{m(n + b'+\tilde{b}')+
(a'+\tilde{a}')(b'+\tilde{b}')}
\xi^{-\tilde{a}' b'} 
\xi^{ab-a'b'} \\
& = \: \prescript{mnp_0q_0}{\tilde{a}\tilde{b}}{\lambda}^{a'b'}_{\tilde{a}'\tilde{b}'} ,\\
\end{split}
\end{align}
where $ind = (m,n,p_0,q_0)$.
\subsection{Impact of Hamiltonian Hermiticity on noise and switching functions}\label{apdx.3.6}
In our noise spectroscopy formalism, we made the assumption that the effective Hamiltonian of the qudit, noises, and spectroscopy pulses is Hermitian. This is equivalent to asserting that the evolution of a qudit, under the influence of noises and subjected to pulses, is unitary. In other words, we consider the whole quantum system of the qudit, environmental noises, and spectroscopy pulses as one closed system.
Now we identify the constraints imposed by the Hermitian characteristic of the effective Hamiltonian on the noise function $\beta_{\mathscr{A}_{\beta}}(t)$ and the switching function $y_{\mathscr{A}_y}(t)$. 
The investigation begins by considering the general format of the effective \textit{\textbf{q}}udit and \textit{\textbf{n}}oise Hamiltonian, i.e. Eq.\eqref{HfullWeyldephasing}, which is reexpressed here as follows:
\begin{align}\label{}
\begin{split}
H^{\mathscr{A^n}_H}_{q,n}(t) =& \sum_{\mathscr{A}_{\sum}\in S} \beta_{\mathscr{A}_{\beta}}(t) Z^{\mathscr{A}^n_Z} X^{\mathscr{A}^n_X},
\end{split}
\end{align}
The conjugate of the above Hamiltonian is computed as follows, and it is expected to be equivalent to the original Hamiltonian itself.
\begin{align}\label{inds}
\begin{split}
H^{*\: \mathscr{A^n}_H}_{q,n}(t) =& \sum_{\mathscr{A}_{\sum}\in S} \beta^*_{\mathscr{A}_{\beta}}(t) X^{-\mathscr{A}^n_X}Z^{-\mathscr{A}^n_Z}\\
=& \sum_{\mathscr{A}_{\sum}\in S} \beta^*_{\mathscr{A}_{\beta}}(t) \xi^{-\mathscr{A}^n_Z \mathscr{A}^n_X} Z^{-\mathscr{A}^n_Z} X^{-\mathscr{A}^n_X}\\
=& \sum_{-\mathscr{A}_{\sum}\in S} \beta^*_{-\mathscr{A}_{\beta}}(t) \xi^{-\mathscr{A}^n_Z \mathscr{A}^n_X} Z^{\mathscr{A}^n_Z} X^{\mathscr{A}^n_X}\\
=& H^{\mathscr{A^n}_H}_{q,n}(t)
\end{split}
\end{align}
Considering the symmetric characteristic of the $S$ set expressed in Eq.\eqref{Sequivalents}, 
we find $-\mathscr{A}_{\sum} = \mathscr{A}_{\sum}$, that yields the following condition on the noise function:
\begin{align}\label{beta_condition-0}
\begin{split}
\beta_{\mathscr{A}_{\beta}}(t) = \xi^{-\mathscr{A}^n_Z \mathscr{A}^n_X} \beta^*_{-\mathscr{A}_{\beta}}(t). 
\end{split}
\end{align}
Now we provide the condition on the switching function that is imposed by the Hermitian constraint. The Hermitian effective Hamiltonian of the the qudit, noises, and the spectroscopy pulses is as follows:
\begin{align}\label{}
\begin{split}
H^{\mathscr{A}_H}_{q\leftarrow(n,p)}(t) =& P_{(i_k,j_k)}^{-1}\Big(  H^{\mathscr{A^n}_H}_{q,n}(t) \Big) P_{(i_k,j_k)} = H^{*\:\mathscr{A}_H}_{q\leftarrow(n,p)}(t)
,\\
=& \sum_{\mathscr{A}_{\sum}\in S} \beta_{\mathscr{A}_{\beta}}(t) y_{\mathscr{A}_y}(t) Z^{\mathscr{A}_Z} X^{\mathscr{A}_X}\\
=& \sum_{\mathscr{A}_{\sum}\in S} \beta^*_{\mathscr{A}_{\beta}}(t) y^*_{\mathscr{A}_y}(t) X^{-\mathscr{A}_X} Z^{-\mathscr{A}_Z}\\
=& \sum_{\mathscr{A}_{\sum}\in S} \beta^*_{\mathscr{A}_{\beta}}(t) y^*_{\mathscr{A}_y}(t) \xi^{-\mathscr{A}_Z \mathscr{A}_X} 
Z^{-\mathscr{A}_Z} X^{-\mathscr{A}_X}\\
=& \sum_{-\A_{\sum} \in S} \beta^*_{-\mathscr{A}_{\beta}}(t) y^*_{-\mathscr{A}_y}(t) \xi^{-\mathscr{A}_Z \mathscr{A}_X} 
Z^{\mathscr{A}_Z} X^{\mathscr{A}_X}\\
=& \sum_{-\A_{\sum}=\mathscr{A}_{\sum}\in S} \Big(\beta_{\mathscr{A}_{\beta}}(t)\xi^{\mathscr{A}^n_{Z} \mathscr{A}^n_{X}}\Big) y^*_{-\mathscr{A}_y}(t) \xi^{-\mathscr{A}_Z \mathscr{A}_X} 
Z^{\mathscr{A}_Z} X^{\mathscr{A}_X}
\end{split}
\end{align}
where we utilized Eq.\eqref{beta_condition-0}. Comparing the second and last lines of the above equation, we find the condition on the switching function as follows:
\begin{align}\label{yconjugate}
\begin{split}
y_{\mathscr{A}_y}(t) =  y^*_{-\mathscr{A}_y}(t) \xi^{\mathscr{A}^n_{Z} \mathscr{A}^n_{X}-\mathscr{A}_Z \mathscr{A}_X}.
\end{split}
\end{align}
For the qutrit and the qudit in Weyl basis, equations \eqref{beta_condition-0} and \eqref{yconjugate} take the following forms:
\begin{align}\label{beta_condition}
\begin{split}
Qutrit:\:\:\:\:\: \beta_{a}(t) =& \beta^*_{-a}(t),\:\:\:\:\:\:\:\:\:\:\:\:\:\:\:\:\:\:\:\:\:\:\:\:\:\:\:\:\: y_{a}(t) =  y^*_{-a}(t).
\\
Qudit:\:\:\:\:\: \beta_{ab}(t)=& \xi^{-ab} \beta^*_{-a,-b}(t),\:\:\:\:\:\:\:\:\:\: y^{ab}_{a'b'}(t) =  y^{*-a,-b}_{-a',-b'}(t) \xi^{ab-a'b'}. 
\end{split}
\end{align}
\subsection{Simplifying specific filter functions}\label{FilterFunctionFormats}
In this section, we convert the integral representation of filter functions for different versions into summation formats. The main definition of the Filter function is provided by Eq.\eqref{Fdefinition}. In our specific scenario, a sequence of $d$ symmetric resonance pulses is applied to a qudit. As a result, the switching function $y_{\A_y}(t)$ remains constant during each $r$-th interval, which is defined by the interval between two similar pulses with a time difference of $\frac{1}{d-1}\frac{T}{r}$ and an energy of $\hbar\omega_{i_hj_h}$. These pulses facilitate the transition of the qudit from energy level $\ket{i_h}$ to $\ket{j_h}$. We denote the constant switching function in interval $h$ as $y^{i_hj_h}_{\A_y}$.
Assuming a resonance pulse sequence represented as Eq.\eqref{symmetricsequence0} i.e. 
$\{ (P^{-1}_{(i_h,j_h)},P_{(i_h,j_h)}) \}_{h=0}^{d-1}$, where $P_{i_h,d}=P_{i_h,0}$ and $P_{d, j_h}=P_{0, j_h}$, we can derive the following relations:
\begin{align}\label{Fsimplified}
\begin{split}
&F_{\A_F}(\omega,t)  =
\int_{t_0}^{t_0+\frac{T}{r}} dt' y_{\A_y}(t') e^{I\omega t'} = 
\sum_{h=0}^{d-1} y^{i_h,j_h}_{\A_y} 
\int_{t_0+t^h_r}^{t_0+t^{h+1}_r} 
e^{I\omega t}dt \\
=& \frac{-I}{\omega}
\sum_{h=0}^{d-1} 
y^{i_h,j_h}_{\A_y} 
\Big(
e^{It^{h+1}_r\omega} - 
e^{It^{h}_r\omega}  \Big)
e^{It_{0}\omega} \\
= &\frac{-I}{\omega}
\sum_{h=0}^{d}  
\Big(
(1-\delta_{h,0})y^{i_{h-1},j_{h-1}}_{\A_y} - (1-\delta_{h,d})y^{i_h,j_h}_{\A_y} 
\Big) 
e^{I(t^{h}_r+t_0)\omega},\\
& t=t_0+\frac{T}{r},\:
t_0=0,\:
t^h_r = h\frac{1}{d}\frac{T}{r}.
\end{split}
\end{align}
where the terms including delta functions are added to maintain the following conditions:
\begin{equation}\label{FsimplifiedCondition}
y^{i_{_{-1}},j_{-1}}_{\A_y} = y^{i_d,j_{d}}_{\A_y}=0.
\end{equation}
Here we simplify the filter function at $t=M\frac{T}{r}$. 
Given that in our formalism the filter function  repeats in each time interval of 
$\Delta t =\frac{T}{r}$, we can have the following:
\begin{align}\label{FsimplifiedIntegral}
\begin{split}
F_{\A_F}(\omega,M\frac{T}{r})  &= \sum_{n=0}^{M-1}\int_{t^n_r}^{t^n_r+\frac{T}{r}} dt' y_{\A_y}(t') e^{I\omega t'} \\
& =  \frac{-I}{\omega}
\sum_{n=0}^{M-1}
\sum_{h=0}^{d-1} 
y^{i_h,j_h}_{\A_y} 
\Big(
e^{It^{h+1}_r\omega} - 
e^{It^{h}_r\omega}  \Big)
e^{I(t^n_rd)\omega} ,
\\
& =  \frac{-I}{\omega}
\sum_{n=0}^{M-1}
\sum_{h=0}^{d}  
\Big(
(1-\delta_{h,0})y^{i_{h-1},j_{h-1}}_{\A_y} - (1-\delta_{h,d})y^{i_h,j_h}_{\A_y} 
\Big) 
e^{I(t^{h}_r+t^n_rd)\omega},
\\
&t^q_r = q\frac{1}{d}\frac{T}{r}, I = \sqrt{-1}.
\end{split}
\end{align}
Now we calculate the filter function at $\omega=0$.
\begin{align}\label{Fsimplifiedw0}
\begin{split}
F_{\A_F}(0,t) 
& =
\int_{t_0}^{t_0+\frac{T}{r}} dt' y_{\A_y}(t') e^{0} = 
\sum_{h=0}^{d-1} 
y^{i_h,j_h}_{\A_y} 
\int_{t_0+t^h_r}^{t_0+t^{h+1}_r} dt = 
\sum_{h=0}^{d-1} y^{i_h,j_h}_{\A_y} 
\Big(t^{h+1}_r - t^{h}_r\Big)
\end{split}
\end{align}
To find the filter function at 
$t =M\frac{T}{r}$ and $\omega=0$, we combine the above equations to achieve the following:
\begin{align}\label{FsimplifiedIntegral-0}
\begin{split}
F_{\A_F}(0,M\frac{T}{r})  
&= \sum_{n=0}^{M-1}
\int_{t^n_r}^{t^n_r+\frac{T}{r}} 
dt' y_{\A_y}(t')  
 =
\sum_{n=0}^{M-1}
\sum_{h=0}^{d-1} 
y^{i_h,j_h}_{\A_y} 
\int_{t^n_r+t^h_r}^{t^n_r+t^{h+1}_r} dt \\
& =  
\sum_{n=0}^{M-1}
\sum_{h=0}^{d-1} 
y^{i_h,j_h}_{\A_y} 
\Big(t^{h+1}_r - t^{h}_r\Big)
 =  
M \sum_{h=0}^{d-1} 
y^{i_h,j_h}_{\A_y} 
\Big(t^{h+1}_r - t^{h}_r\Big)
\\
&t^q_r = q\frac{1}{d}\frac{T}{r}, I = \sqrt{-1}.
\end{split}
\end{align}

\end{document}